\documentclass{mn2e}
\usepackage{graphicx}

\newcommand{\gtsimeq}{\raisebox{-0.6ex}{$\,\stackrel
{\raisebox{-.2ex}{$\textstyle >$}}{\sim}\,$}}

\newcommand{\nb}{Nuclear Bulge}
\newcommand{\xrs}{X-ray source}
\newcommand{\xrss}{X-ray sources}

\newcommand{\chandra}{\textit{Chandra}}
\newcommand{\dr}{UKIDSS DR2 }

\begin{document}
\newfont{\ssq}{cmtt10} 

\title{The UKIDSS Galactic Plane Survey}
\author[Lucas et al.]{
P.W.Lucas$^1$\thanks{email: p.w.lucas@herts.ac.uk}, 
M.G.Hoare$^{2}$, A.Longmore$^3$, A.C.Schr\"{o}der$^{4}$, C.J.Davis$^5$,
\newauthor A.Adamson$^5$, R.M.Bandyopadhyay$^6$, R. de Grijs$^{7,8}$, M.Smith$^9$, 
A.Gosling$^{10}$ \newauthor
S.Mitchison$^9$, A.G\'asp\'ar$^{11}$, M.Coe$^{12}$, M.Tamura$^{13}$, Q.Parker$^{14}$, 
M.Irwin$^{15}$,\newauthor 
N.Hambly$^{16}$, J.Bryant$^{15}$, R.S.Collins$^{16}$, N.Cross$^{16}$, D.W. Evans$^{15}$, 
\newauthor
E.Gonzalez-Solares$^{15}$, S.Hodgkin$^{15}$, J.Lewis$^{15}$, M.Read$^{16}$, M.Riello$^{15}$, 
\newauthor
E.T.W.Sutorius$^{16}$, A. Lawrence$^{16}$, J.E.Drew$^{1}$, S.Dye$^{17}$, M.A.Thompson$^1$\\
$^1$Centre for Astrophysics Research, University of Hertfordshire, 
College Lane, Hatfield AL10 9AB.\\ 
$^2$School of Physics and Astronomy, University of Leeds, Leeds, LS2 9JT.\\
$^3$Astronomy Technology Centre, Royal Observatory, Blackford Hill, EH9 3HJ, Edinburgh.\\
$^4$Department of Physics \& Astronomy, University of Leicester, University Road, Leicester LE1 7RH.\\
$^5$Joint Astronomy Centre, 660 North A'ohoku Place, University Park, Hilo, HI 96720, USA.\\
$^6$Department of Astronomy, University of Florida, 211 Bryant Space Science Center, Gainesville, FL 32611.\\
$^7$Department of Physics \& Astronomy, The University of Sheffield, Hicks Building, Hounsfield Road, Sheffield S3 7RH.\\
$^8$National Astronomical Observatories, Chinese Academy of Sciences, 20A Datun Road, Choyang District, Beijing 100012,
P.R. China\\
$^9$Centre for Astrophysics \& Planetary Science, The University of Kent, Canterbury CT2 7NH.\\
$^10$Department of Physics, University of Oxford, Keble Road, Oxford, OX1 3RH.\\
$^{11}$Steward Observatory, University of Arizona,
933 N Cherry Ave., Tucson AZ 85721-0065, USA.\\
$^{12}$School of Physics and Astronomy, Southampton University, Highfield, Southampton SO17 1BJ.\\
$^{13}$National Astronomical Observatory of Japan, 2-21-1 Osawa, Mitaka, Tokyo 181-8588, Japan.\\
$^{14}$Department of Physics, Macquarie University, Sydney 2109, Australia.\\
$^{15}$Institute of Astronomy, University of Cambridge, Madingley Road, Cambridge CB3 0HA.\\
$^{16}$Scottish Universities' Physics Alliance (SUPA), Institute for Astronomy, School of Physics, University of 
Edinburgh, Royal\\ Observatory, Blackford Hill, Edinburgh EH9 3HJ.
$^{17}$University of Cardiff}

\maketitle
\label{firstpage}

\begin{abstract}
The UKIDSS Galactic Plane Survey (GPS) is one of the five near infrared 
Public Legacy Surveys that are being undertaken by the UKIDSS consortium, using
the Wide Field Camera on the United Kingdom Infrared Telescope. It is
surveying 1868~deg$^2$ of the northern and equatorial Galactic plane
at Galactic latitudes $-5^{\circ}<b<5^{\circ}$ in the J, H and K 
filters and a $\sim 200$~deg$^2$ area of the Taurus-Auriga-Perseus
molecular cloud complex in these three filters and the 
2.12~$\mu$m (1-0) H$_2$ filter. It will provide data on $\sim 2\times 10^9$ 
sources. Here we describe the properties of the dataset and provide a user's guide for 
its exploitation. We also present brief Demonstration Science results from 
DR2 and from the Science Verification programme. These results illustrate 
how GPS data will frequently be combined with data taken in other 
wavebands to produce scientific results. The Demonstration Science 
comprises six studies. (1) A GPS-{\it Spitzer}-GLIMPSE cross match for the star formation 
region G28.983-0.603 to identify YSOs. This increases the number of YSOs identified
by a factor of ten compared to GLIMPSE alone. (2) A wide field study of the M17 nebula,
in which an extinction map of the field is presented and the effect of source confusion
on luminosity functions in different sub-regions is noted. (3) H$_2$ emission in the $\rho$ 
Ophiuchi dark cloud. All the molecular jets 
are traced back to a single active clump containing only a few protostars, which suggests
that the duration of strong jet activity and associated rapid accretion in low mass
protostars is brief. (4) X-ray sources in the Nuclear Bulge. The GPS data distinguishes
local main sequence counterparts with soft X-ray spectra from Nuclear Bulge giant
counterparts with hard X-ray spectra. (5) External 
galaxies in the Zone of Avoidance. The galaxies are clearly distinguished from stars
in fields at longitudes $l$$>$90$^{\circ}$.
(6) IPHAS-GPS optical-infrared spectrophotometric 
typing. The {\it(i'-J) vs.(J-H)} diagram is used to distinguish A-F type dwarfs, G dwarfs,
K dwarfs and red clump giants in a field with high reddening.
\end{abstract}

 
\section{Introduction}

The United Kingdom Infrared Deep Sky Survey (UKIDSS) is a suite of five
public surveys of varying depth and area coverage which began in May 2005 and 
are expected to continue until 2012. These surveys all use the Wide Field
Camera (WFCAM, see Casali et al.2007) on the United Kingdom Infrared Telescope (UKIRT). The 
design and the principal aims of the five UKIDSS surveys are described by Lawrence et al.
(2007), along with a brief example of data from each survey. The data are reduced
and calibrated at the Cambridge Astronomical Survey Unit (CASU) using a dedicated software 
pipeline (Irwin et al.2008) and are then transferred to the WFCAM Science Archive 
(hereafter the WSA, Hambly et al.2008) in Edinburgh. There, the data are ingested 
into a sophisticated SQL database and detections in the different passbands are merged.

The UKIDSS data are released as SQL databases at intervals of six to twelve months. 
The area coverage and data quality of these releases are described by
Dye et al.(2006, the Early Data Release), Warren et al.(2007a, the First
Data Release) and Warren et al.(2007b, the Second Data Release). In addition to the major
releases, smaller batches of pipeline processed images and their associated FITS binary source 
catalogues for individual passbands are released on a faster timescale, but 
with no quality control. The raw data are available through CASU and the ESO archive.
The UKIDSS data are presently public to all professional astronomers in any ESO
member state. World access to the various data releases begins
18 months after each data release to ESO.\footnote{Users can obtain a WSA account through
their local Community Administrator at their home institution. If the institution has
no UKIDSS community then they should contact the Consortium Survey Scientist,
currently Steve Warren (email: s.j.warren@ic.ac.uk)}

The aim of this paper is to provide a guide to the GPS database in the WSA.
We describe the properties of the data in detail, illustrating usage of the 
WSA tools and giving examples of how to combine GPS data with data from other 
wavebands for maximum scientific benefit. This should be regarded as essential 
reading for those wishing to exploit the data. This paper describes the main 
GPS-specific aspects of the UKIDSS database in one place. We note that neither 
the pipeline nor the WSA were 
designed explicitly for the GPS but rather for the UKIDSS surveys as a whole.
Some compromises were made to produce an efficient database in a reasonable
period of time and we describe their effects here.

The paper is structured as follows. In $\S$2 we summarize the survey design,
including recent changes from the original plan described by Lawrence et al.
In $\S$3 we describe the properties of the GPS dataset (depth and stellar populations)
as a function of Galactic location and source confusion. 
In $\S$4 we present Demonstration Science results from the Science 
Verification programme and from DR2 that are designed to illustrate the variety
of uses of GPS data, in combination with data from other wavebands.
Finally, Appendix A contains a detailed guide to the WSA for GPS science. The guide
illustrates how to use the parameters available in the GPS database to best effect 
and describes the strengths and weaknesses of the data that are presently available.

\section{Summary of GPS strategy, area and integration time}

The GPS is designed to cover the entire northern and equatorial Galactic plane 
that is accessible to UKIRT in a 10 degree wide band around the sky. There is also a 
narrower southern extension to the Galactic Centre and a 200 deg$^{2}$ survey of
the Taurus-Auriga-Perseus (TAP) molecular cloud complex, which lies off the plane and follows 
the contours of the $^{13}$CO gas mapped by Ungerechts \& Thaddeus (1987). The southern
extension is narrower because of its poorer accessibility from Mauna Kea. 
In Galactic coordinates, the in-plane region is: $141<l<230$, $-5<b<5$; $15<l<107$, $-5<b<5$;
and $-2<l<15$, $-2<b<2$. The section of the plane at $l$=107 to 141 is not covered
because it lies north of Dec=+60$^{\circ}$, a region of the sky that UKIRT cannot
reach due to fundamental features of the telescope design. 

Observations are made in the J, H and K bandpasses with total on-source integration times of 
80s, 80s and 40s respectively. These integrations are made up of shorter exposures, employing a
sub-pixel dithering strategy to fully sample the image profile (see $\S$A1.2). WFCAM has four spatially
separated arrays, so a tiling strategy is employed to obtain images of contiguous areas of the sky
(see $\S$A1.1). The median 5$\sigma$ 
depths in DR2 given by Warren et al.(2007b) are J=19.77, H=19.00, K=18.05 (Vega system). However, 
the survey depth is spatially variable due to source confusion so these figures should be used with 
caution, see $\$$3. Data in the three filters are generally collected within an 
interval of 20 minutes for any point on the sky, in order to minimise the effects of any photometric
variability on source colours and derived extinction. Exceptions can occur (in DR2 and later releases)
when data that failed quality control in only one or two filters were later repeated in only
those filters. In these cases there will always be near contemporaneous fluxes in at least two
filters which can be used to estimate intrinsic source colours and extinction. The TAP molecular 
cloud component of the survey also includes observations in the 2.12~$\mu$m (1-0) S(1) H$_2$ filter,
with a total on source integration time of 160s.

The data reduction pipeline is described in Irwin et al.(2008), so we do not provide a full
description here. Some of the more significant details that may affect the data in unexpected ways
are described in $\S$A2. One important aspect of the reduction that is worth reiterating here
is the photometric calibration. The photometric zero points are bootstrapped from the 2MASS Point 
Source Catalogue, which is believed to provide a reliable photometric calibration over the whole
sky (Skrutskie et al.2006). If we assume that the 2MASS calibration is perfect, the scatter in the 
derived zero points for the numerous 2MASS stars used in each WFCAM field indicates that the absolute
accuracy of the UKIDSS photometric calibration in the J, H and K bandpasses is 1-2\% in photometric 
conditions and 2-3\% in non-photometric conditions (Hodgkin et al., 2008, submitted).
Since the 2MASS J, H and Ks filters are slightly different from the WFCAM 
filters (see Hewett et al.2006), photometric transformations are used to correct the zero points. From 
DR2 onward, these transformations include extinction terms to correct for the fact that the colours of stars 
in the Galactic plane are influenced by interstellar reddening as well as temperature and gravity
(Warren et al.2007b). These transformations are included in $\S$A2 for completeness.
They should be used in preference to the preliminary transformations given in Hewett et al.(2006).

UKIDSS astrometry is also calibrated using numerous bright 2MASS stars in each field. The precision
of the astrometric solution is typically 0.09 arcsec for GPS data, as determined from the r.m.s. 
residuals of the stars to the fit (see $\S$A1.3 for details of how to determine the precision for
specific fields). Individual stars can of course have much less precise astrometry, owing to source 
confusion or low signal to noise ratio.

Owing to source confusion the survey depth is spatially variable in the more crowded parts 
of the plane (see $\S3.1$). Hence, in many regions of the plane the GPS represents the 
deepest survey possible with ground based, seeing limited wide field instrumentation.
Deeper images can be obtained by using higher resolution cameras with smaller pixels
in conditions of good seeing, but for the foreseeable future such cameras will not be able
to cover large areas of the Galactic plane.

The sky coverage in the various data releases so far is described in the references given in 
$\S1$. Those areas of the plane at $-2<l<107$ with three band coverage can be viewed as an 
interactive three colour mosaic at 
http://surveys.roe.ac.uk:8080/wsa/gps\_mosaic.jsp, (credit M.Read) where the finest spatial scale 
has 1\arcsec pixels.

\subsection{Changes to the GPS Programme}

In early 2007 there was a review of the UKIDSS project in response to a call for additional
large observing proposals. This led to changes and reductions in the UKIDSS programme in order
to accomodate the new observing campaigns.
The original GPS design (Lawrence et al.2007) also included two additional epochs of data in the K band
only, which were intended to provide a long time baseline for proper motions
and to detect high amplitude variable stars. Following the 2007 review, further progress on these 
K-only scans of the survey area has been deferred 
until at least 2010. Some 500 deg$^2$ of the plane have already been observed in the K-only mode during
conditions of thin cirrus cloud (defined as $<$20\% drop in throughput), covering regions not yet observed in 
the main three band survey. Since the UKIDSS photometry is calibrated using numerous 2MASS sources in every 
field there is only a very small loss of precision for relative photometry (at the millimagnitude level) even 
when observing through thin cirrus (Hodgkin et al.2008). The main JHK survey has been conducted in photometric 
conditions, in order to ensure maximum depth, but thin cirrus conditions ($<$10\% absorption) may be accepted 
from late 2008 in order to assist timely completion of the survey.

It was originally planned to observe a region 300~deg$^2$ of TAP molecular clouds
three times: once in the J, H, K and H$_2$ filters and twice more in the K and H$_2$ filters,
in order to measure the velocities and accelerations of protostellar jets. However this
part of the survey was cancelled in the 2007 review of UKIDSS, at which point 
$\sim$150~deg$^2$ of the region had been observed. Nonetheless, the remainder of the TAP region will observed
in the Z, Y, J, H and K filters as part of the ongoing UKIDSS Galactic Clusters Survey,
which included the same region. Observations of the TAP region with the H$_2$ filter are continuing
via Directors Discretionary Time and small observing programmes, with the aim of completing
one pass of the denser cloud cores, with a total area of $\sim$200~deg$^2$. The TAP area coverage 
is described in more detail in Davis et al.(2008), and the reduced data can be viewed at 
www.jach.hawaii.edu/UKIRT/TAP.\footnote{To gain access to this web page, UKIDSS members should request the 
user name and password from Chris Davis, email: c.davis@jach.hawaii.edu. The data become world public
18 months after release to UKIDSS members.}



\section{Populations and data properties as a function of Galactic coordinates}

In this section we present magnitude histograms, colour magnitude diagrams and two colour diagrams
for representative fields throughout the survey region. We discuss the survey depth as a function
of Galactic coordinates and then describe the nature of the main Galactic (and extragalactic) populations that 
are detected. The Besan\c{c}on Galactic stellar population model\footnote{The Besan\c{c}on models are available 
on-line at bison.obs-besancon.fr/modele} (Robin et al.2003) is used to help interpret the diagrams and we then 
compare some diagrams for the synthetic population with the real data.

Figures 1 to 13 show the results for 12 GPS fields. The fields at $l<$90$^{\circ}$ are square boxes
that are 0.2 degrees on a side in $l$ and $b$. The fields at $l>$90$^{\circ}$ are square boxes that are 0.5 
degrees on a side in $l$ and $b$, in order to provide similar sized samples to reveal the nature of the 
populations in these less crowded regions. All fields include data from at least 2 spatially 
separate WFCAM multiframes and usually more. Any errors in global photometric calibration would add to the 
scatter in these plots. However the quality of the 2MASS-based photometric calibration is such that we see no 
obvious increase in scatter compared to data from individual WFCAM arrays.

\subsection{Survey sensitivity and the effect of source confusion}

Figure 1 shows the magnitude histograms. Visual inspection of the images shows that the fields at $l>$90$^\circ$ 
have less source confusion than the other fields and source confusion is insignificant for the two fields near 
$l$=170$^{\circ}$. The corresponding histograms represent the maximum intrinsic depth of the survey.
We see that 
the modal depths in uncrowded fields are approximately K=17.75 to 18.0, H=18.5 to 18.75, J=19.4 to 19.65. The 
variation is due to the variable seeing, sky brightness, thermal background and atmospheric transparency during 
the survey. These modal depths are slightly shallower than the 5$\sigma$ depths quoted in $\S$2: 
K=18.05, H=19.00, J=19.77. The reasons for this are described in $\S$A3.
By visually extrapolating the histograms and assuming that the decline at faint magnitudes is entirely 
due to limited sensitivity we estimate that the typical 90\% completeness 

\begin{figure*}
\begin{center}
\vspace{-3cm}
\includegraphics[width=1.05\textwidth]{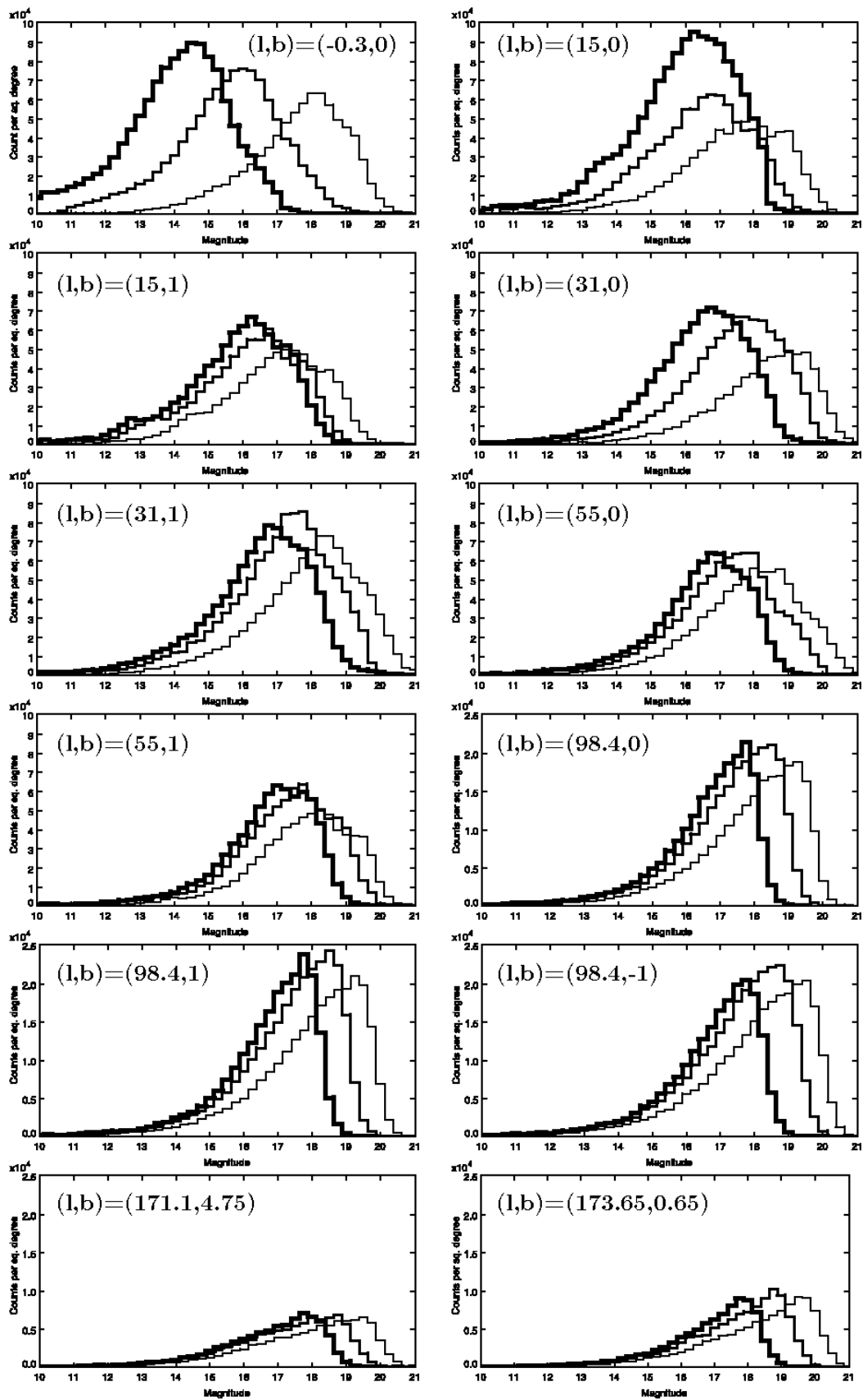}

\end{center}
\vspace{-8mm}
\caption{Magnitude histograms in a selection of GPS fields. Thick lines: K band;
medium lines: H band; thin lines: J band. A linear scale is used rather than the conventional
log scale in order to better illustrate the observed variation in modal magnitude between
fields. Source counts per square degree are given, using bins of width 0.25 magnitudes. A general
decline in source density is seen with increasing longitude and increasing distance from the mid-plane, though
there are some exceptions to this trend.}
\end{figure*}

\begin{figure*}
\begin{center}
\begin{picture}(200,160)

\put(0,0){\includegraphics{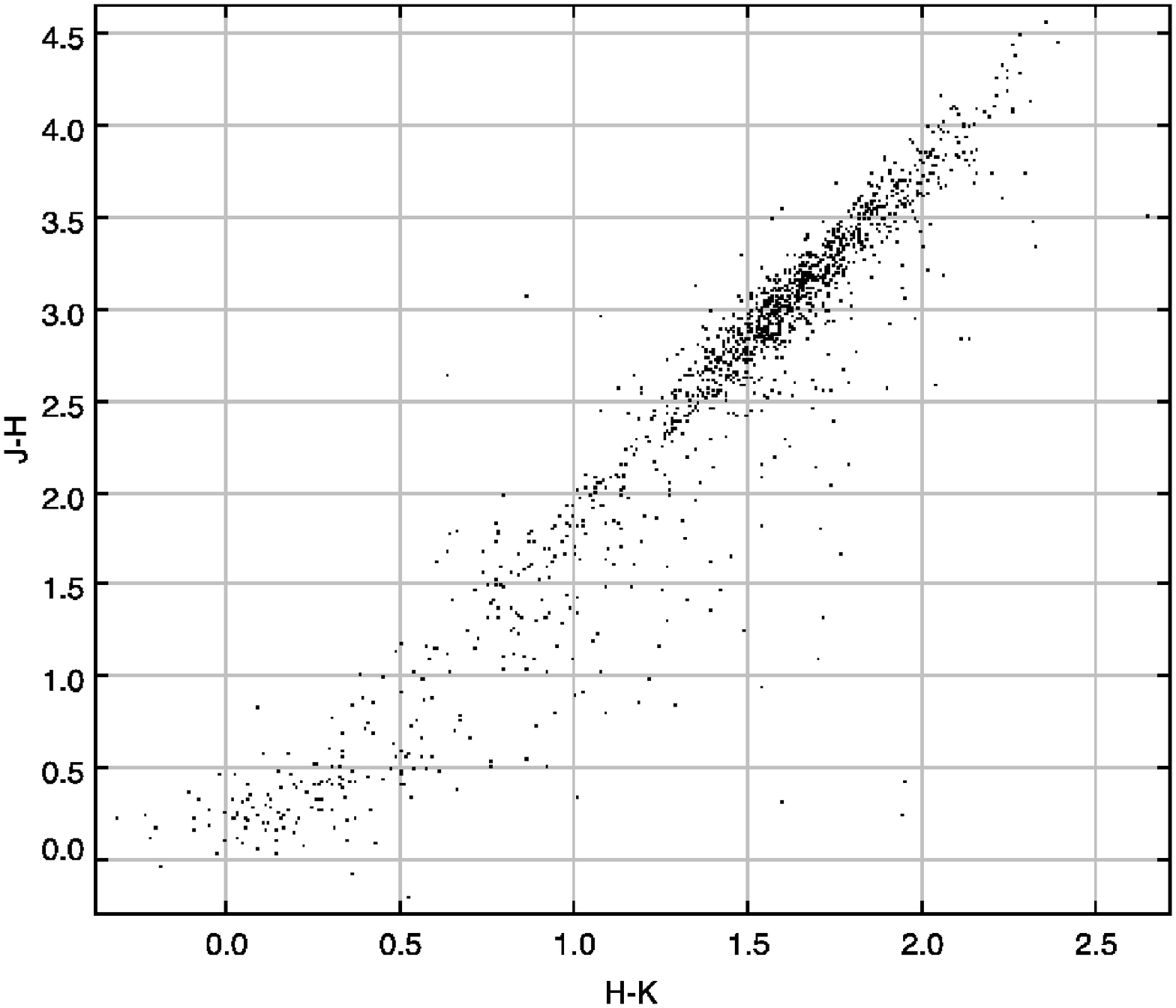}}

\put(0,0){\includegraphics{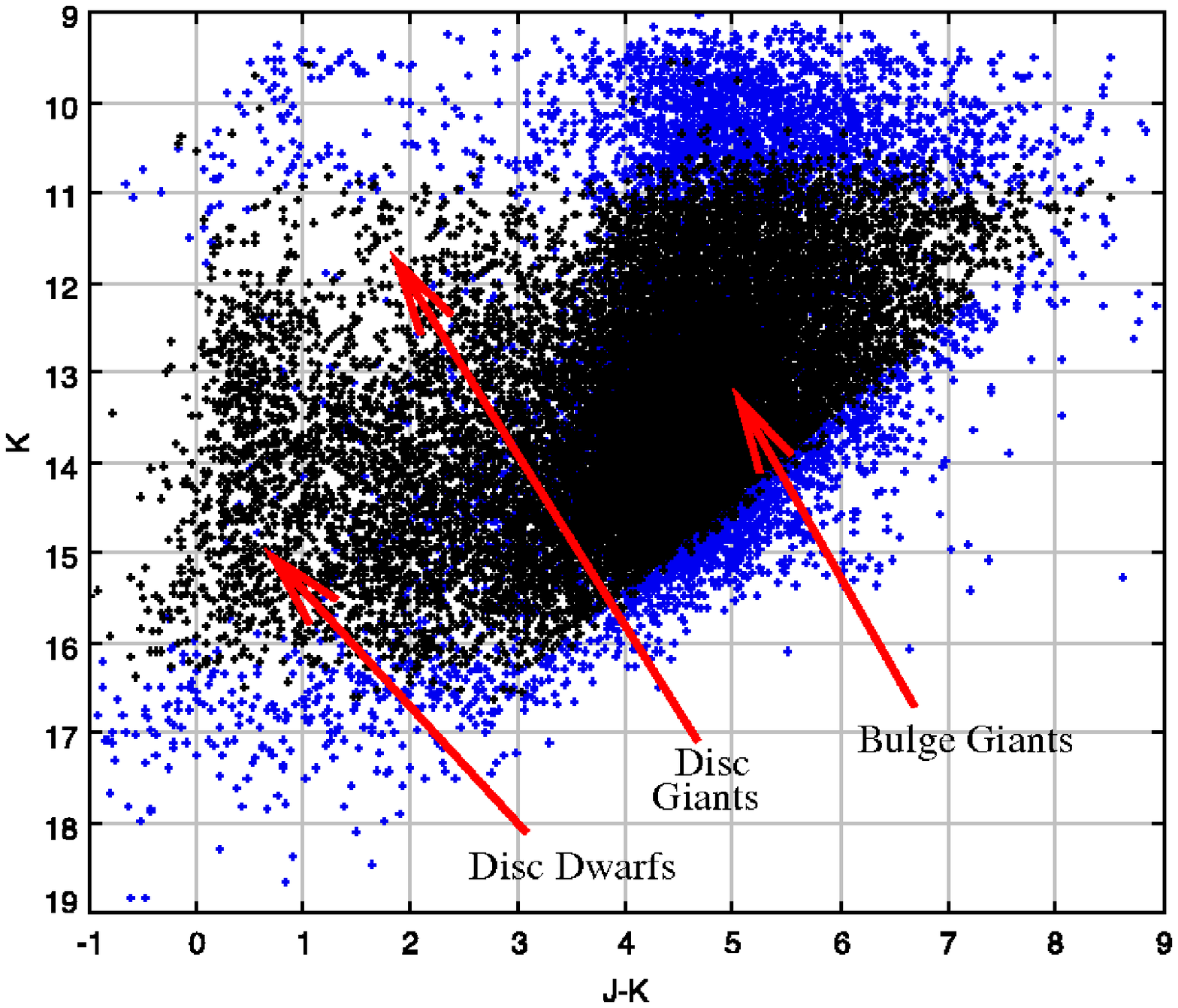}}

\bf

\end{picture}
\end{center}
\vspace{1.1cm}
\caption{($l,b$)=(-0.3,0). Colour magnitude and two colour diagrams for a 0.2$\times$0.2$^{\circ}$ field
adjacent to the Galactic centre. The two colour diagram, like those in figures 3 to 13, shows only sources with 
the ``most reliable, low completeness'' selection described in $\S$A3. By contrast, the colour magnitude diagram,
as in figures 3 to 13, shows all sources except ``noise'' detections. Sources with ``unreliable
photometry'' (see text $\S$3.2) are marked in blue. Both diagrams are dominated by the highly reddened Bulge population.
The less reddened sequences of disk dwarfs and disk red clump giants are also visible in the colour magnitude 
diagram but less clearly than in figures 3 to 13.}
\end{figure*}

\begin{figure*}
\begin{center}
\begin{picture}(200,160)

\put(0,0){\includegraphics{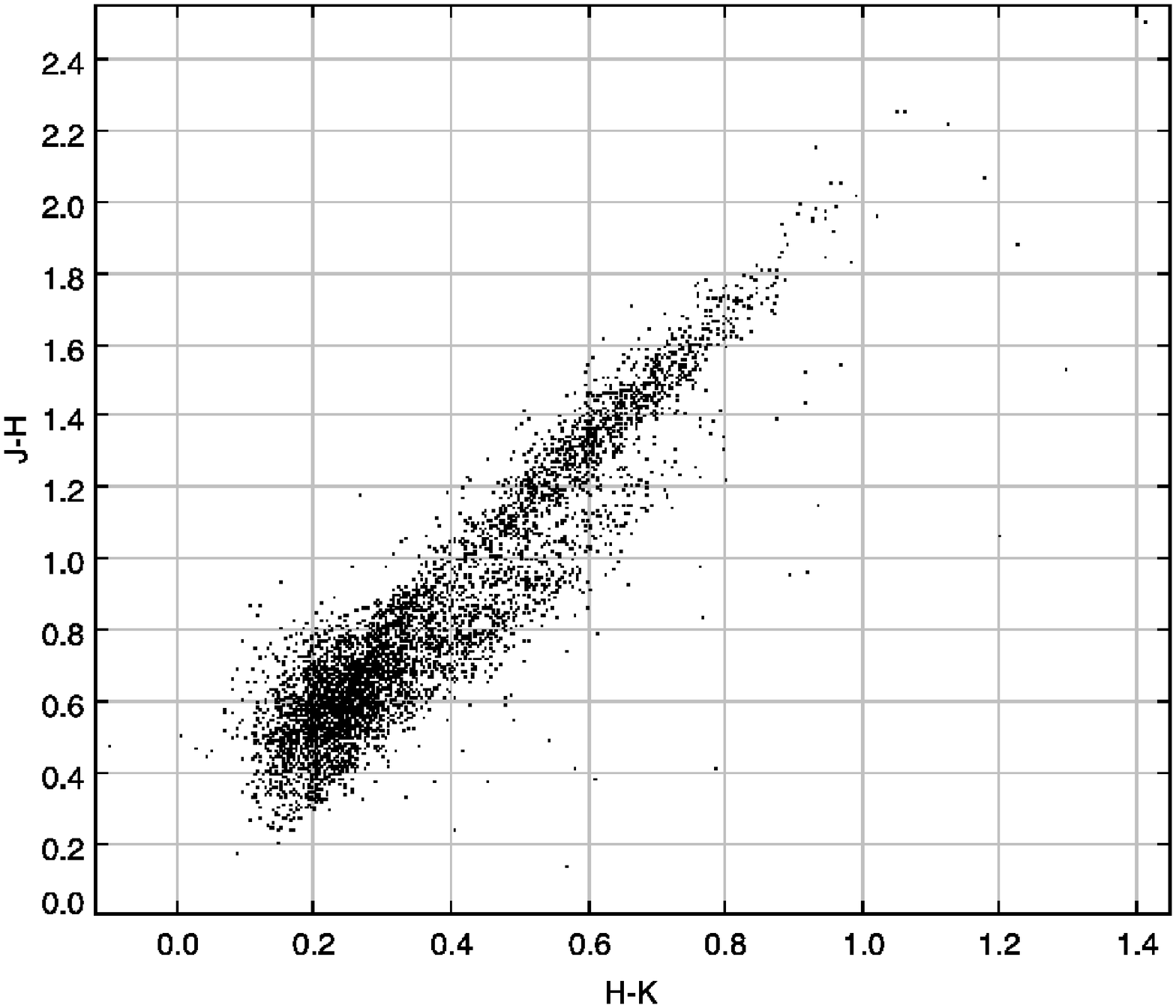}}

\put(0,0){\includegraphics{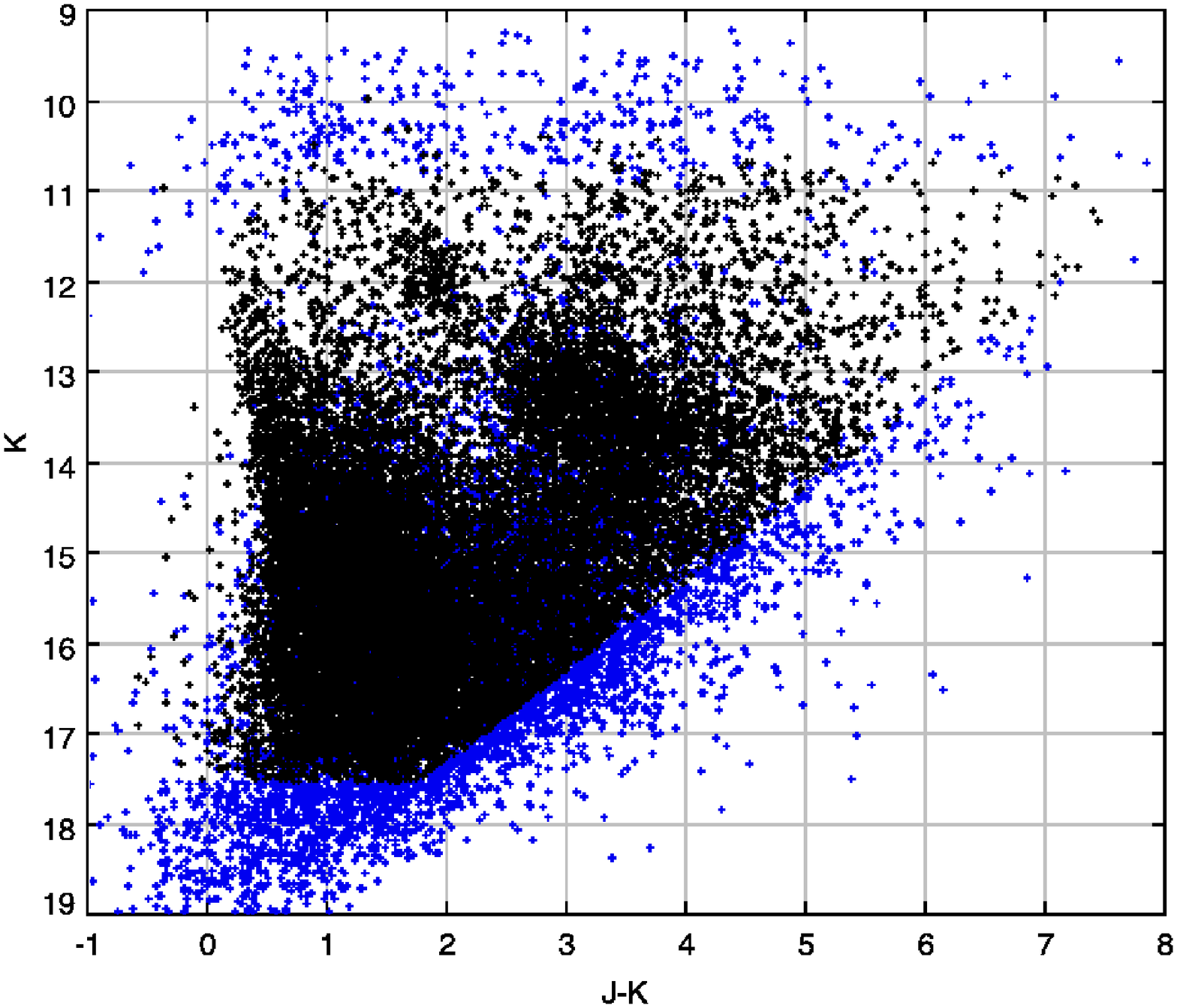}}

\bf
\rm

\end{picture}
\end{center}
\vspace{1.2cm}
\caption{($l,b$)=(15,0). This crowded field in the inner Galactic mid-plane shows a 
well defined dwarf sequence at the left of the colour magnitude diagram and a diagnonal red clump giant sequence to 
the right of it. The two colour diagram shows a distinction between giants and dwarfs. Both luminosity classes form 
sequences parallel to the reddening vector but giants have redder {\it (J-H)} colours than the dwarf sequence 
underneath (see text $\S$3.2). The points plotted are selected in the same way as in figure 2.}
\end{figure*}

\begin{figure*}
\begin{center}
\begin{picture}(200,160)

\put(0,0){\includegraphics{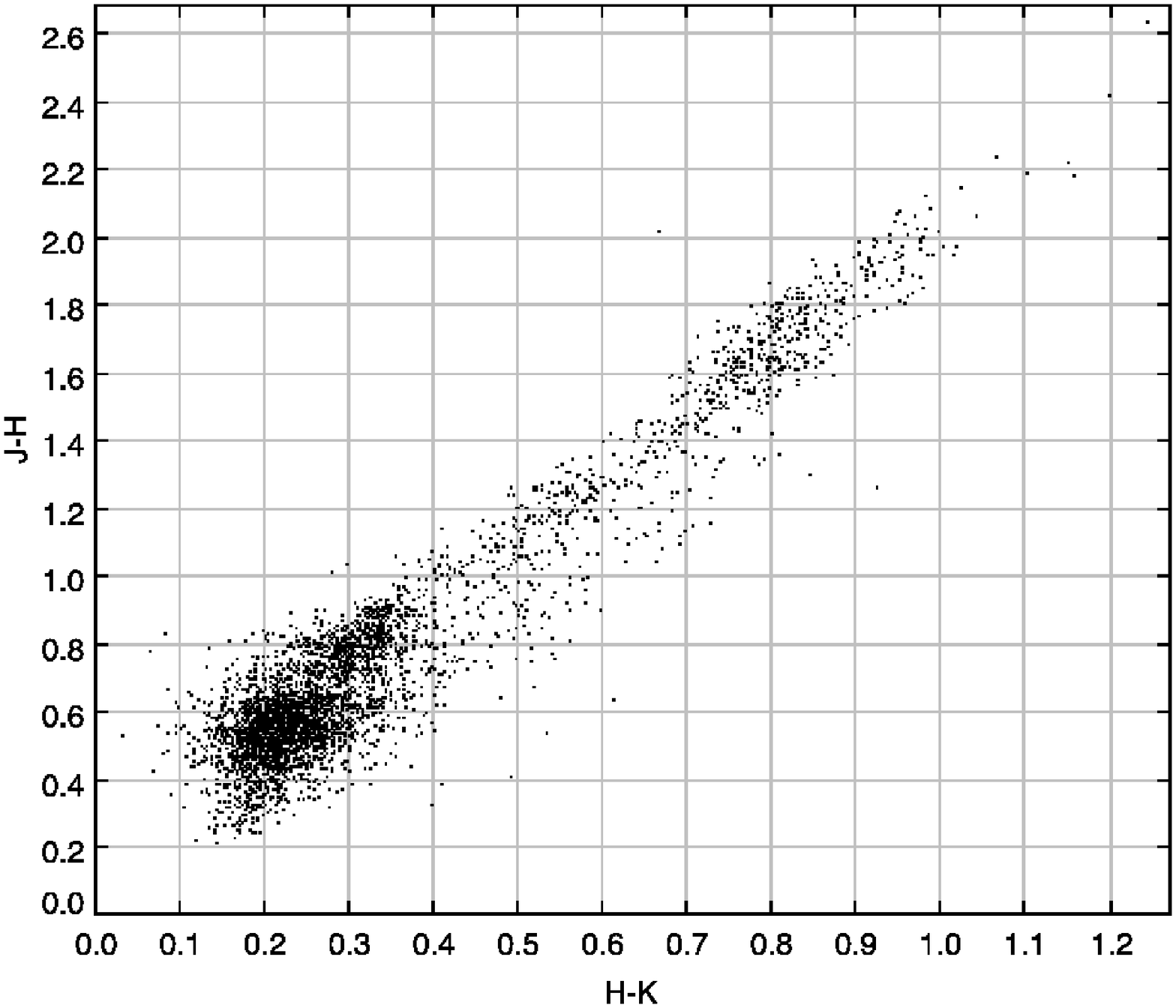}}

\put(0,0){\includegraphics{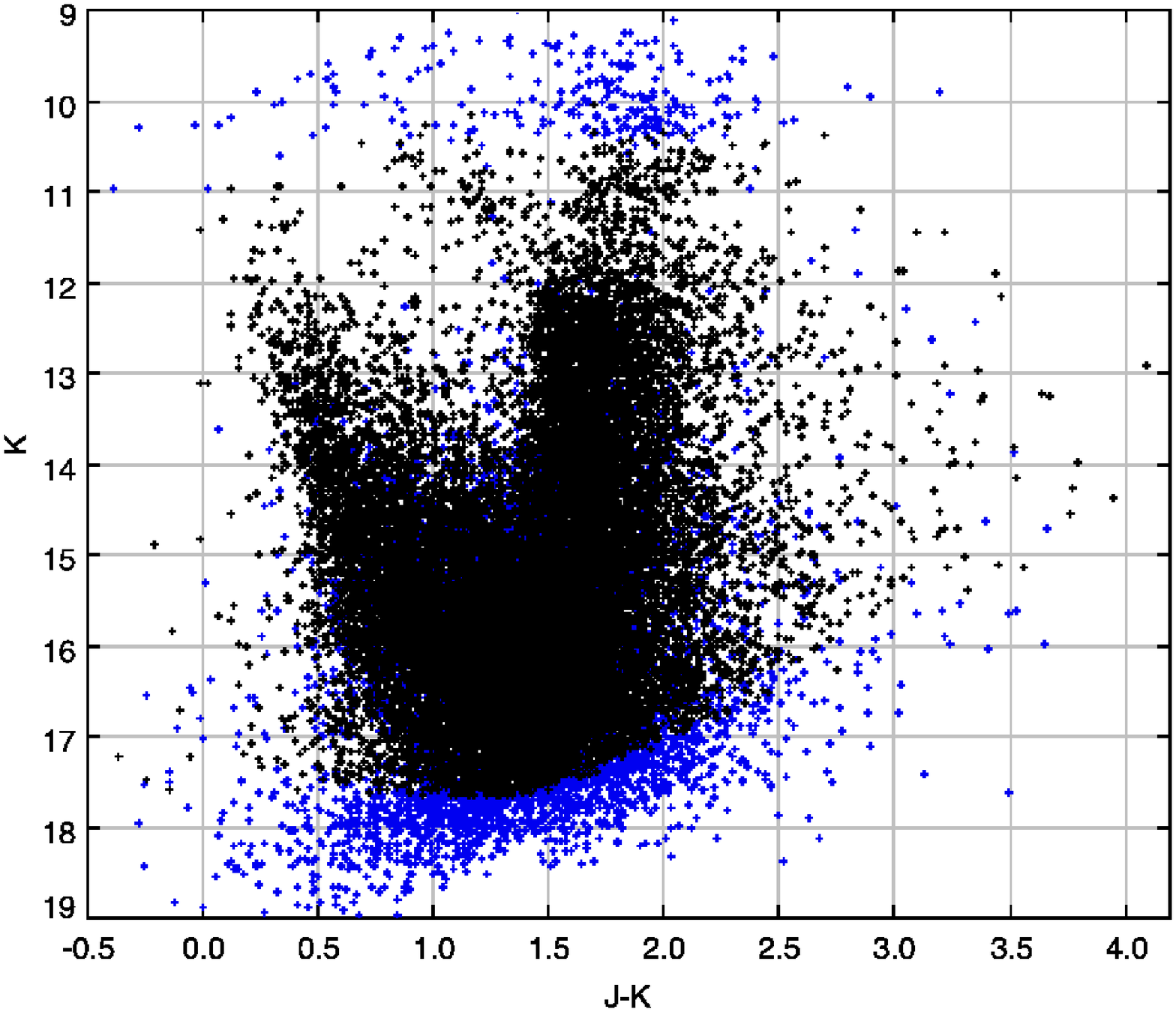}}

\bf
\rm

\end{picture}
\end{center}
\vspace{1.0cm}
\caption{($l,b$)=(15,1). The red clump giants form a diagonal sequence in the colour magnitude 
diagram, from m$_K$=10.5 to 12, {\it(J-K)}=0.8 to 1.7, which then becomes approximately vertical at fainter magnitudes. 
This field is on a line of sight that gradually rises above the Galactic plane at large heliocentric distances. 
This leads to reduced extinction at large distances (see text $\S$3.2 and figure 15), which explain the change in slope
of the giant sequence at faint magnitudes. A substantial Bulge giant population is also seen to the right of the 
diagonal red clump giant sequence at m$_K < 12$. The two giant populations appear to merge at fainter magnitudes.
The reddening sequences in the two colour diagram are shorter than in figure 3. The points plotted are selected in the 
same way as in figure 2.}
\end{figure*}

\vspace{1.5cm}

\begin{figure*}
\begin{center}
\begin{picture}(200,160)

\put(0,0){\includegraphics{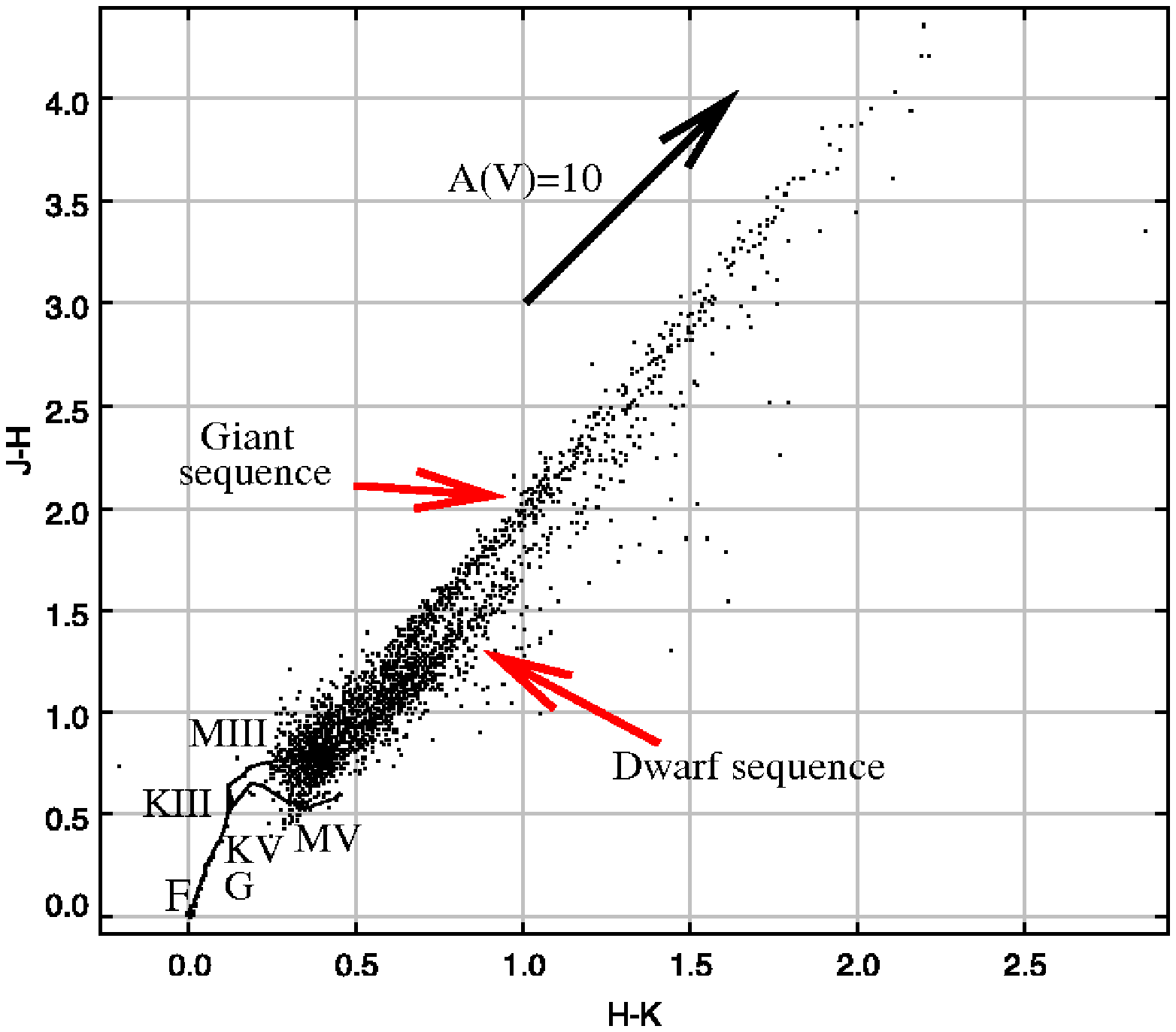}}

\put(0,0){\includegraphics{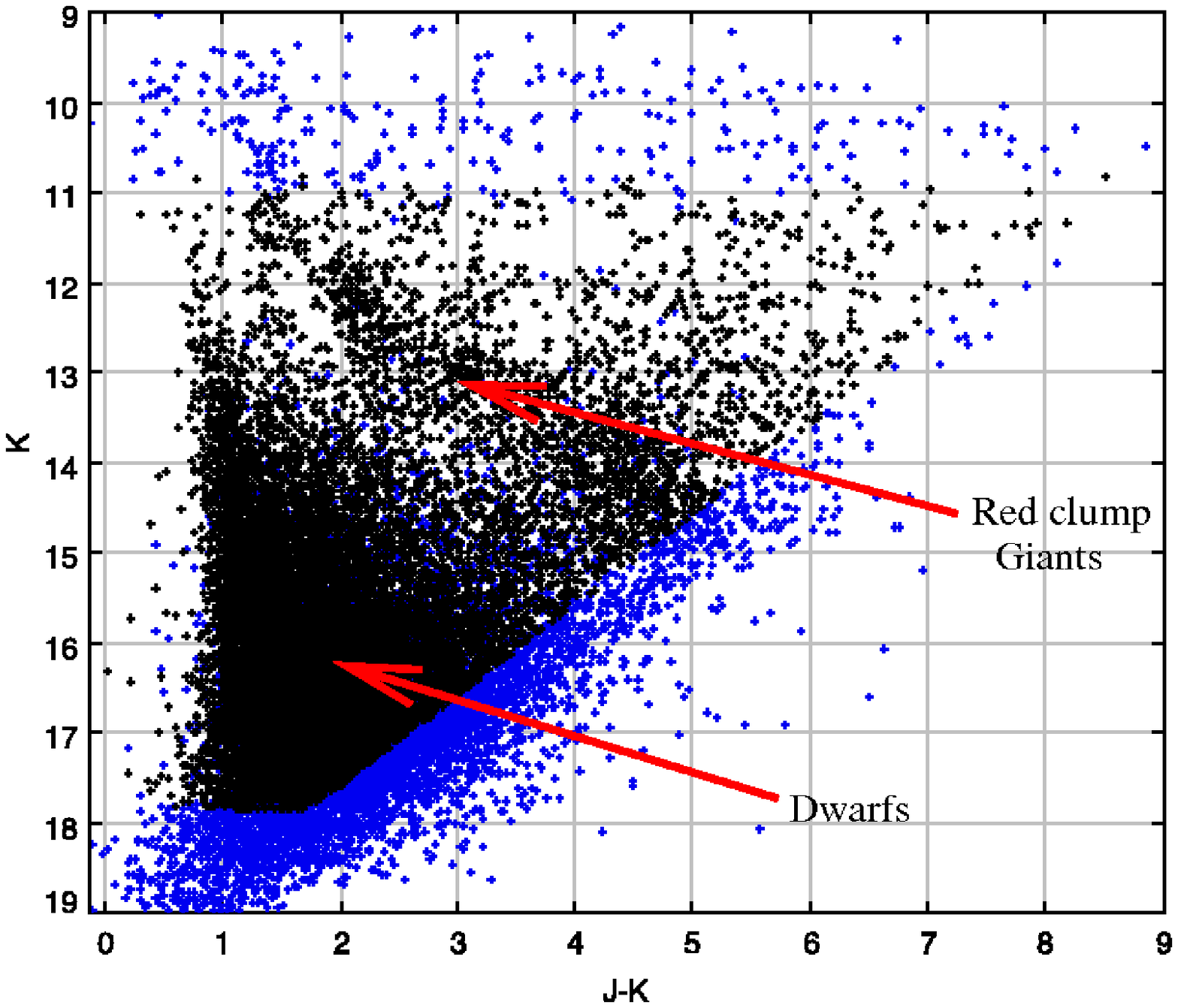}}

\bf
\rm

\end{picture}
\end{center}
\vspace{1.2cm}
\caption{($l,b$)=(31,0). The field shows very clearly defined dwarf and giant reddening sequences in the
two colour diagram. The colours of unreddened stars from Hewett et al.(2006) are indicated by the solid lines,
with approximate spectral types marked. The reddening vector of Rieke \& Lebofski (1985) is overplotted.
The colour magnitude diagram shows that the Bulge giant population has disappeared at this 
longitude, while the disk red clump giant population follows a winding diagonal sequence, indicating variable
reddening along the line of sight.  The points plotted are selected in the same way as in figure 2.}
\end{figure*}

\begin{figure*}
\begin{center}
\begin{picture}(200,160)

\put(0,0){\includegraphics{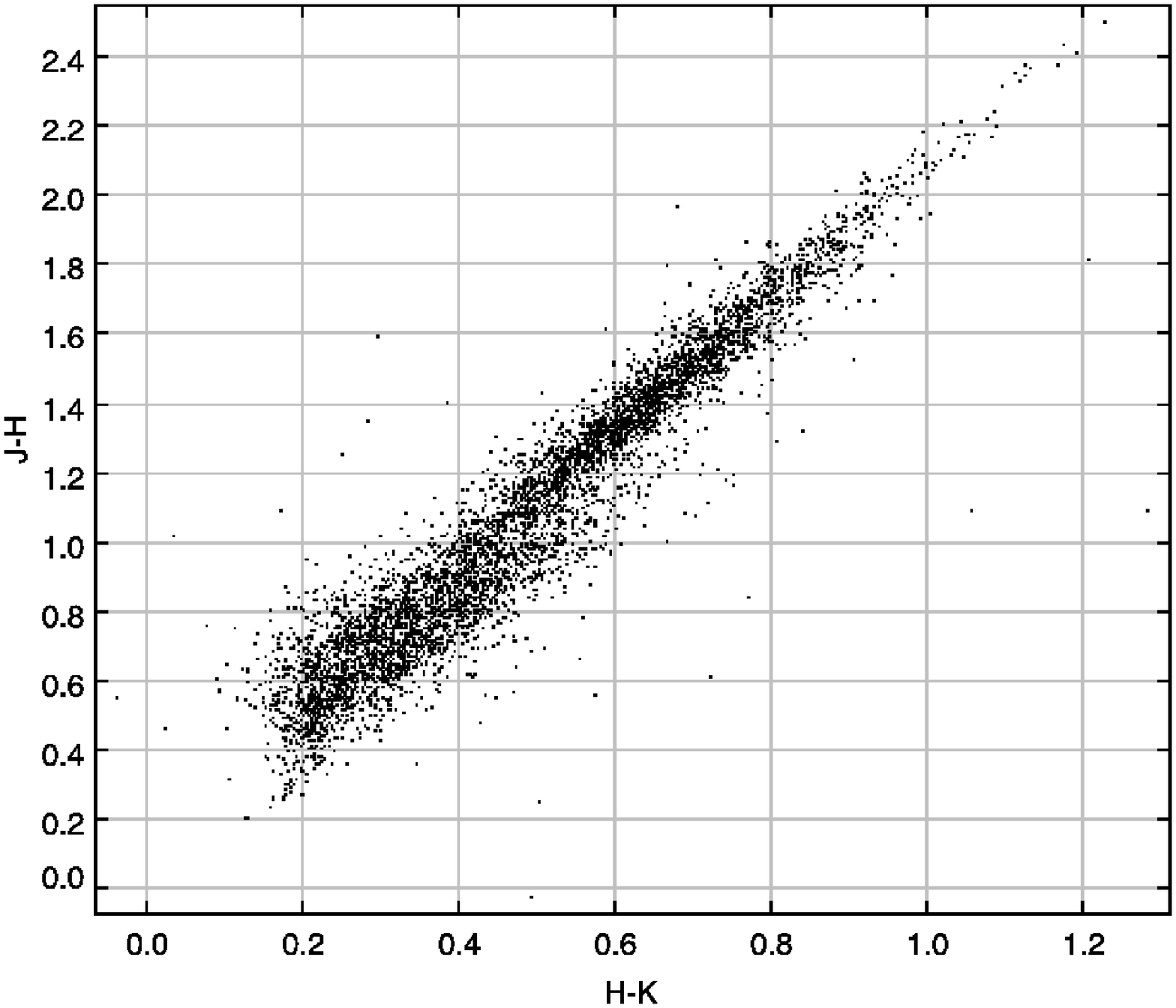}}

\put(0,0){\includegraphics{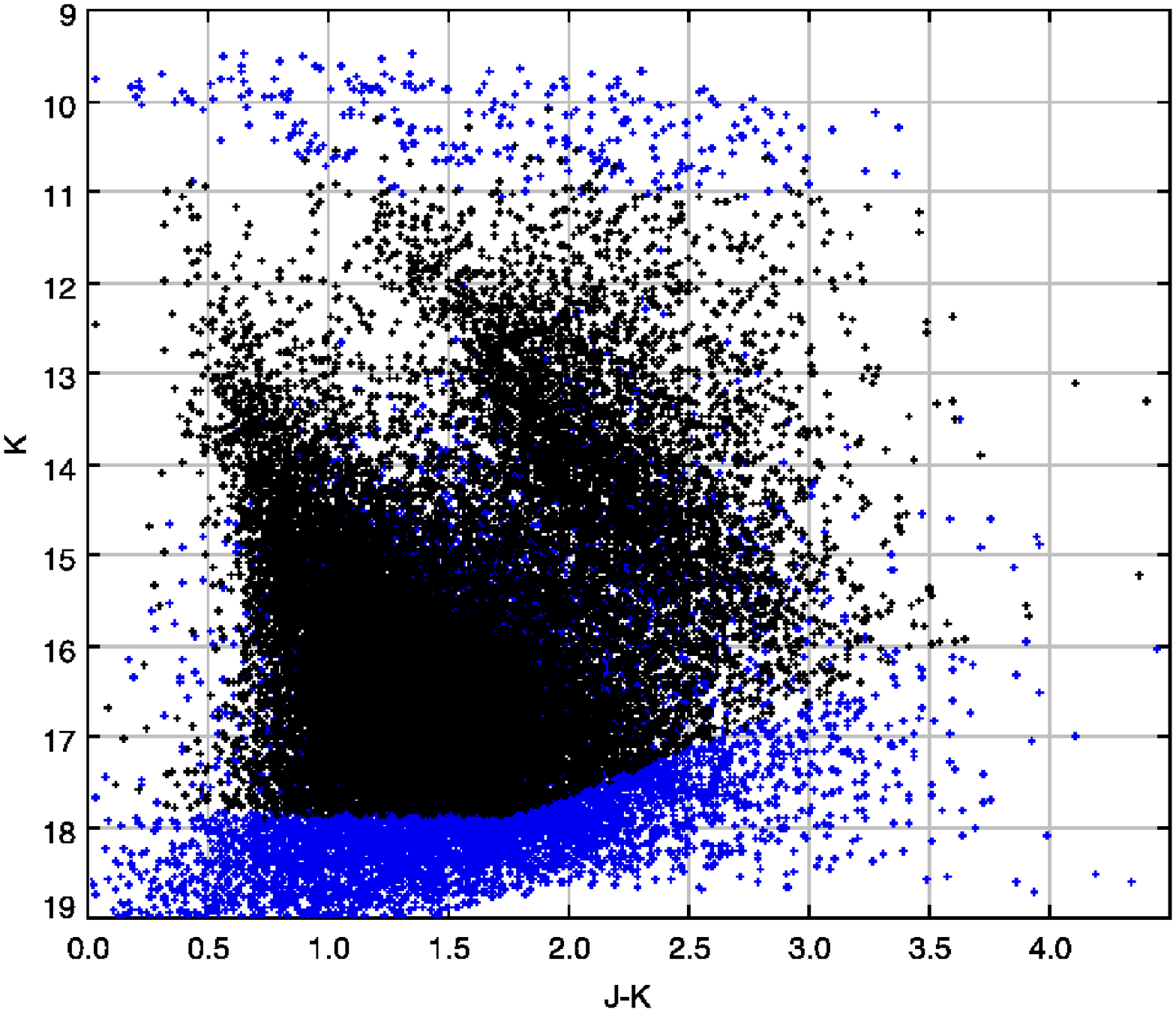}}

\bf
\rm

\end{picture}
\end{center}
\vspace{1.1cm}
\caption{($l,b$)=(31,1). The two colour diagram is similar to that for $(l,b)$=(15,1). However, the
colour magnitude diagram shows a diagonal red clump giant branch that does not become vertical at faint
magnitudes. This suggests that this line of sight passes through a region of the galaxy in which stars
are located above the plane, perhaps due to a warp. The points plotted are selected in the same way as in figure 2.}
\end{figure*}

\vspace{1.2cm}

\begin{figure*}
\begin{center}
\begin{picture}(200,160)

\put(0,0){\includegraphics{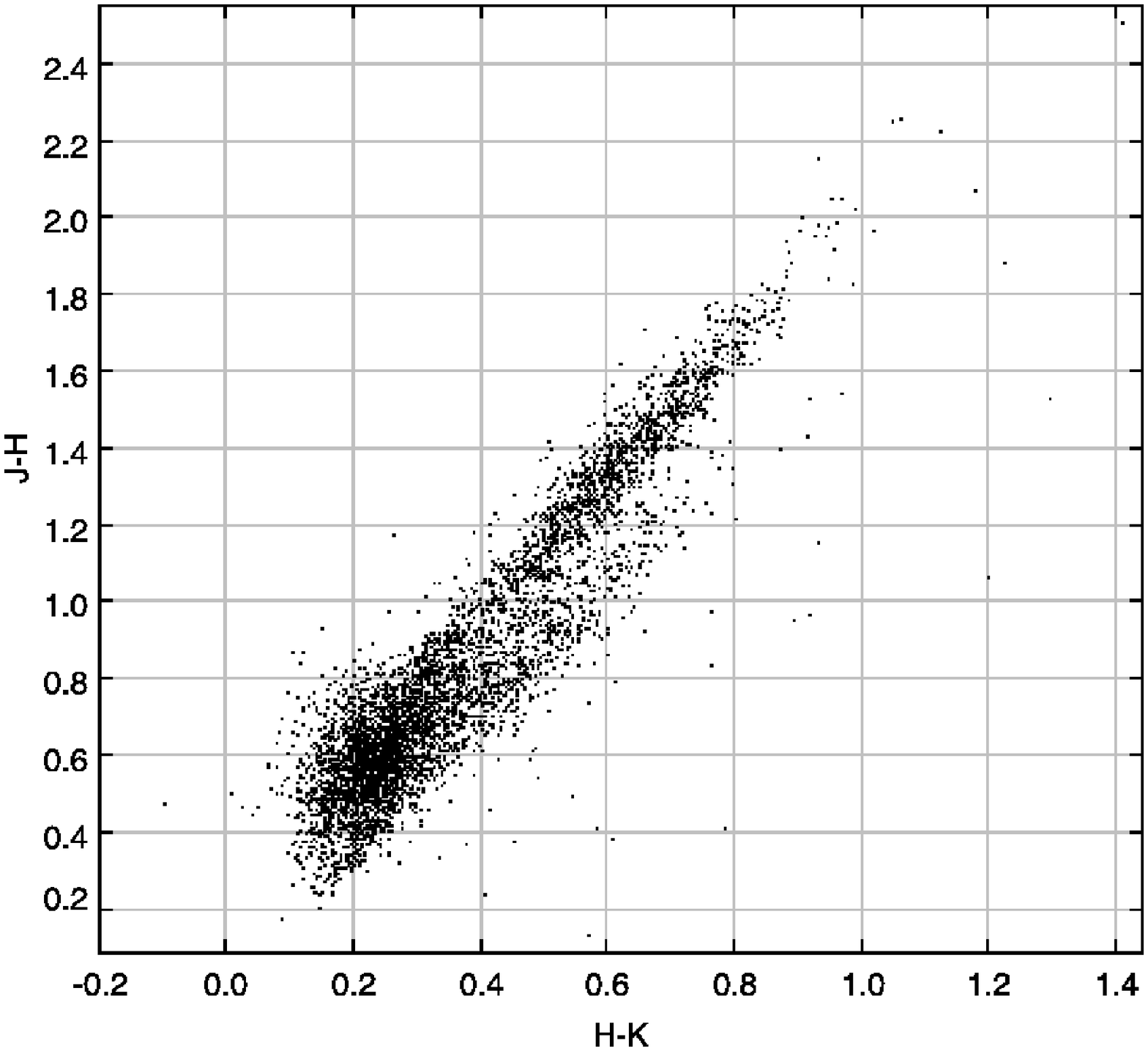}}

\put(0,0){\includegraphics{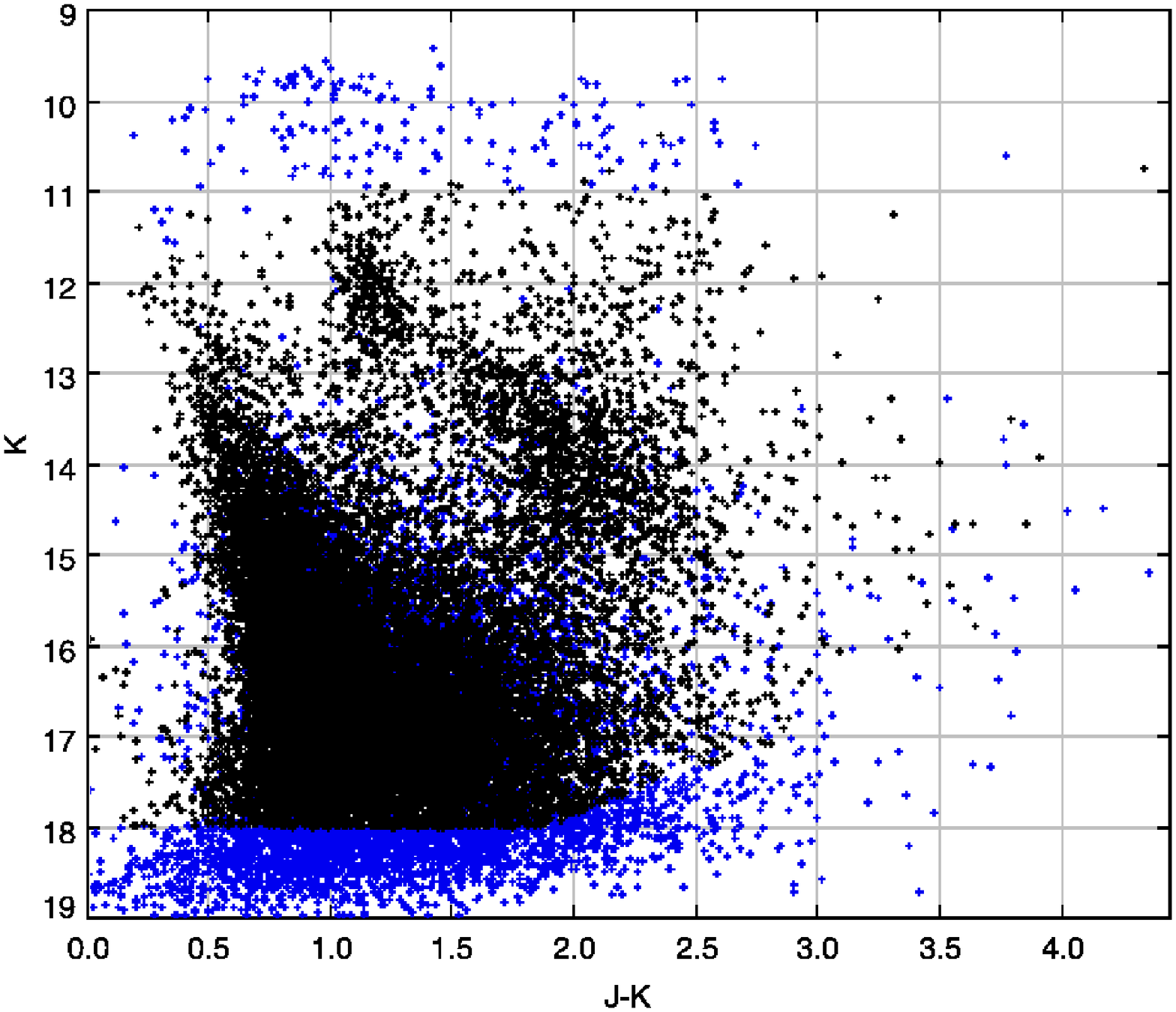}}

\bf
\rm

\end{picture}
\end{center}
\vspace{1.2cm}
\caption{($l,b$)=(55,0). This field shows an abrupt change in the slope of the disk red clump giant branch
in the colour magnitude diagram at m$_K$=13, indicating the presence of a discrete cloud of dense interstellar
gas and dust. The points plotted are selected in the same way as in figure 2.}
\end{figure*}


\begin{figure*}
\begin{center}
\begin{picture}(200,160)

\put(0,0){\includegraphics{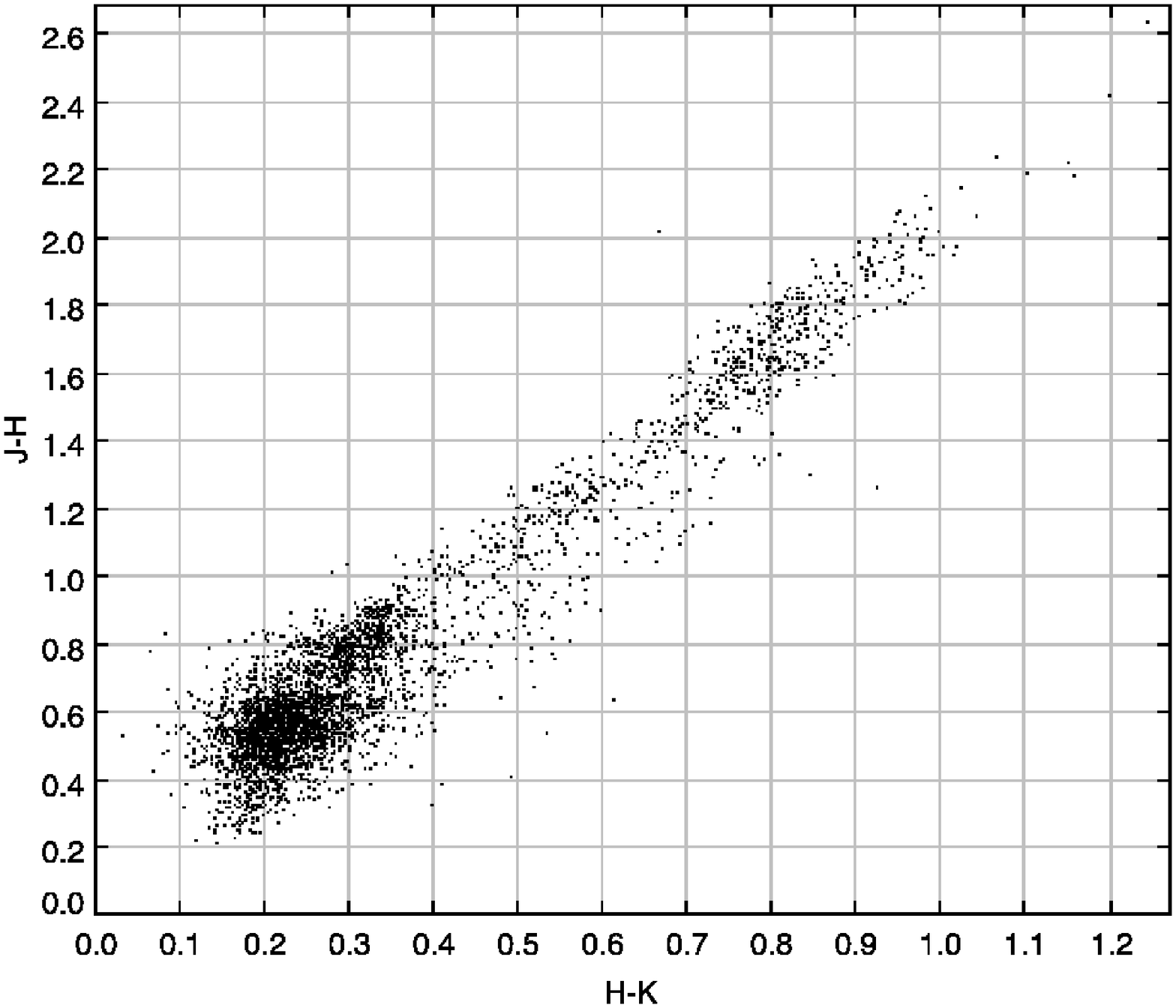}}

\put(0,0){\includegraphics{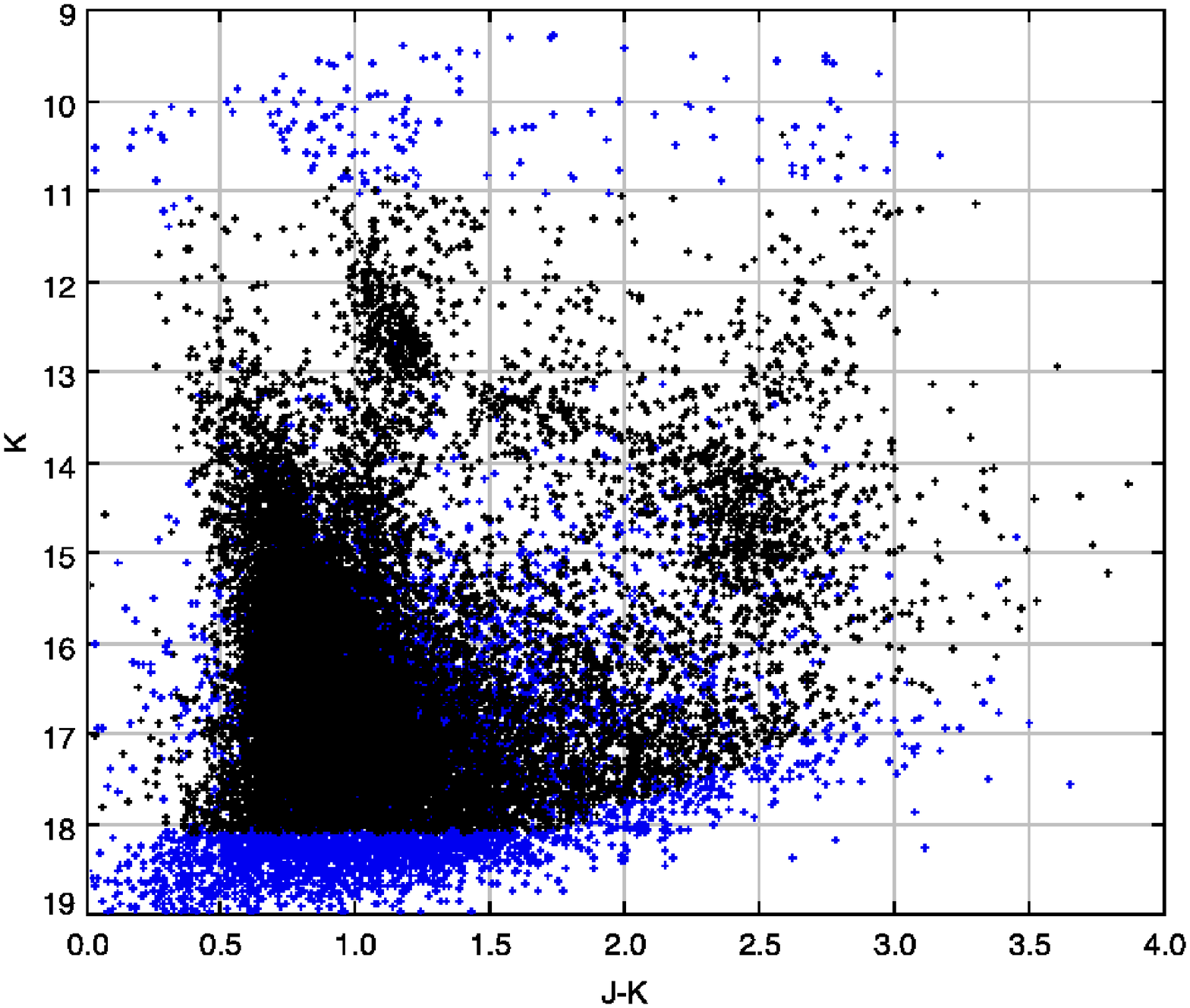}}

\bf
\rm

\end{picture}
\end{center}
\vspace{1.1cm}
\caption{($l,b$)=(55,1). This field shows a very strong discontinuity in the slope of the disk red clump giant branch
in the colour magnitude diagram at m$_K$=13 to 14, where the {\it (J-K)} colour increases from 1.2 to 2.4. This
indicates the presence of a discrete cloud of dense interstellar gas and dust (see text $\S$3.2). This may be related to the 
weaker feature in figure 7. At fainter magnitudes the giant sequence become vertical, indicating that the line of 
sight has risen above the plane to a region of low interstellar extinction. Note that the sequence of 
late K and M dwarfs (see text $\S$3.2) becomes visible at m$_K$=13 to 14.5, {\it(J-K)}=1.0. It protrudes from the bulk of 
the earlier type disk dwarfs at the left hand side and extends up toward the giant branch. The points plotted are selected 
in the same way as in figure 2.}
\end{figure*}

\begin{figure*}
\begin{center}
\begin{picture}(200,160)

\put(0,0){\includegraphics{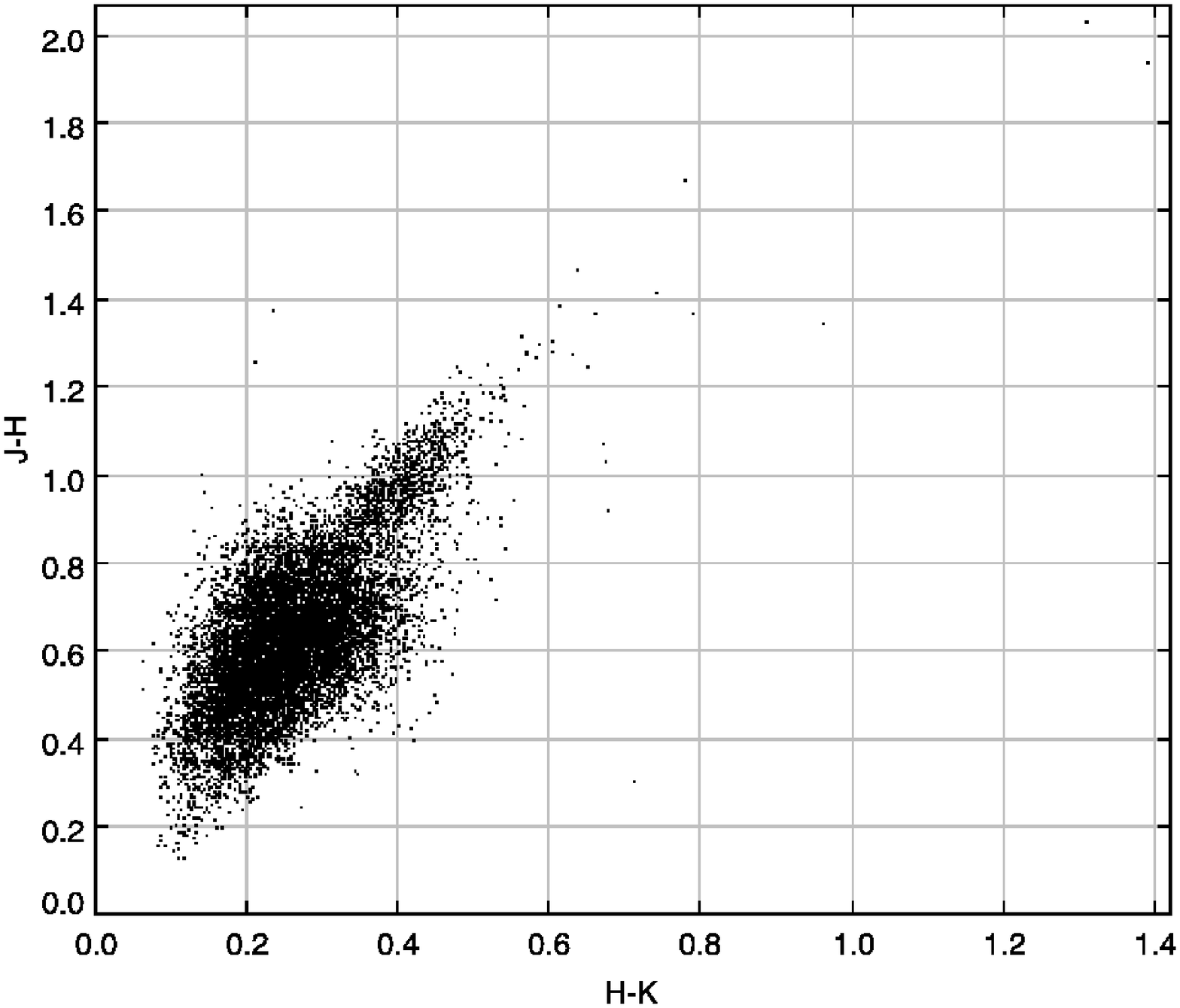}}

\put(0,0){\includegraphics{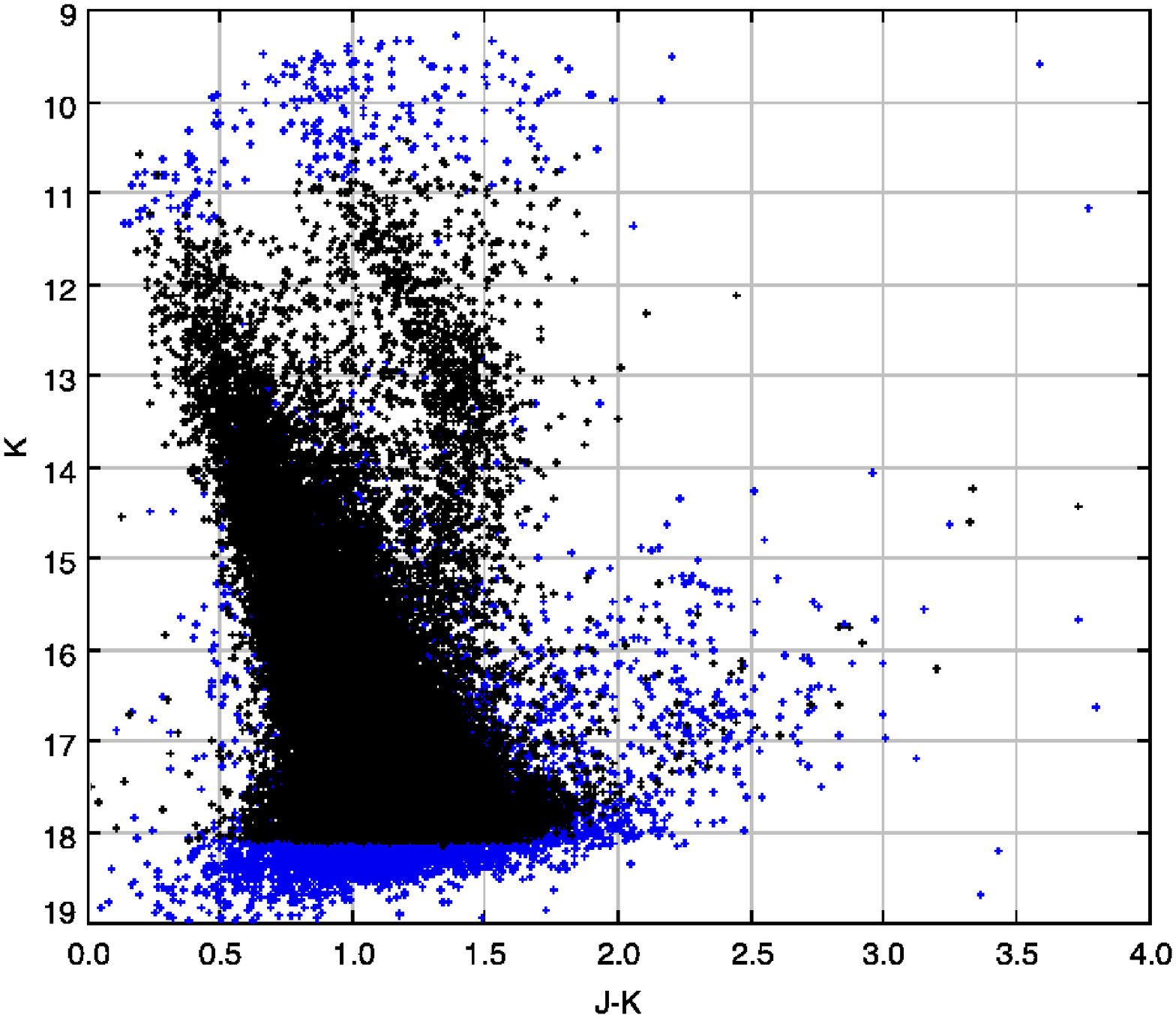}}

\bf
\rm

\end{picture}
\end{center}
\vspace{1.0cm}
\caption{($l,b$)=(98.4,0). The field lies in the outer galaxy and the extinction is now clearly less than
in figures 2 to 8, as evidenced by the shorter reddening sequence in the two colour digram. A 0.5$\times$0.5
degree field is plotted here and for the other outer Galaxy fields in order to maintain sufficient source 
counts to define the populations clearly. The giant branch
in the colour magnitiude diagram becomes vertical at m$_K>$13. This indicates a lack of additional extinction at 
large heliocentric distances in the mid-plane, which is attributed to the northern Galactic warp. Note the 
population of external galaxies that is visible at lower right ({\it (J-K)}$>$2). These are mostly marked in 
blue because they are spatially resolved, and therefore fall into our ``unreliable photometry'' category, even 
though their fluxes in the 2\arcsec aperture are well measured.
The points plotted are selected in the same way as in figure 2.}
\end{figure*}

\vspace{1cm}

\begin{figure*}
\begin{center}
\begin{picture}(200,160)

\put(0,0){\includegraphics{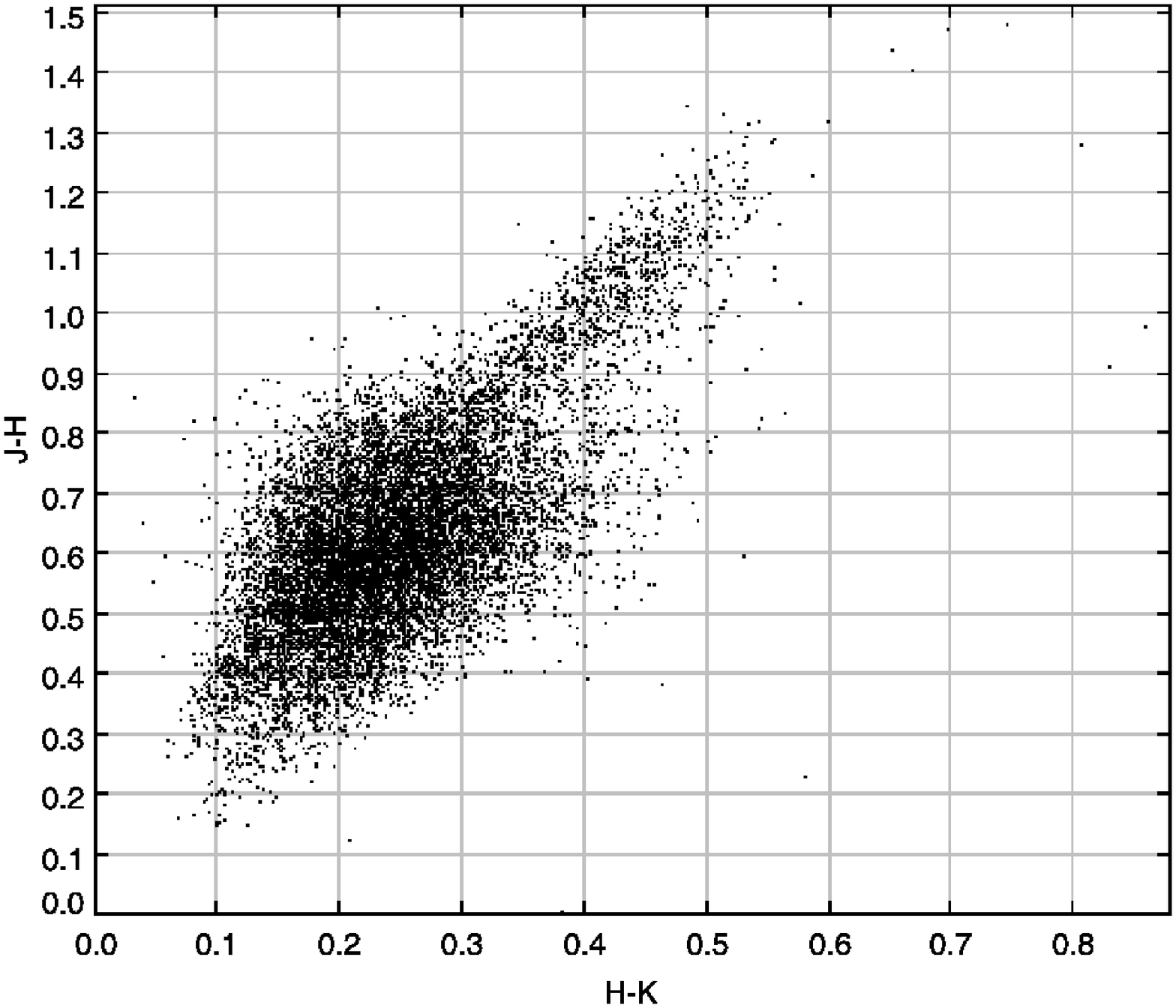}}

\put(0,0){\includegraphics{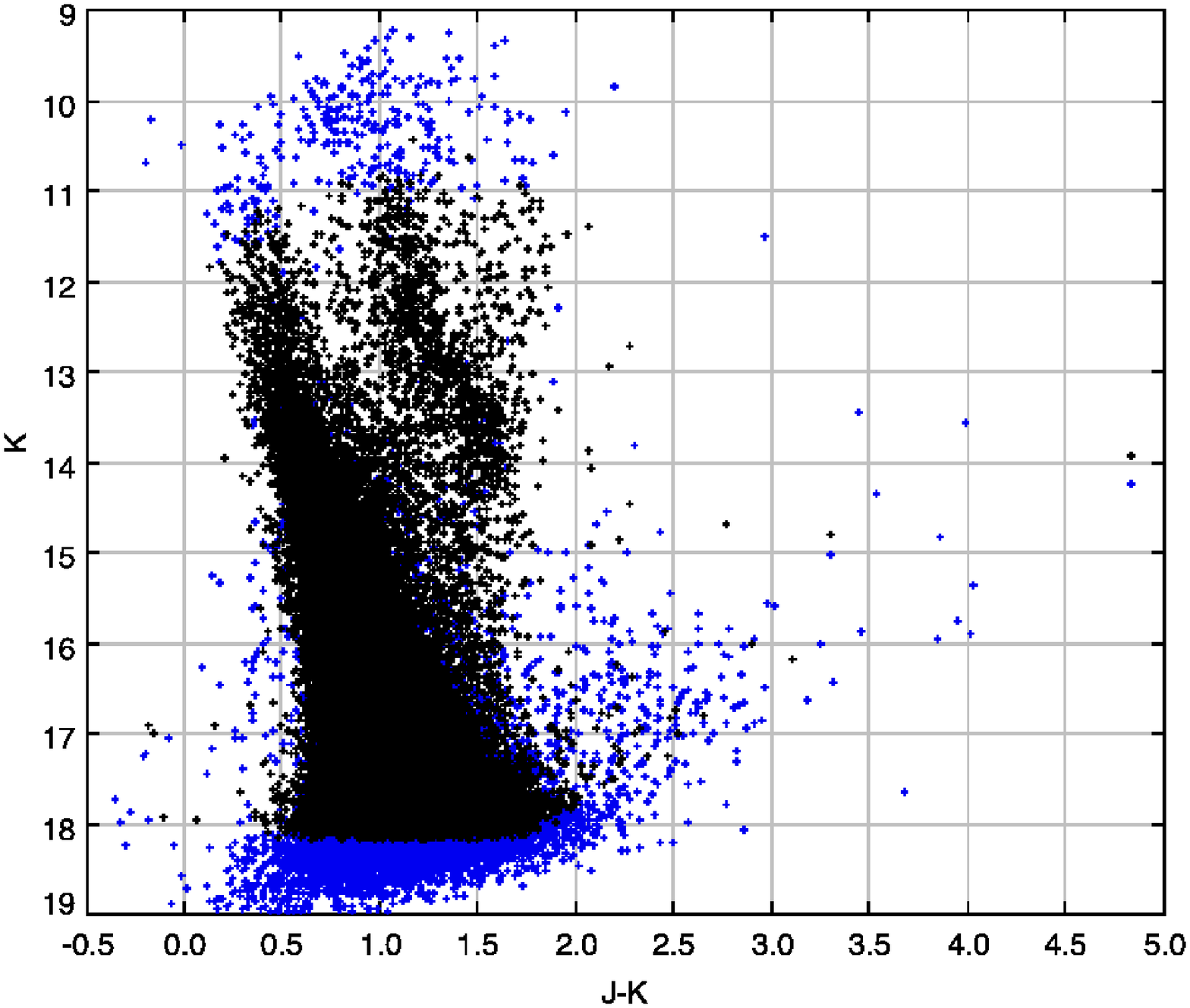}}

\bf
\rm

\end{picture}
\end{center}
\vspace{1cm}
\caption{($l,b$)=(98.4,1). The diagrams for this field appear very similar to those for ($l,b$)=(98.4,0),
in term of reddening. This is consistent with expectations due to the presence of the northern warp, which 
creates an overdensity of stars and interstellar extinction above the plane at large Galactocentric distances, 
relative to what would be expected if the outer Galaxy were centred in the plane.
The points plotted are selected in the same way as in figure 2.}
\end{figure*}


\begin{figure*}
\begin{center}
\begin{picture}(200,160)

\put(0,0){\includegraphics{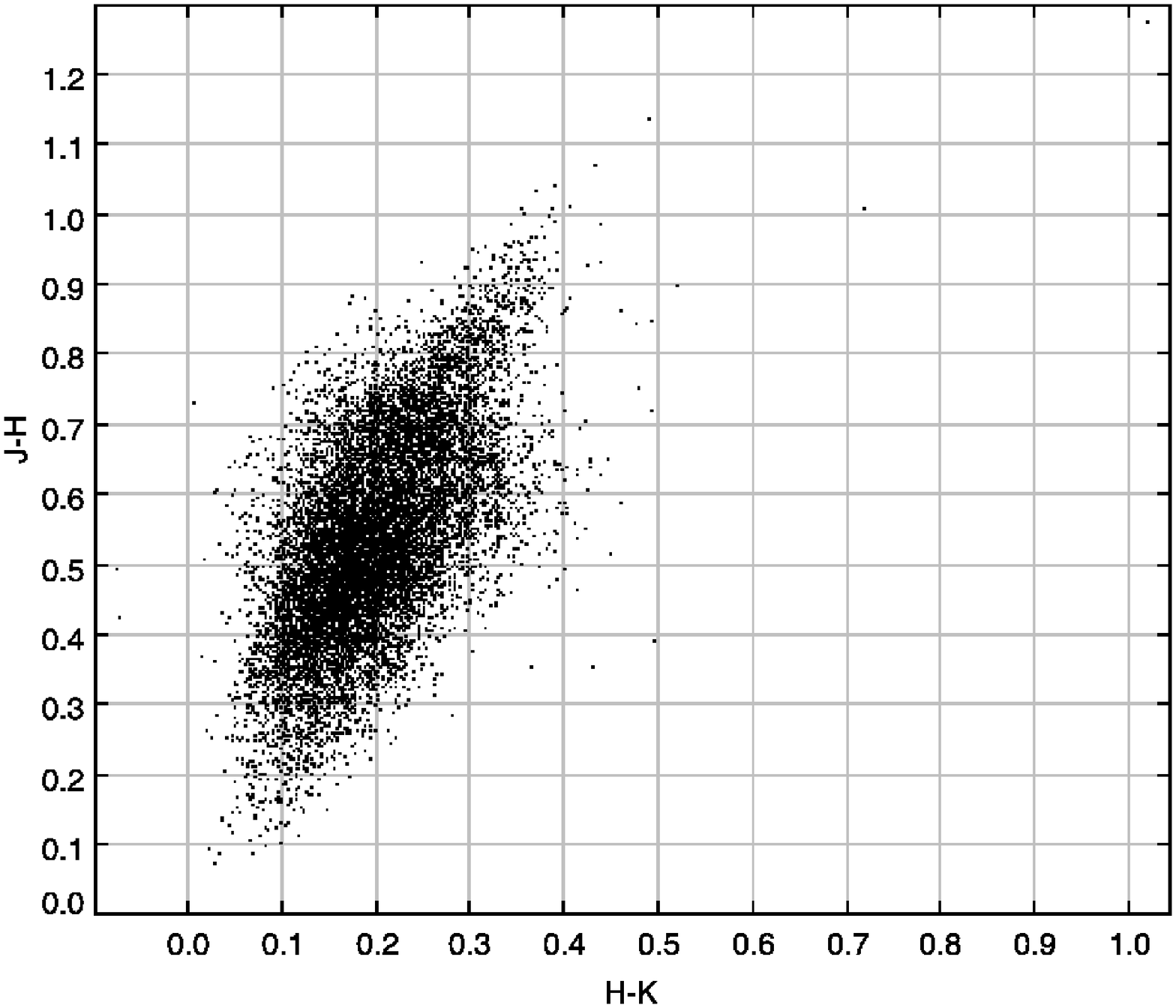}}

\put(0,0){\includegraphics{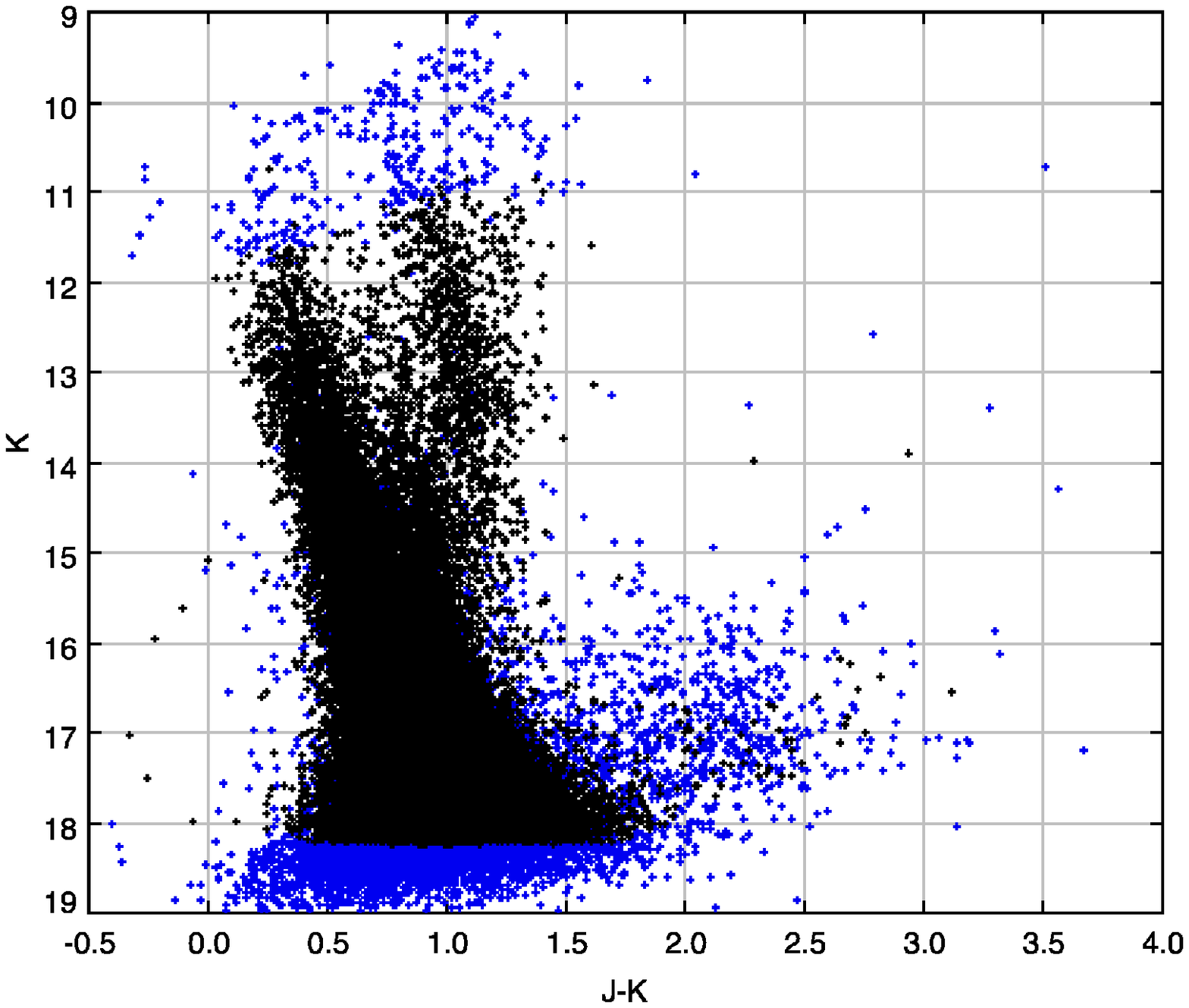}}

\bf
\rm

\end{picture}
\end{center}
\vspace{1cm}
\caption{($l,b$)=(98.4,-1). This field on a line of sight below the mid-plane shows reduced reddening compared 
to figures 9 and 10, since the overdensity due to the warp is located at positive latitudes. The external
galaxy population remains visible at lower right in the colour magnitude diagram and the late K and M dwarf dwarf 
sequence is faintly visible at m$_K$=12.6 to 14.5, {\it(J-K)}=0.7, between the bulk of the FG dwarf sequence and the 
giant branch. The points plotted are selected in the same way as in figure 2.}
\end{figure*}

\vspace{1cm}
\begin{figure*}
\begin{center}
\begin{picture}(200,380)

\put(0,0){\includegraphics{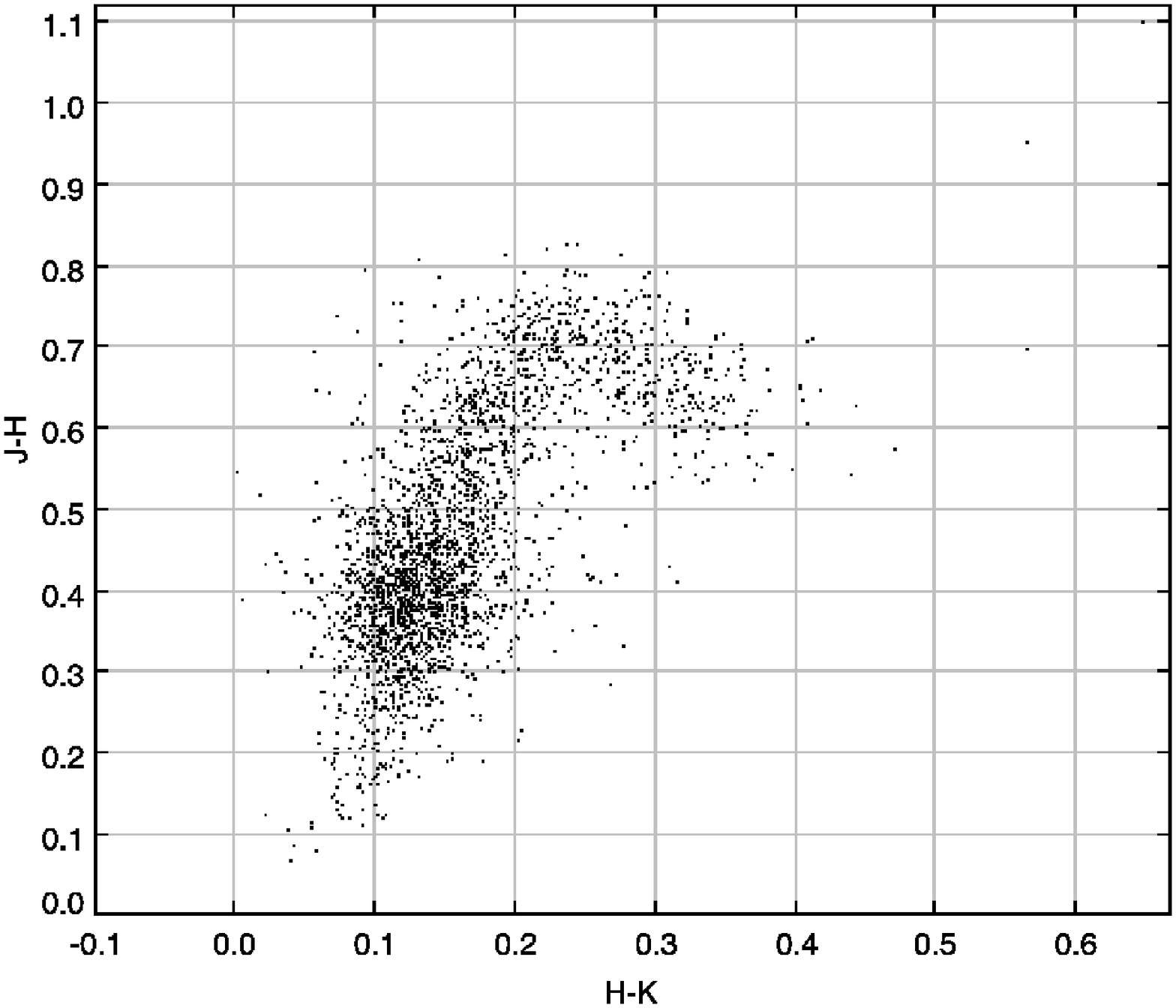}}

\put(0,0){\includegraphics{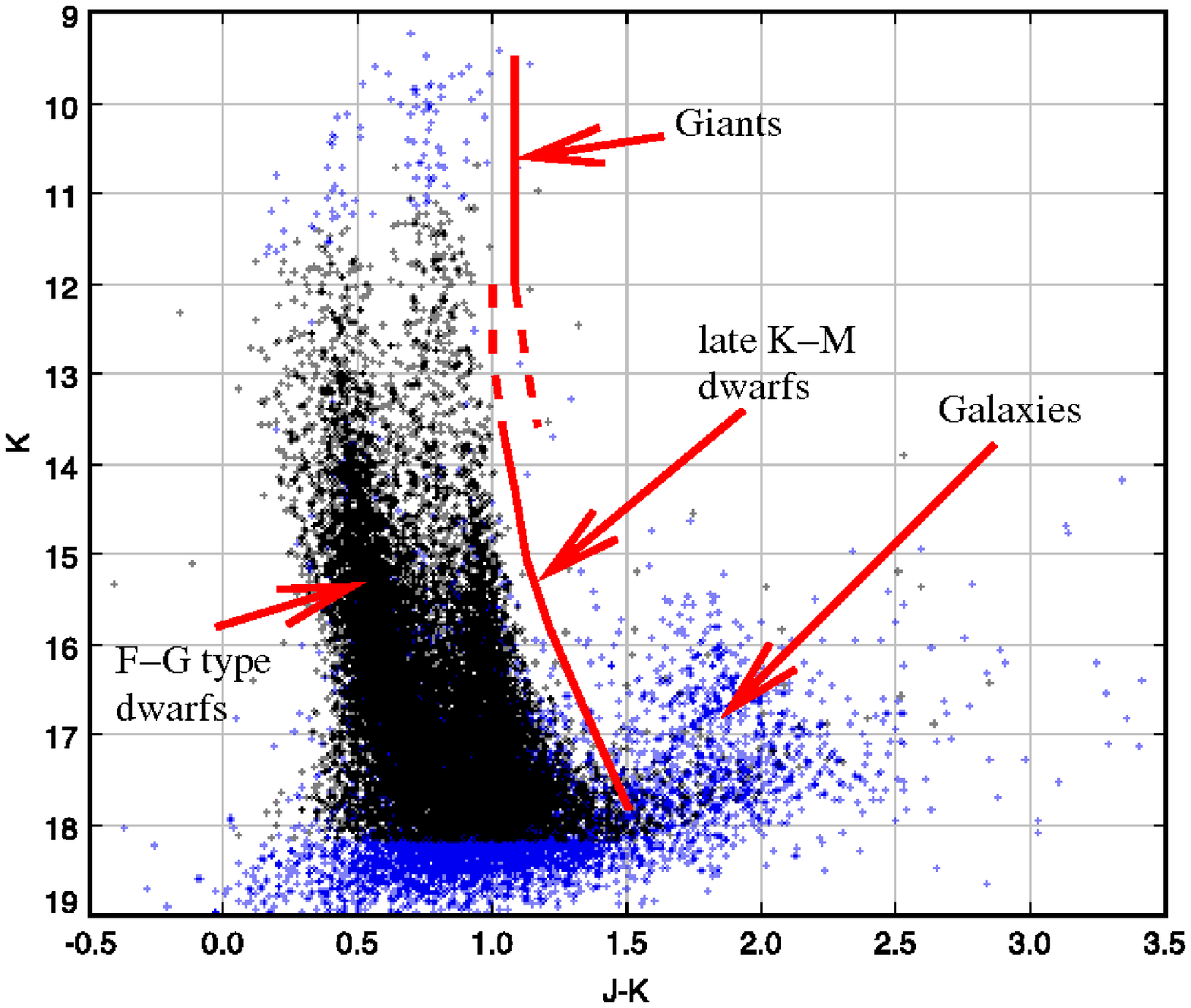}}

\put(0,0){\includegraphics{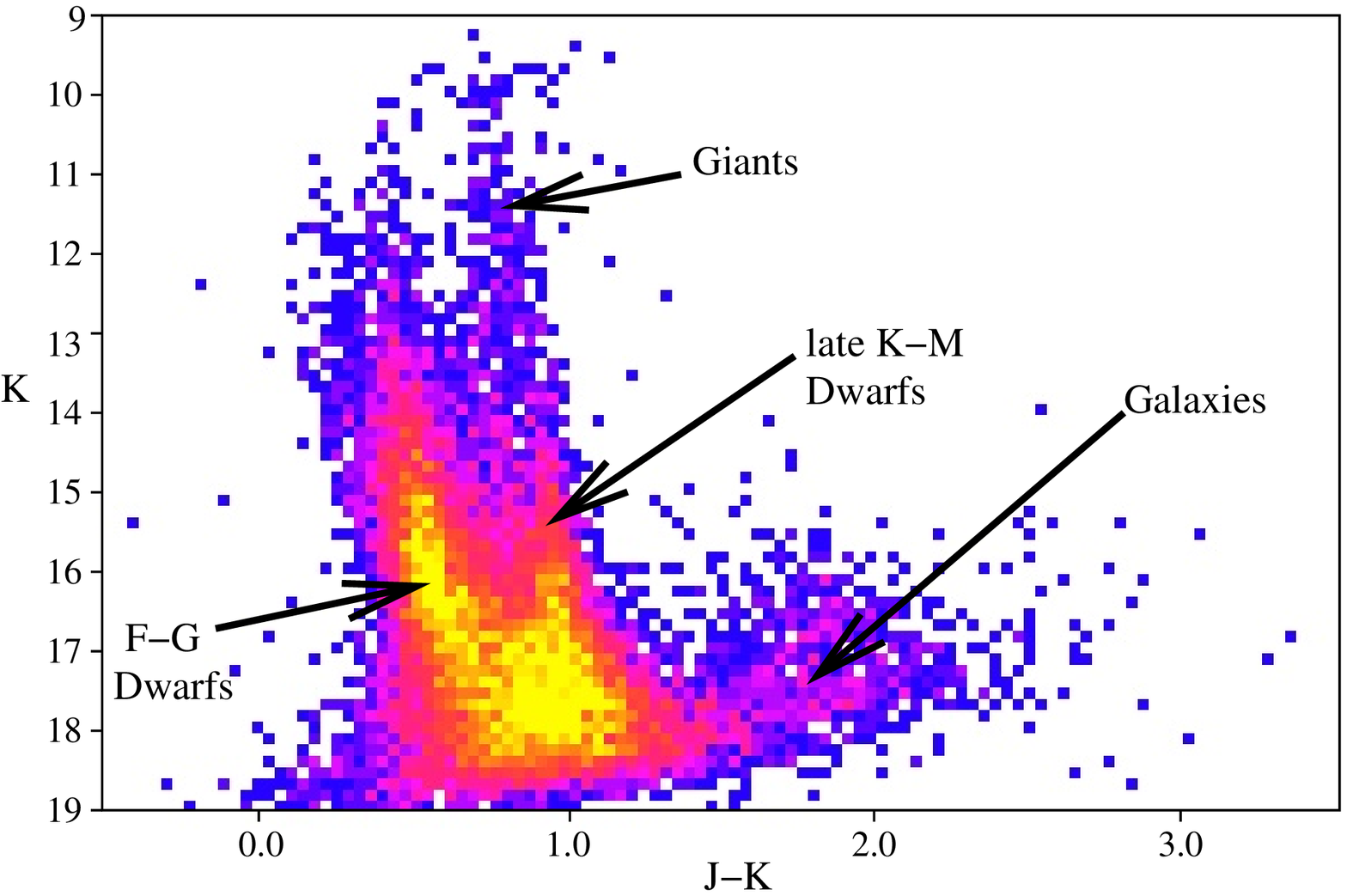}}

\bf
\rm

\end{picture}
\end{center}
\caption{($l,b$)=(171.1,4.75). This field lies on a line of sight almost directly opposite the Galactic centre
and 4.75$^{\circ}$ off the plane. The stellar populations are very different from those in figures 2 to 10:
the right hand sequence of stars in the colour magnitude diagram is now dominated by nearby late K and M dwarfs, as 
opposed to red clump giants. Giants are more numerous than late K and M dwarfs only at m$_K\la$13, but they are not 
clearly distinguished from the dwarfs in the colour magnitude diagram. The left hand sequence is dominated by F and G 
dwarfs. The rapid change in {\it(J-K)} colour with spectral type among late G and early K dwarfs causes the gap 
between the sequences (see figure 14). The lower panel shows a density map of the colour magnitude diagram
(a ``Hess diagram''). It shows that the two sequences join at faint magnitudes, where the early 
K types are more numerous and there also is a larger population of late K and early M dwarfs. The two colour diagram 
shows that there is little reddening in this field. The points plotted are selected in the same way as in figure 2.}
\end{figure*}

\vspace{1.3cm}

\begin{figure*}
\begin{center}
\begin{picture}(200,160)

\put(0,0){\includegraphics{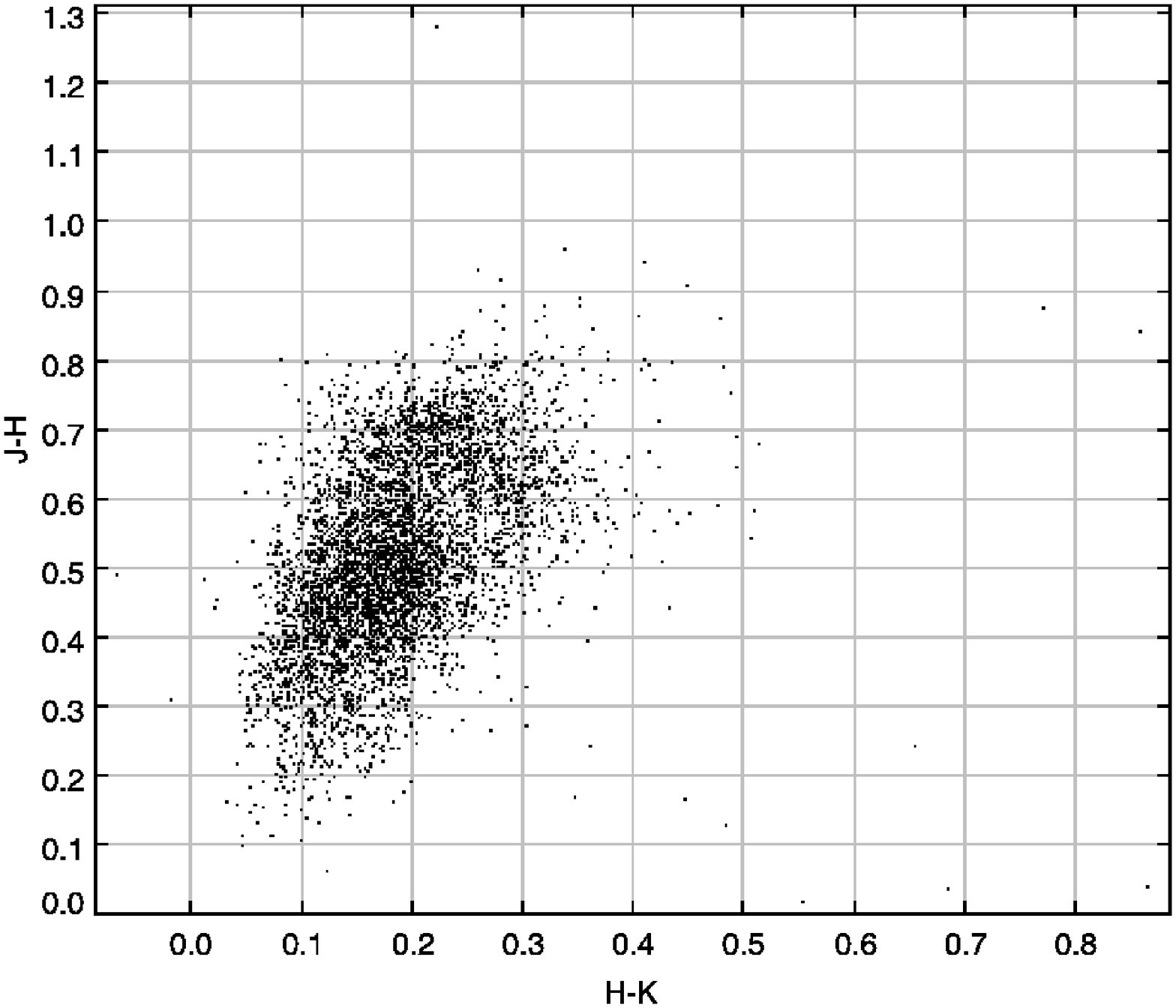}}

\put(0,0){\includegraphics{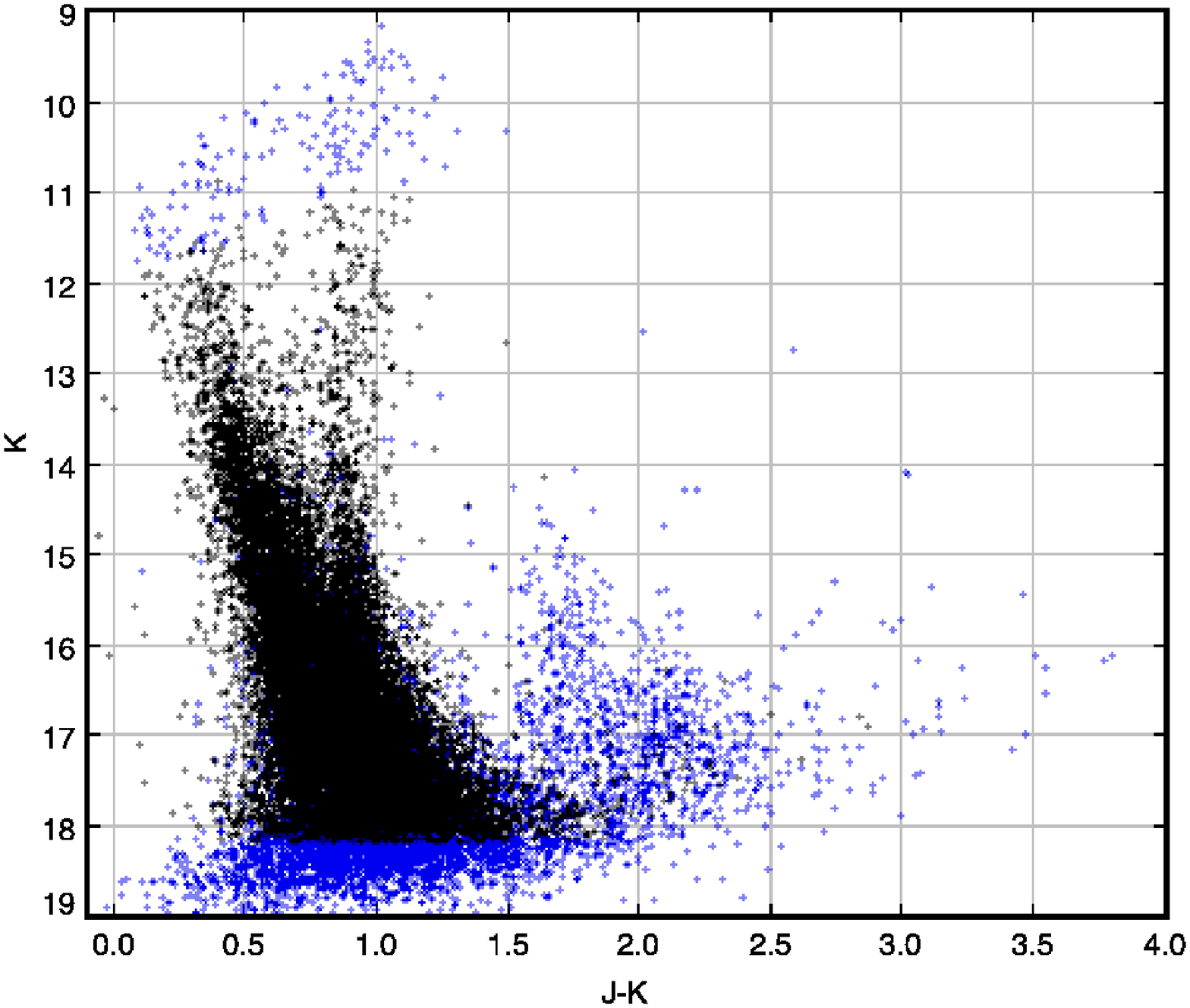}}

\bf
\rm

\end{picture}
\end{center}
\vspace{1.0cm}
\caption{($l,b$)=(173.65,0.65). The colour magnitude diagram for this field shows some similarity to figure 12, but 
exhibits greater extinction due to the location near the mid-plane. The principal sequences observed remain F and 
G dwarfs at the left of the diagram and late K and M dwarfs slightly to the right (mixed with red clump giants at 
m$_K\la$ 14). The population of external galaxies is prominent at {\it (J-K)}=1.5 to 2.5, m$_K>$15. The two
colour diagram appears different to that in figure 12 because there is sufficient extinction to redden the relatively
distant population of F and G dwarfs and move them close to the location of the nearer, less reddened, late K and 
early M dwarfs. This creates a poorly defined mass of stars in which different spectral types are poorly 
distinguished. The points plotted are selected in the same way as in figure 2.}
\end{figure*}

\vspace{1cm}
\begin{figure*}
\begin{center}
\begin{picture}(200,220)

\put(0,0){\includegraphics{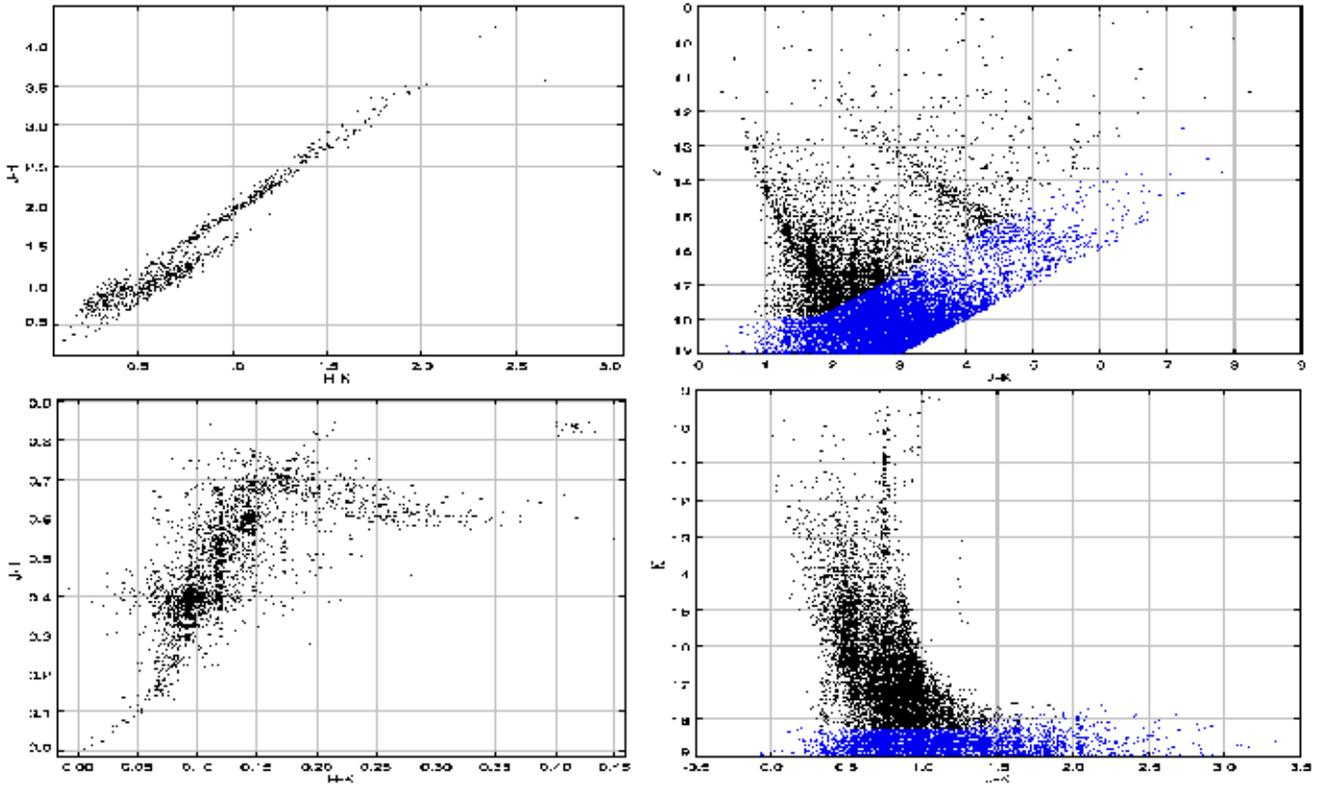}}





\bf
\rm

\end{picture}
\end{center}
\vspace{3.2cm}
\caption{Besan\c{c}on model results with synthetic errors. 
Upper panels: ($l,b$)=(31,0). Lower panels ($l,b$)=(171.1,4.75). These models reproduce the data fairly well,
although our use of synthetic errors based on the archival values has underestimated the true errors at
bright magnitudes (see $\S$A2), causing the various stellar sequences to appear narrower than in the real
data. The small, well defined sequence of rather red white dwarfs with m$_K \sim $14 in the plots for 
($l,b$)=(171.1,4.75) is not observed in figure 12 (see text $\S$3.3). The A type stars near the origin in the lower
left panel are not measured by the GPS because they are saturated. Points with synthetic errors $>$0.2 mag
are plotted in blue.}
\end{figure*}

\begin{figure*}
\begin{center}
\begin{picture}(200,120)

\put(0,0){\includegraphics{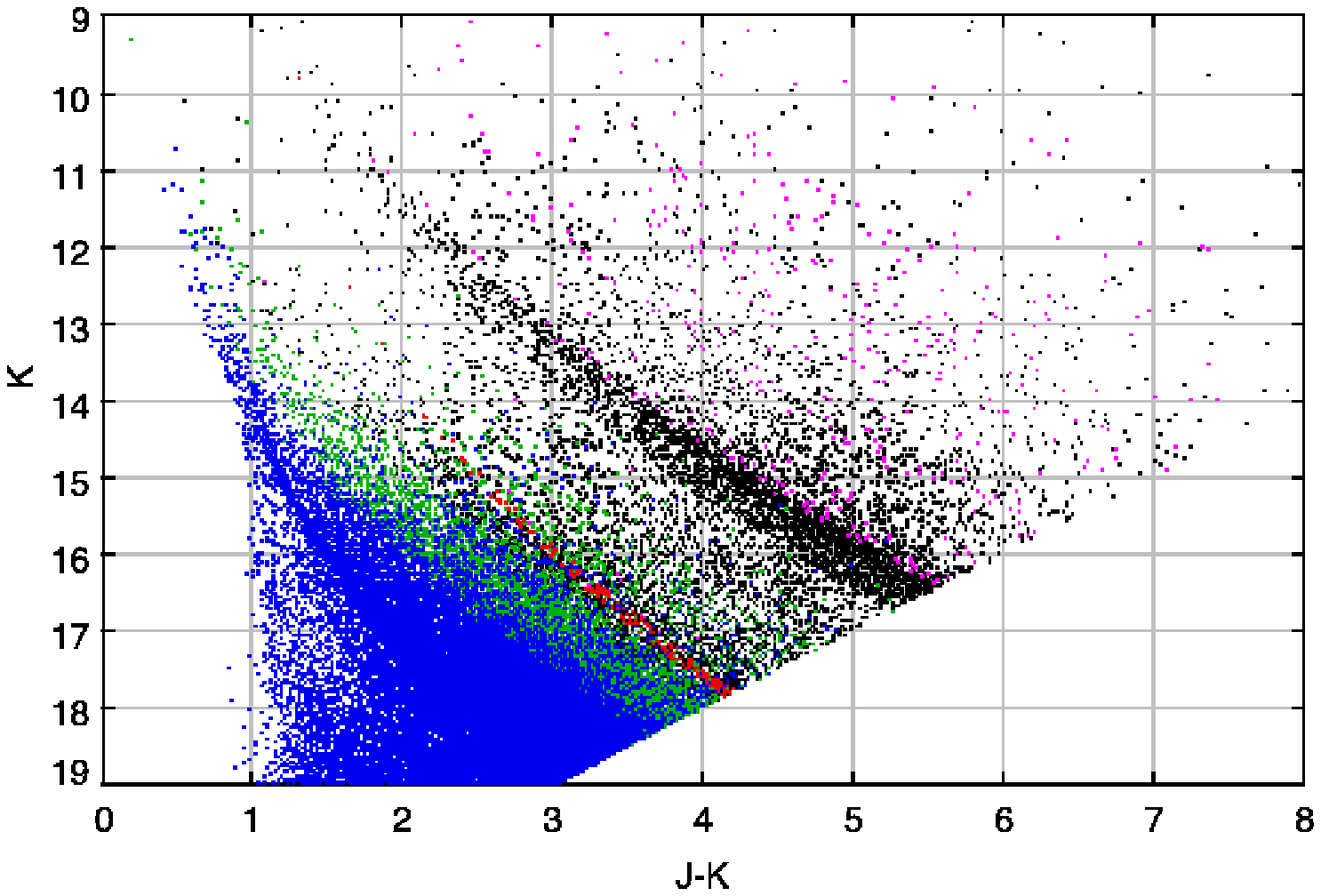}}

\put(0,0){\includegraphics{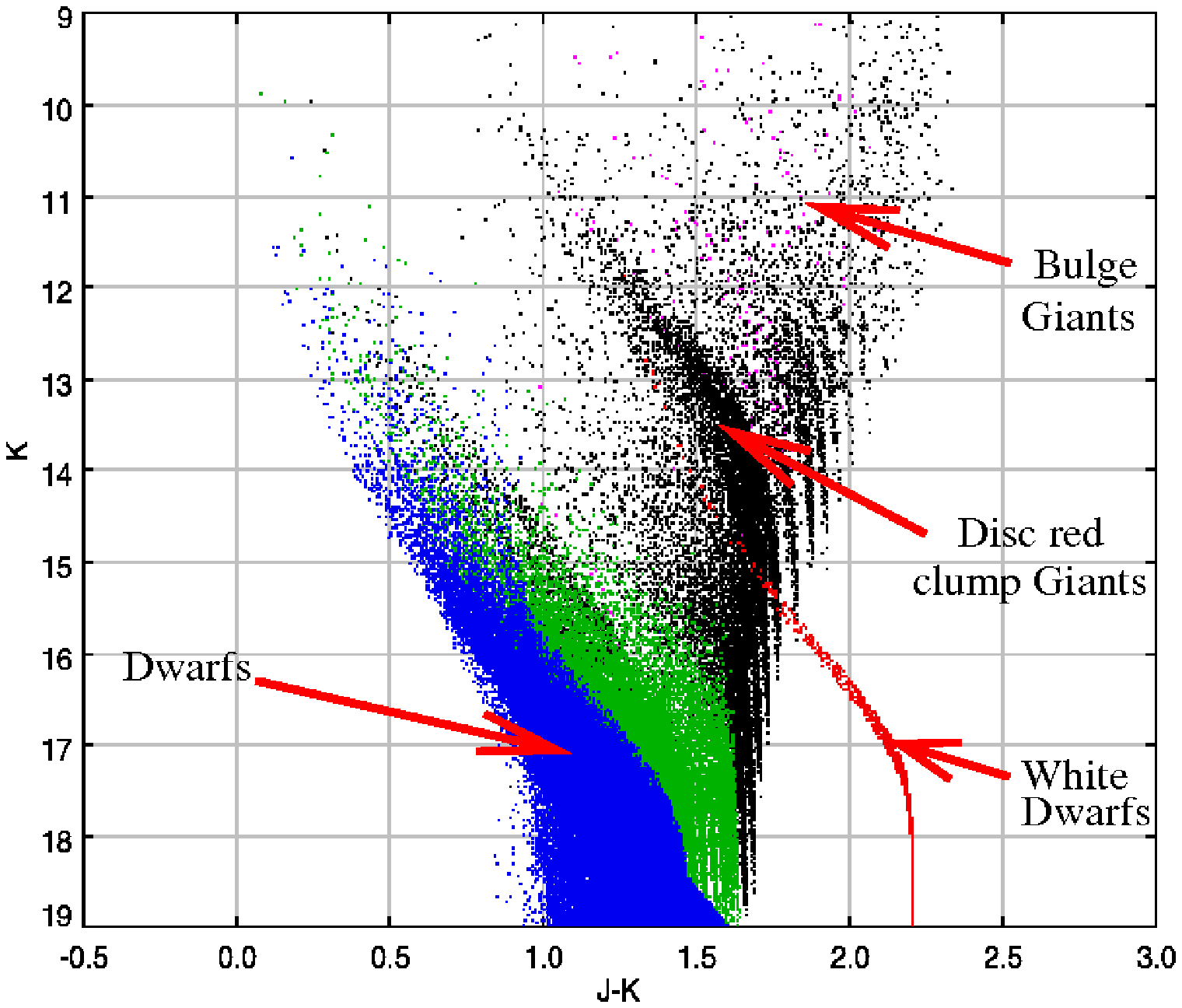}}

\bf
\rm

\end{picture}
\end{center}
\vspace{1.6cm}
\caption{Besan\c{c}on model $K$ vs.{\it(J-K)} diagrams with no errors. Main sequence stars are shown in blue,
luminosity class III giants are shown in black, subgiants (class IV) are shown in green, white dwarfs are shown
in red and luminous giants (classes I and II) are shown in pink.
Left panel: ($l,b$)=(31,0), showing a fairly simple division between dwarfs and subgiants on the left and the red 
clump giant branch to the right.
Right panel: ($l,b$)=(15,1), showing the numerous sub-sequences of giant stars which comprise the red giant clump 
giant branch. The sequence of white dwarfs at the lower right is not observed.}
\end{figure*}

\hspace{-6mm}limits in uncrowded fields are located at 
K=18.0, H=18.75, J=19.5, with uncertainties of about 0.2 magnitudes. Individual fields in the outer Galaxy should 
always be within 0.3 magnitudes of these typical values, since only a modest range of observing conditions (seeing
and transparency) is permitted, as described in the data release papers. We do not attempt a more precise 
quantification of the mode and spread of the sensitivity at this stage since there is limited data for the outer 
Galaxy in DR2, and the sensitivity in future data releases is likely to change slightly. The seeing conditions 
required to begin an observation at $l>$107$^\circ$ were relaxed from a maximum FWHM of 0.8 arcsec (measured in
the K bandpass) to 1.0 arcsec from February 2007 in order to speed the survey progress.

It is clear that the survey is much less sensitive near the Galactic centre than in uncrowded fields. However
the 90\% completeness limits and modal magnitudes are usually degraded by no more than 1 magnitude in fields
at $l>$30$^{\circ}$, even in the mid-plane at $b$=0$^{\circ}$.

The effects of extinction on the histograms are apparent in the change in the relative numbers of J, H  and K
band detections in different parts of the Galaxy. The K band source density is the highest in the fields at 
$l<$30$^{\circ}$. At $l>$30$^{\circ}$ the H band source density is higher (at least for the fields at 
$|b|<$1$^{\circ}$ for which multi-band data presently exist). The source density in the J band becomes 
comparable to the H band at $l>$90$^{\circ}$. 

\subsection{Stellar and extragalactic populations}

The two colour diagrams and colour magnitude diagrams in figures 2-13 allow us to clearly identify the
main stellar populations detected in the GPS and detect changes in extinction as a function of distance 
along different lines of sight. Some readers will be familiar with similar diagrams
constructed from data in the DENIS and 2MASS surveys (e.g. Ojha et al.2000) or from surveys of the stellar 
populations in external galaxies. In general, the populations detected by the GPS consist of less luminous stars 
than those in most previously published wide field infrared surveys. Consequently, the diagrams look quite
different and their interpretation is often less obvious than might be supposed.
We have chosen to plot $K$ vs. {\it(J-K)} colour magnitudes since the other near infrared colours
({\it(J-H)} and {\it(H-K)}) provide less separation between the various stellar populations and add
little to this discussion. In general, near infrared diagrams can separate the nearby late dwarf, distant early 
dwarf, and red giant populations, with more complex behaviour depending on the sightline.

In this section we first summarise the general properties of the two colour diagrams and then discuss the colour 
magnitude diagrams, beginning with the inner Galactic fields.

The two colour digrams for the inner Galactic fields in figures 2 to 8 all show two parallel sequences of stars 
extending from the lower left to the upper right. The upper sequence is composed mainly of giants (mostly of the type
known as ``red clump giants'') while the lower sequence consists of dwarf stars. These long sequences are due to 
interstellar reddening: they lie {\it approximately} parallel to the reddening vector and can be used to estimate the 
interstellar extinction toward each star by comparing the observed colours with those of unreddened stars.
The colours of a sample of unreddened main sequence and giant stars from Hewett et al.(2006) are 
plotted in figure 5 for illustration, along with the reddening vector of Rieke \& Lebofsky (1985). Late type giants 
have $\sim$0.15-0.3 mag larger {\it(J-H)} colours than dwarfs of the same spectral type, due to the effect of their low 
surface gravities on the opacity of the H$^{-}$ molecular ion. The more distant dwarfs also tend to have earlier spectral 
types than the giants, which also contributes to the difference in {\it(J-H)} colours.

A careful inspection of the reddened sequences of stars in the two colour diagrams shows that they are not
quite straight lines. This is due to two effects. The first is that the nature of the detected stellar population 
shifts towards more luminous stars with increasing distance and reddening. The clump of stars seen at the lower left
of most of the two colour diagrams is composed mainly of nearby G, K and early M type dwarfs, which have a high space
density. As we move further along the dwarf sequence the population shifts towards early G, F and A type dwarfs,
which have a lower space density but are detected at larger distances. The dwarf sequence ends at a lower reddening 
than the giant sequence because the dwarfs are less luminous. This change in populations means that the the dwarf sequence
has a slightly shallower gradient than the reddening vector. The second reason that the gradient is not linear is that
the effective bandpasses of the filters change with increasing reddening when observing a continuum source, so 
reddening vectors become curved as the extinction increases (e.g. Kaas et al.1999). The precise form of the
curve differs for sources with different spectra, so there is no universal reddening vector, see e.g. Drew et al.(2005).
The giant sequence generally begins at a slightly higher reddening than the clump of nearby late type dwarfs, since 
giant stars have a lower space density. This population can in principle be used to measure the reddening curve
for red clump giants. In general a standard linear interstellar reddening vector is 
adequate for most applications and this approach is used in this work. A more sophisticated approach using a model of the 
instrumental transmission and stellar model atmospheres will be developed in a future paper on Galactic structure.

All of the two colour diagrams show samples selected using the least complete, most reliable selection described in 
$\S$A3, with errors $<$0.05 mag in the colours on each axis, in order to distinguish the closely spaced 
stellar populations. Including fainter sources with less precise photometry simply increases the scatter in the 
diagrams, as shown in figure A1(a), and extends the dwarf and giant sequences to slightly higher reddening values 
at the upper right. 

By contrast, the colour magnitude diagrams show all the three band detections within the chosen axis ranges, except
those classified as ``noise'' detections (see $\S$A3).
Sources with more reliable photometry are shown in black, while sources with less precise photometry are shown 
in blue. Sources are put in the ``less 
precise'' selection if they fail to satisfy any of the following criteria. (1) Photometric uncertainties 
$<$0.2 mag in the two bands used in each diagram. The corresponding less precise sources are faint objects 
lying in a group at the bottom of each diagram. (2) {\ssq ppErrbits}$<$256. This parameter is decribed in
$\S$A3. Most sources failing this criterion
do so because they are flagged as close to saturation, and therefore lie in a group at the top of each diagram. 
Inspection of the diagrams shows that they nonetheless often play a valuable role in defining the different 
stellar populations, indicating that the errors in their photometry need not be extreme.  
(3) {\ssq pstar}$>$0.9. Sources only fail this criterion if they appear significantly non-stellar in at least
one of the three passbands, usually because they are either part of a marginally resolved stellar pair
or a resolved external galaxy. Marginally resolved stellar pairs are evenly spread among the sources with more reliable 
photometry but extend further to the left and right sides of the diagrams. The resolved galaxies are located
in a group at the lower right of the colour magnitude diagrams. The extragalactic group
is clearly seen only in the fields at $l>90^{\circ}$.

The colour magnitude diagrams all show at least two principal sequences of stars. On first inspection, one 
might suppose that these correspond to luminosity class V dwarfs on the left side of the diagrams and luminosity
class III red clump giants on the right. This is indeed the case for the inner Galactic fields, which we will 
now discuss. However, the situation is more complicated in the outer Galaxy, as we shall see later.

The main feature of the colour magnitude diagrams for $l$=15, 31, 55 and 98.4 (see figures 3 to 11) is the division into 
two principal sequences corresponding to dwarfs on the left and giants on the right. The Besan\c{c}on model indicates 
that the majority of the giants are K1III-K2III types. These stars occupy a sharp and well defined peak in the 
absolute magnitude function of Class III giants, with absolute magnitude $M_K$=-1.65 (e.g. Lopez-Corredoira et al.2002),
though we caution that both flux and colour have some dependence on metallicity and age (Salaris \& Girardi 2002; 
Luck et al.2007; Gullieuszik et al.2007). These giants form the well defined ``red clump'' in H-R diagrams for globular
clusters and galaxies toward which the extinction is fairly uniform. They are therefore a useful tracer of stellar 
distance, extinction and population density, (see Lopez-Corredoira et al.2002) but only in fields in which they can be 
distinguished from dwarfs and other types of giant.

The dwarf and giant sequences do not always show a smooth increase in reddening with distance. The giant 
sequences in the $(l,b)$=(55,1) and (55,0) fields (see figures 8 and 7) show an abrupt increase in reddening between 
$m_K$=13 and $m_K$=14. Such features can be explained by discrete clouds of interstellar dust. A similar feature occurs 
in the $(l,b)$=(15,0) field (figure 3), where there is a sudden increase in reddening at $m_K$=12.75, {\it(J-K)}=2.25. 
The distance to the cloud and the extinction in front of it and within it
can be simply determined from the known luminosity and colour of the red clump giants and by adopting 
an extinction law that is appropriate for the diffuse interstellar medium (e.g. Rieke \& Lebofsky 1985; Cardelli,
Clayton \& Mathis 1989). (Here we neglect the effect of the change in effective bandpasses at high extinction that were 
mentioned above). In the case of the field at $(l,b)$=(55,1) the cloud is located at a heliocentric distance of 8~kpc and 
it introduces a visual extinction of A$_V$=7.5 magnitudes. Since the feature at $m_K$=13-14 is well populated with
red clump giants it is likely to be caused by a large cloud of atomic gas rather than a relatively small dense
molecular cloud. Inspection of the $^{13}$CO data cube from the BU-FCRAO survey (see e.g. Shah et al.2004) shows weak 
but uniform emission across the field, with much brighter emission outside the field on both sides. The emission
is too weak by more than an order of magnitude to explain the infrared extinction, which tends to support the 
interpretation of the cloud as an atomic feature rather than a molecular one.
 
An advantage of the near infared determination of extinction (as opposed to optical determinations) is that a 
``universal'' extinction law applies to both the diffuse interstellar medium and 
molecular clouds (Cardelli et al.1989). This is due to the fact that the dust grains responsible for extinction
have a size much smaller than the wavelength of observation for wavelengths $\lambda \ga 1~\mu$m (except in extreme
environments such as the accretion discs around protostars). Small variations in the extinction law can still occur,
e.g. due to variations in grain composition (and hence refractive index). This may explain why some authors have found 
values of the colour excess ratio E(J-H)/E(H-K) ranging from 1.32 to 2.08 in some dark clouds (e.g. Moore et al.2005; 
Racca, G\'{o}mez \& Kenyon 2002), bracketing the value of 1.70 found in Rieke \& Lebofsky (1985). However, most 
measurements, e.g. Naoi et al.(2006) find that E(J-H)/E(H-K) lies in the range 1.6 to 1.7, and they note that some of 
the reported variations are attributable to differences in photometric systems.
 
Some of the colour magnitude diagrams show an abrupt change in the slope of the giant branch for another reason.
For example, the plots in figures 4 and 9 for $(l,b)$=(15,1) and (98.4,0) show diagonal sequences of red clump giants at
$m_K<12$ and $m_K<13$ respectively, and then a change to an almost vertical giant sequence at fainter magnitudes.
This change is due to the red clump giant population leaving the main body of the Galactic disk
at large distances. The position of stars along the red clump giant sequence is determined mainly by
distance. At heliocentric distances $d>$12~kpc, the sequence in the $(l,b)$=(15,1) plot rises far enough above the dusty 
Galactic Plane for extinction to become almost independent of distance. The similar behaviour of the $(l,b)$=(98.4,0) 
plot is attributed to the northern Galactic warp, which moves the dusty plane away from $b$=0 at large 
Galactocentric radii. In figure 15 we show a Besan\c{c}on model for $(l,b)$=(15,1) which reproduces and clarifies the 
behaviour at $d>$12~kpc. We see that the red clump giants are split into a number of faint parallel sub-sequences. 
These correspond to the various evolutionary groups: core helium burning horizontal branch giants, fainter hydrogen 
shell burning stars ascending the red giant branch and asymptotic giant branch stars with hydrogen and helium shell 
burning (e.g. Robin 1989). Each evolutionary group is also split into multiple sequences due to binning by 
spectral sub-type in the Besan\c{c}on model. Each sequence has a slightly different effective temperature and some, 
e.g. the K type red giant branch sequences, have lower luminosities. At $d>$12~kpc the sub-sequences continue to 
become slightly redder at fainter 
magnitudes (i.e. larger distances) since there is a small amount of dust above the Galactic plane. However the 
sub-sequences all end $\sim$2 magnitudes below the change in slope, as the old disc population declines to zero 
at large distances above the plane. At fainter magnitudes, new 
sub-sequences of slightly fainter giants with slightly bluer {\it(J-K)} colours (which we believe are stars ascending
the red giant branch) come to dominate the giant branch. In the GPS data the adjacent sub-sequences merge together, 
with the result that in the vertical part of the giant sequence the stars become slightly bluer in {\it(J-K)} colour 
at fainter magnitudes.

The existence of less luminous hydrogen shell burning stars in the red clump indicates that the red clump method
for determining distances, extinction and stellar population density should be used with caution at faint magnitudes.
The Besan\c{c}on model catalogue displayed in figure 15 for (l,b)=(15,1) indicates that at m$_K$\ga 15.5 
these less luminous giants dominate the red clump. 

The field near the Galactic Centre at $(l,b)$=(-0.3,0) has a rather different colour magnitude diagram
from the other fields. Three principal sequences are apparent in figure 2. The largest population is giants in the 
Galactic bulge, with $3<${\it(J-K)}$<7$. According to the Besan\c{c}on model these have a wide range of spectral 
types and metallicities. However most suffer similar reddening, since they are all at distances of 7-10~kpc 
(according to the model). The other populations are the nearby disc dwarfs at the left side of the diagram and the 
disc red clump giants, which form a diagonal sequence with {\it(J-K)} colours intermediate between the other two 
groups. 

Moving to the outer Galactic fields, we consider first the diagrams in figure 12, for $(l,b)$=(171.1, 4.75). This 
shows a field with negligble reddening, which can be discerned from the corresponding two colour diagram. Two principal 
sequences of stars are apparent in the corresponding colour magnitude diagram, mostly with reliable photometry. A more 
diffuse clump of sources with ``less reliable'' photometry is situated near {\it(J-K)}$\approx$1.8, m$_K$=17. This clump
is composed of external galaxies ({\ssq mergedClass=+1, pstar$\ll$ 0.9}), and the corresponding luminosity 
function increases down to the completeness limit (see $\S$4.5).
Inspection of the Besan\c{c}on model catalogue for these coordinates (plotted in the lower panels of figure 14) indicates 
that the overwhelming majority of the sources in both of the two 
principal sequences are actually luminosity class V dwarfs. The left hand sequence is composed of F and G type dwarfs
(and some subgiants of luminosity class IV). The right hand sequence is composed mainly of M and late K type
dwarfs, with only a small proportion of giant stars.  Luminosity class III giants are restricted to the upper 
portion of the right hand sequence with $m_K\la$13.5. The dwarf component of this sequence begins at $m_K\approx$12 
and extends down to the faintest sources at the bottom of the diagram. The overlapping dwarf and giant populations in the
right hand sequence are not distinguishable in the magnitude range 12$<m_K<$13.5. The two colour
diagram also fails to separate them, although it should in principle reveal any large population of early K-type
red clump giants as an overdensity near {\it(J-H)}=0.57, {\it(H-K)}=0.12. 

The reason that two sequences of dwarf stars are detected at $(l,b)$=(171.1, 4.75) is not immediately obvious.
Inspection of the density map of the colour magnitude diagram shown in the lower panel of figure 12 (a ``Hess diagram'')
and the Besan\c{c}on model shows that there is a connection between the two sequences at 17$<m_K<$18. However, 
the M type dwarfs, which are the most numerous population in the diagram, extend above and 
below this connection. Nearby late K and early M-type dwarfs have a narrow range of colours ({\it(J-K)}$\approx$0.9) and 
they are sufficiently numerous to define the right hand dwarf branch at 12$<m_K<$17. Similarly the distant F and early
G-type dwarfs observed in this magnitude range also have a narrow range of colours. By contrast the late G and 
early K type dwarfs are spread over a wider range of {\it(J-K)} colours due to the steeper colour-temperature relation
for these types. Their numbers are similar to the earlier type dwarfs in this magnitude range, so the rapid
change in colour with temperature creates the sparsely populated gap between the two branches and the bimodal colour 
distribution. At $m_K$$>$17 the late G and early K type dwarfs are sufficiently numerous to fill the gap.

The two sequences of dwarfs are present in all the colour magnitude diagrams plotted
but they are less obvious at $l<100^{\circ}$. This is due to a combination of factors: (i) detectable 
F and G type dwarfs greatly outnumber M dwarfs near the mid-plane at $l<100^{\circ}$; (ii) there is a trend of 
increasing reddening with decreasing Galactic longitude, which causes the distant highly reddened F and G 
dwarf sequence to merge with K and M dwarf sequence;
(iii) the population of distant red clump giants is much larger and produces a more obvious second sequence of 
stars to the right of the dwarf sequence that is clearly separated from the common F, G, K and M type dwarfs by 
high reddening. The K and M type dwarf branch is still detectable in some of the colour magnitude diagrams for 
$l$=98.4 and $l$=55 (see figures 8, 9, 10 and 11) appearing as a faint protuberance above the mass of dwarfs at 
$m_K \approx 14$.

The clump of external galaxies seen in the colour magnitude diagrams for the $l>90^{\circ}$ fields is not detected at 
smaller longitudes, owing to the combination of higher extinction through the plane and the reduced sensitivity
caused by source confusion.

The two colour diagrams show clearly the trend of increasing extinction with decreasing Galactic longitude and 
latitude. The dependence on Galactic latitude is apparent even in the outer Galaxy, where extinction is relatively 
low. In the $(l,b)$=(171.1, 4.75) field (see figure 12) the densest cluster of sources near {\it(J-H)}=0.4 is 
composed of F and G type dwarfs. In the $(l,b)$=(173.65,0.65) field (see figure 13), we see that the F and G types have 
moved up to {\it(J-H)}=0.5, owing 
to the significant extinction present in the mid-plane toward these relatively distant sources (1 to 6 kpc). By contrast,
the K and M type dwarfs with {\it(J-H)}=0.5 to 0.7 lie in the same place in both diagrams, owing to
the much lower extinction toward these nearer and intrinsically fainter sources
($<$3~kpc distance). The effect of the northern Galactic warp is apparent in the two colour diagrams
for $l$=98.4. The extinction at $(l,b)$=(98.4,1) is similar to that at $(l,b)$=(98.4,0) and clearly
greater than that at $(l,b)$=(98.4,-1), as shown in figures 9, 10 and 11.

\subsection{Further comparison with the Besan\c{c}on Models}

The Besan\c{c}on stellar population model has been used to aid the interpretion
of the GPS data discussed above. In figures 14 and 15 we show some of the diagrams produced by the
model in order to illustrate the quality and the limitations of such comparisons. The results in figure 14 have 
synthetic errors included in the simulation. The error curve used was based on a fit to the errors as a 
function of magnitude given in the WSA for each passband in the appropriate field. As noted in $\S$A2, this 
will tend to underestimate the true errors (particularly for bright stars), but it is adequate for a simple 
comparison.

The visual extinction per kiloparsec in the diffuse interstellar medium (ISM) is a free parameter in the 
Besan\c{c}on models. Discrete clouds may also be added to model strong extinction features, as noted earlier.
The ISM extinction parameter was adjusted by trial and error until the best combined match by eye to the observed 
luminosity function and the K vs.{\it (J-K)} colour magnitude diagram was produced. 
The derived extinction was A$_V$=2.25 mag/kpc for the ($l,b$)=(31,0) simulation and A$_V$=0.7 mag/kpc for
the ($l,b$)=(171.1,4.75) simulation. The models use the extinction law of Mathis (1990), assuming R$_V$=3.1, so 
the equivalent K band extinctions are A$_K$=0.242 mag/kpc and A$_K$=0.075 mag/kpc.
The default value of A$_V$=0.7 mag/kpc is appropriate for the diffuse ISM
and will usually be correct in fields with low extinction. Higher values are required to fit fields
with high extinction. In reality high extinction is expected to be caused by numerous discrete molecular 
clouds, but often the individual clouds have too small an effect to be resolved in the data.
We note that some authors quote a slightly higher value for the extinction in the diffuse ISM, e.g.
Whittet (1990) quotes 0.8 mag/kpc. This difference is likely to be due to the choice of distance scale
in the Besan\c{c}on models, which places the Galactic centre at $d$=8.5~kpc from the sun. The distance to the 
supermassive black hole at the Galactic Centre has been reliably determined by matching the observed angular 
size of stellar orbits to the corresponding stellar radial velocities: $d$=7.62$\pm$0.32~kpc (Eisenhauer et al.
2005). Hence some users may prefer to scale the Besan\c{c}on distances and extinction values accordingly.

The two colour diagrams in figure 14 for ($l,b$)=(31,0) and (171.1,4.75) are a good quantitative match to the 
data. The blue end of the model dwarf sequence for ($l,b$)=(171.1,4.75) appears sharper than the observed
sequence in figure 12, since the photometric errors of these hot and bright stars are underestimated in 
the model. The other noticeable discrepancies in the (171.1,4.75) plot are: 
(i) the appearance of a cluster of sources at the upper
right of the diagram ({\it(H-K)}=0.45, {\it(J-H)}=0.85), which is not observed; and (ii) the sparesely populated
sequence of stars extending off the top of the diagram at {\it(H-K)}=0.2-0.25. The first group represents 
K type white dwarf stars in the model, which all have very similar masses and spectral types. The 
data suggests that the real white dwarf population has less uniform parameters than in the model. The second group 
represents late K-type
giants in the model. These are not included in figure 12 because they are too bright (i.e. heavily 
saturated sources). The two colour diagram for the ($l,b$)=(31,0) model shows a similar small population of 
luminous giants at {\it(H-K)}=2.5-3.0 (late M-types in that case) that is not seen in figure 5 for the same 
reason. Such bright populations are better studied with the 2MASS or DENIS surveys.

The colour magnitude diagram for the ($l,b$)=(31,0) model is in close agreement with the data in figure 5.
The only difference is that slope of the red clump giant sequence at the upper right is
more constant in the model than in reality. This reflects the patchy nature of interstellar extinction.
The colour magnitude diagram for the ($l,b$)=(171.1,4.75) model is also in fair agreement with the corresponding
data in figure 12. The following differences are obvious however. (1) The external galaxies at the bottom right 
of the plot in figure 12 are not in the Besan\c{c}on model. (2) The two sequences of dwarf stars discussed earlier 
appear as three sequences at 0.45$<${\it(J-K)}$<$0.9 in the model. The sparsely populated middle sequence at 
{\it(J-K)}=0.65 corresponds to early K dwarfs in the model. This may well be a real feature but it is not apparent 
in figure 12 due to the larger photometric errors in the real data. (3) The short vertical sequence of red sources 
at {\it(J-K)}=1.25 seen in the model are the K-type white dwarfs; these are not seen in the data, as noted
earlier.

In figure 15 we show Besan\c{c}on synthetic colour magnitude diagrams with no errors added, for
$(l,b)$=(31,0) and (15,1). These illustrate the discrete binning of sources in the Besan\c{c}on model
grid. The synthetic plots for $(l,b)$=(31,0) appear very similar in figures 14 and 15 since distance and interstellar 
reddening introduce numerous degeneracies in magnitude and colour between different spectral types.
In contrast, the plot for $(l,b)$=(15,1) clearly shows the numerous red clump giant sub-sequences (see discussion
in $\S$3.2). These appear well defined in the model because each sub-type of giant star has a unique luminosity, effective
temperature and colour in the model and there is no additional extinction to merge the sub-sequences in the dust 
free region above the Galactic plane. Some of the numerous sub-sequences correspond to the various evolutionary 
stages on the HR diagram, as noted earlier. In addition, the effect of binning by spectral sub-type in the model creates 
an artificial division which further multiplies the number of sub-sequences. We note that the more luminous giants 
above the most populated part of the red giant clump appear to be more numerous in the data than the model.

\section{Demonstration Science}

\subsection{Identification of Young Stellar Objects with the UKIDSS GPS and the {\it Spitzer} GLIMPSE survey}

\begin{figure*}
\begin{center}
\vspace{-4mm}
\includegraphics[width=0.85\textwidth]{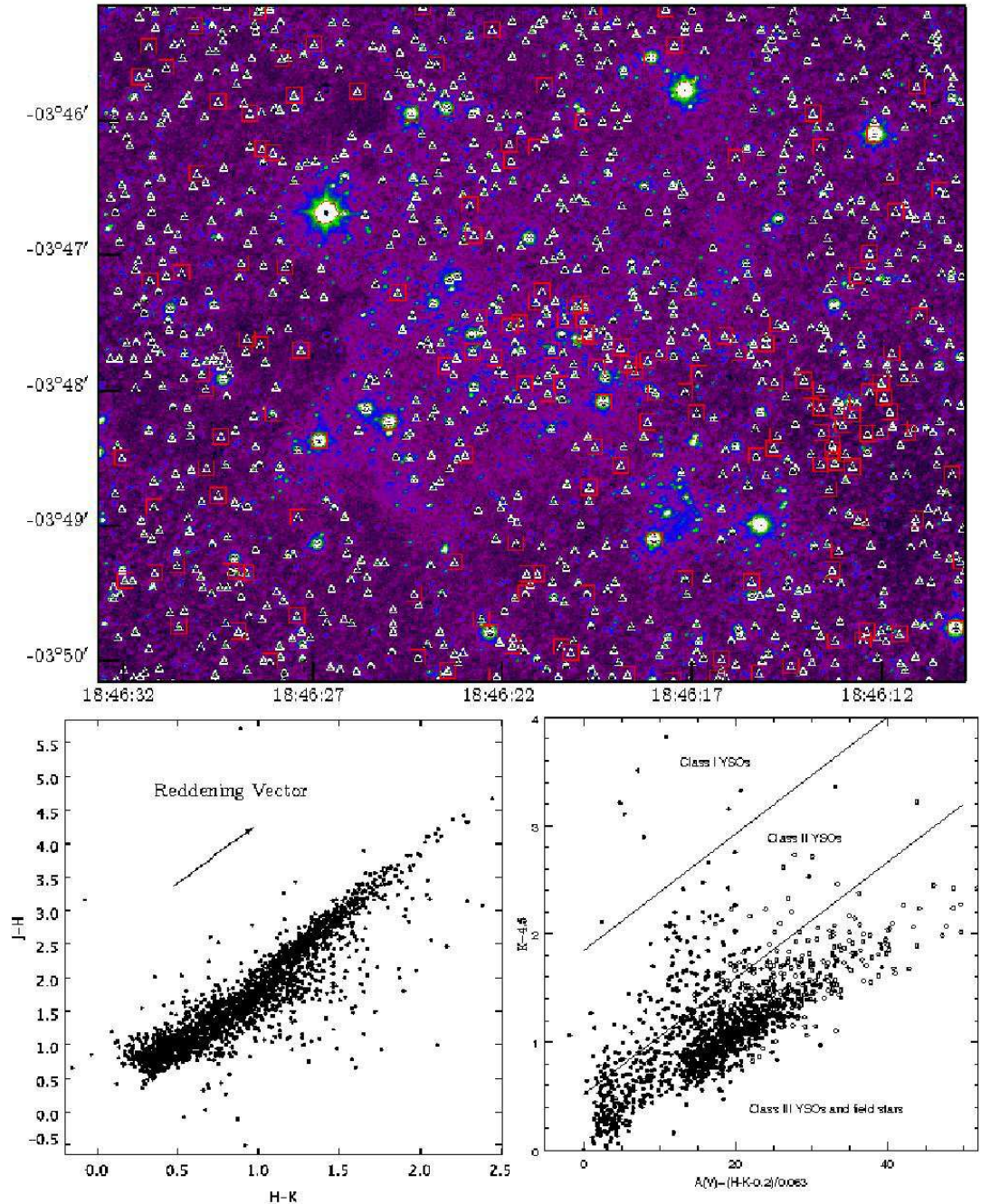}


\end{center}
\vspace{0cm}
\small
\caption{The synergy of UKIDSS-GPS and {\it Spitzer}-GLIMPSE data. ({\it top}) K band image of the 
central parts of a star formation region in the mid-plane: G28.983-0.603. Sources detected by GLIMPSE
at both 3.6 and 4.5~$\mu$m are marked with white triangles. A 0.5 arcsec matching radius was used.
Sources identified as Class I or Class II YSOs in the lower right panel are also marked with red 
squares in the image. The YSOs show a slight concentration towards the centre of the cluster.
The lower panels show how the combination of near IR data from the GPS with mid-IR data from the 
{\it Spitzer} GLIMPSE legacy survey can be used to efficiently identify and classify YSOs by their 
K-4.5~$\mu$m excesses. JHK detections are plotted as filled circles in the lower right panel, while
sources undetected in the J band are plotted as open circles. The near infrared reddening vector of Rieke \& 
Lebofsky (1985) is plotted in the lower left panel. The lines in the lower right panel that divide the
YSO classes are parallel to the reddening vector of Flaherty et al.(2007).}
\end{figure*}

One of the principal science goals of the GPS is to detect and characterise large numbers of star formation
regions, in order to examine the effect of environment on star formation. Young Stellar Objects (YSOs) can
be identified and classified by the excess emission from hot dust around the central protostar at wavelengths 
$\lambda \ga 2~\mu$m. The GPS can detect this excess in some YSOs by using the {\it(J-H)} vs.{\it(H-K)} two colour 
diagram. However, the excess is larger and therefore more easily measured at mid-IR wavelengths. Mid-IR data
from the {\it Spitzer Space Telescope} alone can be used to identify and classify YSOs (e.g. Whitney et al.2004; 
Churchwell et al.2004; Allen et al.2004; Megeath et al.2004). However, this requires a detection in three or four 
{\it Spitzer}-IRAC passbands and the relatively poor 
sensitivity and spatial resolution of the longer wavelength passbands at 5.8 and 8.0~$\mu$m limits the number of YSOs 
that can be classified. There is therefore a natural synergy between the GPS and the mid-IR {\it Spitzer} GLIMPSE 
legacy survey of the mid-plane (Benjamin et al.2003), particularly since the 2MASS survey generally has insufficient
depth to adequately match GLIMPSE in star formation regions.

This is illustrated in figure 16 for the young cluster G28.983-0.603, which is coincident with an HII region
observed by Lockman (1989). This cluster was identified by Bica et al.(2003)
using 2MASS data and is listed as cluster 10 in their catalogue. 
A fair number of YSOs are identified via their K band excesses in the near IR two colour diagram
(lower left panel), lying to the right of the principal sequence of reddened field stars and the more evolved 
``discless'' Class III YSOs. However, the median {\it (H-K)} colour excess of YSOs is only $\sim0.2$ magnitudes 
(e.g. Wilking, Greene \& Meyer 1999), so it is hard to distinguish cluster members with modest K band excesses from field 
stars. The lower right hand panel shows that the YSOs are far more clearly identified by using the (K-4.5)
colour (many showing excesses $>$0.5~magnitudes) where the 4.5~$\mu$m flux comes from {\it Spitzer}/IRAC band 2. The 
horizontal axis in the lower right panel gives the approximate visual extinction, estimated from the {\it(H-K)} colour. 
In the lower right panel we plot the 1345 sources within a 6$\times$6 arcminute box centred on the cluster that
are detected in all of the $J$, $H$, $K$, 3.6 and 4.5~$\mu$m bands as filled circles. Sources which lack
a $J$ band detection are plotted as open circles. A 0.5 arcsec matching radius was used to combine the GPS and GLIMPSE
data, since the default IPAC matching radius of 1.2 arcsec was found to create numerous false matches in this
crowded field. We note that there are only 128 four band {\it Spitzer}-IRAC detections in this field, so the
near IR/mid-IR combination increases the sample size by an order of magnitude. There is also an order of magnitude 
increase in the sample size for the YSOs.

The sources in the lower right panel of figure 16 are divided into Class I YSOs, Class II YSOs and Class III/field
stars according to their $K$-4.5~$\mu$m excess by straight lines that are parallel to the reddening vector.
The Class III/field stars form a continuous sequence parallel to the reddening vector but show an abrupt
increase in the source density at A$_V \ga 12$. Sources with a ($K$-4.5) excess are also located mostly
at A$_V \ga 12$ (typically at $12 \la A_V \la 30$) so we can infer that the cluster lies behind 12 magnitudes of 
visual extinction and that there is significant extinction within the cluster. Thus the near IR data allow us to 
identify stars with $A_V \ga 12$ as the most likely cluster members.
Lockman (1989) measured a radial velocity of 52.6$\pm$1.3 km/s for the HII region. Adopting a Galactic rotation model
with a rotation speed of 220 km/s and 7.62~kpc as the distance to the Galactic centre (Eisenhauer et al.2005) we derive
a kinematic distance of 3.1~kpc or 10.2~kpc. Inspection of the $K$ vs. {\it (J-K)} colour magnitude diagram for stars 
in the same WFCAM field indicates a preference for the closer distance. This is because of stars in the red giant branch
with A$_V$$\approx$12 have m$_K$=12-14, corresponding to distances of 2.9-7.3~kpc. The large spread is due to 
highly variable extinction across the field. The cluster lies in a extensive region of high extinction oriented east-west
across the field, so m$_K$$\approx$12 and d$\sim$3~kpc is a reasonable distance to derive using the red giant branch
and this is consistent with the nearer kinematic distance. The GPS data are fully complete to m$_K$$\approx$16.0 in this
crowded field and are $\sim$50\% complete to m$_K$=17.25. At d=3.1~kpc, m$_K$=17.25 corresponds to very low mass
pre-main sequence stars with masses as low as M$\sim$0.095~M$_{\odot}$ (using the NextGen isochrone of Baraffe et al.1998 
for an age of 1 Myr) in the absence of extinction. Such sources would be detected if they were ejected from the
region of high extinction in which most of the cluster lies. For cluster members with a more typical extinction of
A$_V$=21 (A$_K$=2.35), the 50\% completeness limit corresponds to low mass stars with M$\sim$0.5~M$_{\odot}$. 

The locations of the dividing lines were determined using information from Gutermuth (2005) in which 
Class I, Class II and Class III YSOs (with classifications determined from 4 band IRAC detections) were 
plotted in the dereddened ($K$-3.6) vs. (3.6-4.5) two colour diagram.
Gutermuth (2005) found that the different YSO classes were easily located in that diagram since the colours correlated
excellently with the colours in the (3.6-4.5) vs.(5.8-8.0) IRAC two colour diagram. The near IR and mid-IR passbands
with the best sensitivity to reddened young stars are $H$, $K$, 3.6~$\mu$m and 4.5~$\mu$m. Gutermuth (2005) therefore
devised a sophisticated scheme to classify the numerous YSOs detected in only these 4 passbands. To briefly summarise
the method, he dereddened sources with $J$ band detections to the classical T Tauri locus of Meyer, Calvet \& 
Hillenbrand in the {\it(H-K)} vs. {\it(J-H)} diagram and thereby determined the individual extinction values.
He used that information to plot YSOs in a dereddened {\it(H-K)} vs. (3.6-4.5) two colour diagram and find the
the best fit locus of YSO colours. The larger sample of sources which lack $J$ band detections were then
dereddened to this locus in the observed {\it(H-K)} vs. (3.6-4.5) diagram, taking advantage of the fact that
the locus is inclined to the reddening vector by a large angle in that diagram.
Finally, he used the extinction values for the full sample to classify YSO candidates using their location in the 
dereddened ($K$-3.6) vs. (3.6-4.5) diagram. This diagram was found to be the best diagram for clasifying YSOs, 
owing to the significant infrared excesses of protostars on both axes. 

In the lower right panel of figure 16 we have used a scheme which is simpler in some ways but is more
effective in crowded GPS/GLIMPSE fields. We find that the dereddened ($K$-3.6) vs. (3.6-4.5) diagram does not readily 
detect the YSOs in this field due to the far larger number of field stars and the large uncertainties in the 
3.6 and 4.5~$\mu$m fluxes, which cause many stars to scatter into the region occupied by YSOs. These large
uncertainties are due to the fact that the GLIMPSE legacy survey is shallower than most {\it Spitzer} GO observing
programmes: it was designed to provide quite literally a glimpse of a very large area of the Galactic plane. 
In comparison the GO programmes also usually have fewer field stars contaminating the sample since the nearby
star formation regions in the Gould belt often lie off the Galactic plane.
We use a simpler formula for extinction: ($A_V=(H-K-0.2)/0.063$), which is based on the extinction law of Rieke \& 
Lebofsky (1985), with {\it(H-K)}=0.2 adopted as the typical intrinsic colour of most stars. This formula does not
take account of the fact that YSOs will have an {\it(H-K)} excess due to hot circumstellar dust. However, by avoiding
use of the GLIMPSE fluxes in calculating extinction we greatly reduce the scatter in the lower right panel of figure 16.
Like Gutermuth, we use the extinction law of Flaherty et al.(2007) to determine the extinction in the mid-IR bandpasses, 
relative to the K band. Flaherty et al.(2007) report a possible difference in the extinction law for dark clouds and
for the interstellar medium, the latter being calculated by reanalysing the data of Indebetouw et al.(2005). The
difference could increase the gradient of the dividing lines in the lower right panel of figure 16 by 5\%, leading
to a different classification for a small number of sources. We note that the common method of dereddening to the 
Classical T Tauri star (CTTS) locus of Meyer et al.(1997) runs into difficulties in GPS datasets containing a 
variety of dwarf and giant field stars that do not lie on the locus. In any case, the CTTS locus requires fluxes in
the $J$ bandpass, which are often unavailable.

We therefore recommend the use of diagrams similar to the lower right panel of figure 16 for simple
classification of YSO candidates in GPS data. The extinction axis helps to identify stars that are likely
to lie in front of or behind the cluster and the ($K$-4.5) colour contains most of the information
in the dereddened ($K$-3.6) vs. (3.6-4.5) diagram. The use of the {\it(H-K)} colour to determine extinction
will lead to a systematic overestimate of the extinction towards YSOs, but a typical {\it(H-K)} excess of
0.2 magnitudes will lead an overestimate of only 3 magnitudes in the extinction, which is small in the context 
of figure 16. After the likely YSOs have been identified this bias can be allowed for.

We caution that there are limits outside which these YSO clasifications become unreliable.
Gutermuth (2005 and private comm.) noted that his classification system was calibrated using low mass
protostars. It is not calibrated for more massive YSOs (Herbig Ae/Be stars) or for proto-brown dwarfs,
which the Class I/II/III system of Lada et al.(1984) was not designed for. 
Herbig Ae/Be stars are likely to be the most numerous types of YSO detected in distant GPS clusters so
further work will be needed to devise a classification scheme for them.

\subsection{GPS data in nebulous star formation regions: M17}

\begin{figure*}
\begin{center}
\begin{picture}(200,350)

\put(0,0){\includegraphics{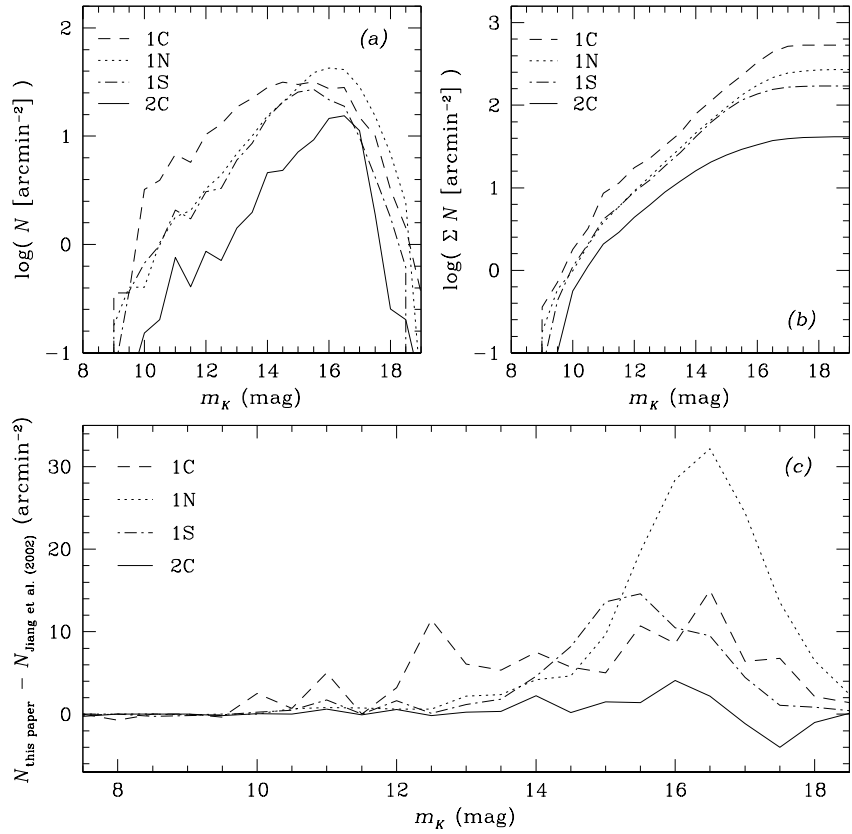}}

\end{picture}
\end{center}
\vspace{-1.6cm}
\small
\caption{K band Luminosity Functions for M17 in the regions ``1C'', ``1N'', ``1S'' and 2C
defined by Jiang et al.(2002). The top left panel shows the LFs, the top right panel shows
the cumulative source counts and the bottom panel compares the source density with that
measured in the same regions by Jiang et al.(2002).}
\end{figure*}

\begin{figure*}
\begin{center}
\begin{picture}(200,250)

\put(0,0){\includegraphics{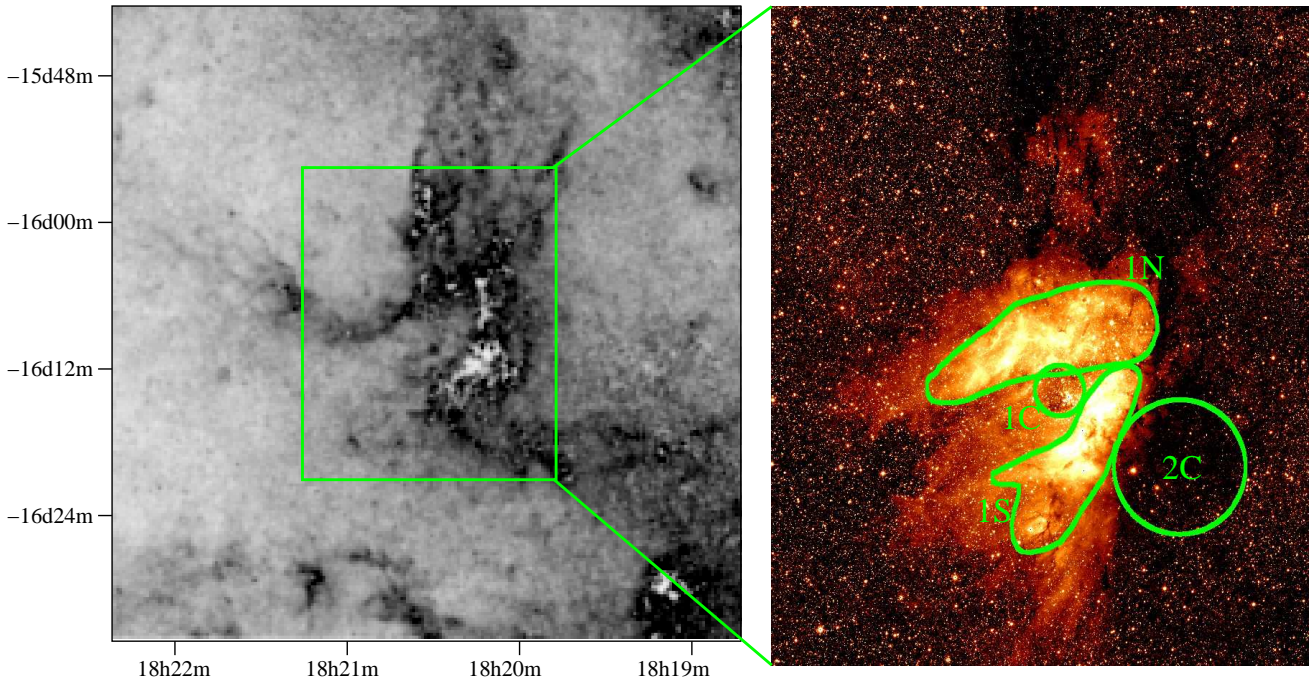}}

\end{picture}
\end{center}
\small
\caption{{\it (left)} Extinction map of the M17 field, illustrating the ``Omega'' shape of the nebula. 
The greyscale runs from A$_V$=0 (white) to A$_V$=26 (black). The extinction for each star was 
derived from the {\it (H-K)} colour and the results were averaged in 15 arcsecond bins (see text $\S$4.2).
The white regions surrounded by black are places with very high extinction where only foreground stars 
are detected. {\it (right)} K band image of the central portion of the field, illustrating the location
of the regions used in figure 17.}
\end{figure*}

\begin{figure*}
\includegraphics[scale=0.88,angle=0]{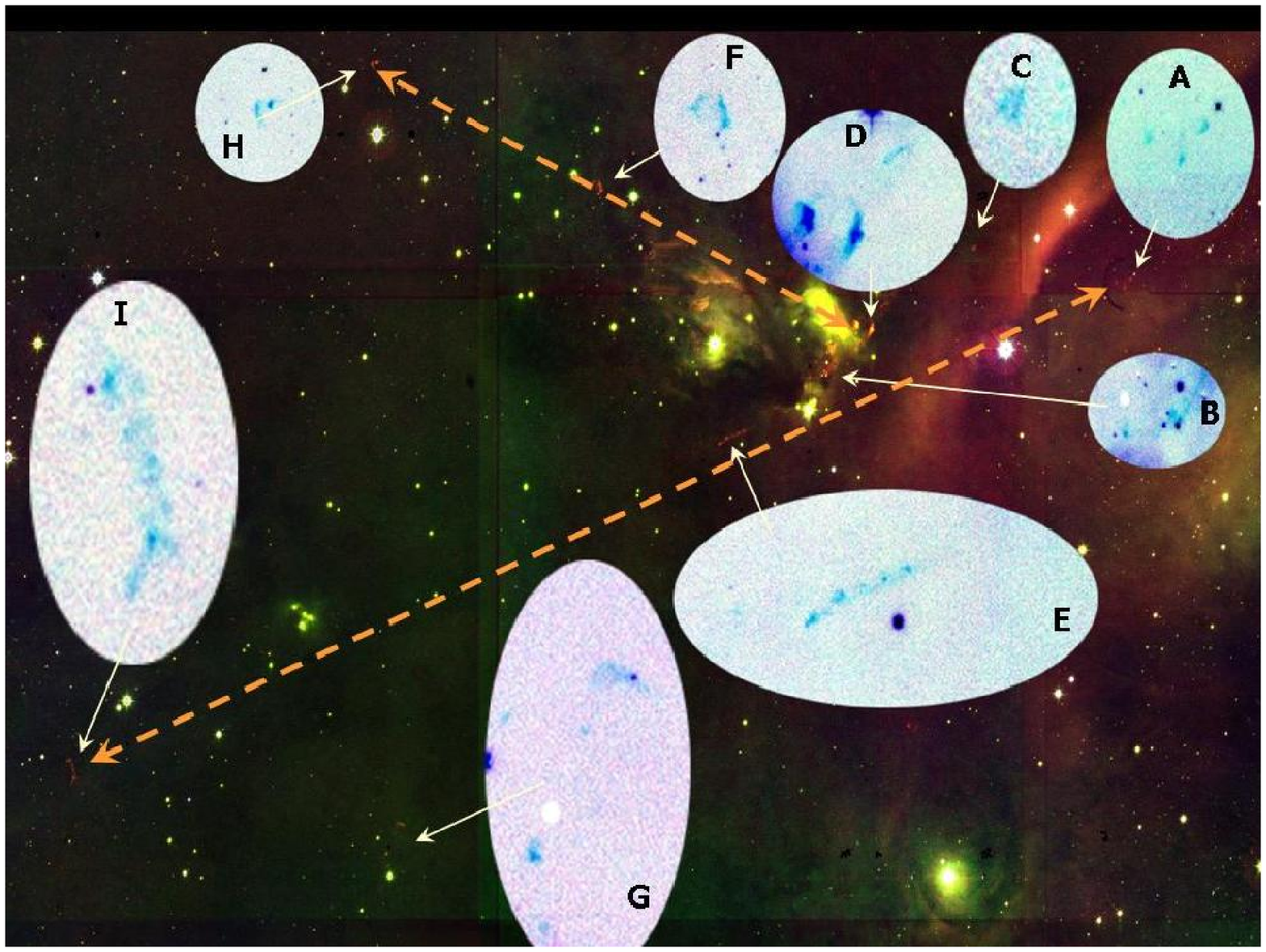}
\vspace{1cm}
\small
\caption{A three colour JKH$_2$ image centred on the L1688 dark cloud in $\rho$ Ophiuchi. Red patches show
H$_2$ emission where the outflows impact the denser environment, creating compressed layers
of shocked emission revealed in the 2.12~$\mu$m filter. Blue and green colours correspond to emission in the 
J and K bands respectively.  The image size is 31\arcmin\,40\arcsec $\times$  18\arcmin\,44\arcsec. The 
insert images show close-ups of several of the H$_2$ shocks (2.12~$\mu$m images with no continuum subtraction) 
with the following identifications from G\'{o}mez et al.(2003) (labelled G) and Khanzadyan et al.(2004) 
(labelled f). Insert A: New feature; B: G14; C: G8a; D: G11,G12,G13, HH313; E: G20, f10-04;
F: G24, HH79a; G: G3, f09-01; H: G25, HH711; I: f08-01a}
\end{figure*}

The well known high mass star formation region M17 was included in the UKIDSS Science Verification
programme undertaken in April-May 2005, before the start of survey operations. A single WFCAM tile
was observed in the three near IR passbands, using the same total exposure times as the GPS. The 
central portion of the K band image is shown in figure 18. It was noted in Lawrence et al.(2007)
(see also $\S$A2) that the WSA fails to include almost all the stars that are apparent in the regions of 
brightest nebulosity. This is due to a limitation of the source detection algorithm in the CASU reduction pipeline
that is planned to be fixed from DR5 onward. 
 Here we show the results of our own crowded field profile fitting photometry 
of the images from the WSA with {\sc IRAF/DAOPHOT} and a comparison with the published data of 
Jiang et al.(2002), which covered a much smaller part of the same field. Only the central 4 arrays out of the 16 
in the WFCAM tile contain bright nebulosity, and then only in a minority of each array. We used a 5$\sigma$ 
threshold for the source detection with {\sc DAOFIND} for each array in each filter. Visual inspection showed 
that this produced very few spurious detections. Image profiles were generated using $\sim$15 isolated
stars in each array, assuming a spatially independent profile. The photometry was calibrated by 
comparison with the WSA catalogue for each array, using the median zero point correction
for the sources common to both catalogues. Finally, the photometry for these four arrays was band 
merged in {\sc TOPCAT} with a 0.5 arcsec pair matching radius and then merged with the
WSA catalogue for the other 12 arrays in the tile. A total of 788841 sources were detected in the 
GPS field. 

In figure 17(a-b) we show the linear and cumulative $K$-band luminosity functions (LFs) of regions 
1C, 1N, 1S and 2C defined by Jiang et al. (2002) (these are illustrated in the right hand panel of 
figure 18.)  While the overall LFs exhibit similar characteristics to 
those in figure 7 of Jiang et al., figure 17(c) shows quantitatively to what extent 
both sets of LFs are in agreement. In general, the WFCAM-derived LFs contain more objects at
fainter $K$-band magnitudes than those of Jiang et al., and reach a
similar limiting magnitude. The number of sources detected in each region in the GPS data
is (1047, 7402, 3658 and 1711) for (1C, 1N, 1S and 2C) respectively. This compares with
figures of (528, 2584, 1588 and 1367) for the same regions in Jiang et al.(2002).
The excess of objects in our LFs is most likely due to the superior spatial resolution of UKIRT, 
relative to the set up used by Jiang et al. Our WFCAM observations are characterised by about 
0.8 arcsec FWHM and have 0.2 arcsec pixel sampling after microstepping. This compares very favourably 
with the 0.45 arcsec pixels of the SIRIUS camera on the Infrared Survey Facility 
used by Jiang et al., which is located at a site in South Africa with a typical
seeing of 1.5 arcsec FWHM. This is supported by figure 17(c): region 1C exhibits the
highest stellar density, and shows a systematically greater number of
objects in our LFs compared to those of Jiang et al., for $K \ga 12$
mag down to the detection limit of both sets of observations. Regions
1N and 1S, coinciding with the northern and southern nebular regions
near the core of M17, also have fairly high-density subregions. As
such the excess of fainter sources ($K \ga 14$ mag) compared to Jiang
et al. -- in particular in region 1N -- is not surprising if we assume
that the difference in spatial resolution is the driving force behind
this behaviour. 

An effect of the larger number of detections in the UKIDSS data is to shift the peak
of the LFs in the crowded regions 1C and 1S to fainter magnitudes than in Jiang et al.(2002).
In that publication the authors were unsure whether the different locations of the peaks
in the different regions were due to an actual property of the regions or to issues of
depth and spatial resolution. It now seems clear that the differences are caused by spatial
resolution effects, and still deeper observations with even higher resolution will be required
to search for any genuine differences.

In figure 18 we show an extinction map of the region. This was generated by calculating the
extinction towards each individual star and averaging the results in 15 arcsecond bins, then
gaussian smoothing the results with a FWHM of 1.5 bins. The region of high
extinction clearly follows the ``Omega'' shape of M17 (also known as the Omega Nebula).
In regions of very high extinction the extinction derived by this method is low, evinced 
by the white regions with the darkest parts of the map. This is because the densest part
of the molecular cloud obscures all the background stars that dominate the star counts in 
most other parts of the map, so only foreground stars with low extinction are detected.
The extinction towards individual stars was derived from the measured $H$ and $K$ fluxes using 
the simple formula A$_V$=($H$-$K$-0.2)/0.063, as in $\S$4.1. Comparison with the K band
image in the right hand panel of figure 18 shows that the $\Omega$ shape of high extinction
(i.e. the Giant Molecular Cloud) follows the outer rim of the HII region, which is ionised by the 
cluster located in region 1C. Regions with high extinction located further north and west
of the HII region are seen as dark areas in the K band image.

A full analysis of the M17 data will be presented in a future paper.

\subsection{H$_2$ emission in the $\rho$ Ophiuchi dark cloud}

Figure 19 shows results of WFCAM H$_2$ imaging of the $\rho$~Ophiuchi dark cloud.
These data were taken as part of UKIDSS GPS Science Verification in April 2005 with the same 
integration times as the TAP survey (see $\S2$), covering a 52$\times$52 arcminute field.
The $\rho$ Ophiuchi cloud has been the focus of 
many programmes to understand the formation of low-mass stars. While it is difficult
to detect the youngest YSOs (Class 0 protostars) because their near IR radiation is obscured by
the material from which they form, prodigious outflowing streams of molecules
are often visible at near infrared wavelengths 

\begin{figure*}
  \begin{center}
    \includegraphics[width=0.48\textwidth]{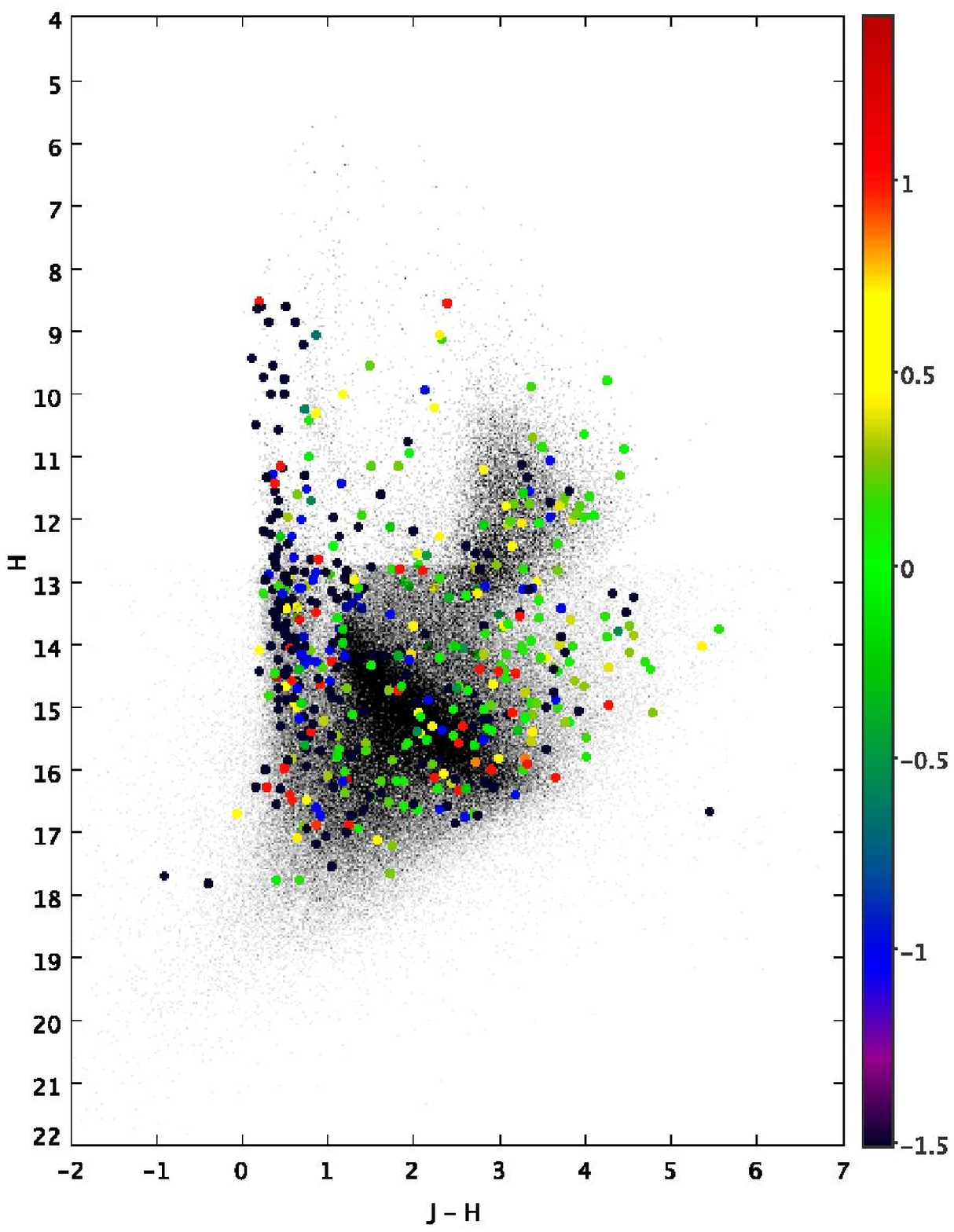}
    \includegraphics[width=0.48\textwidth]{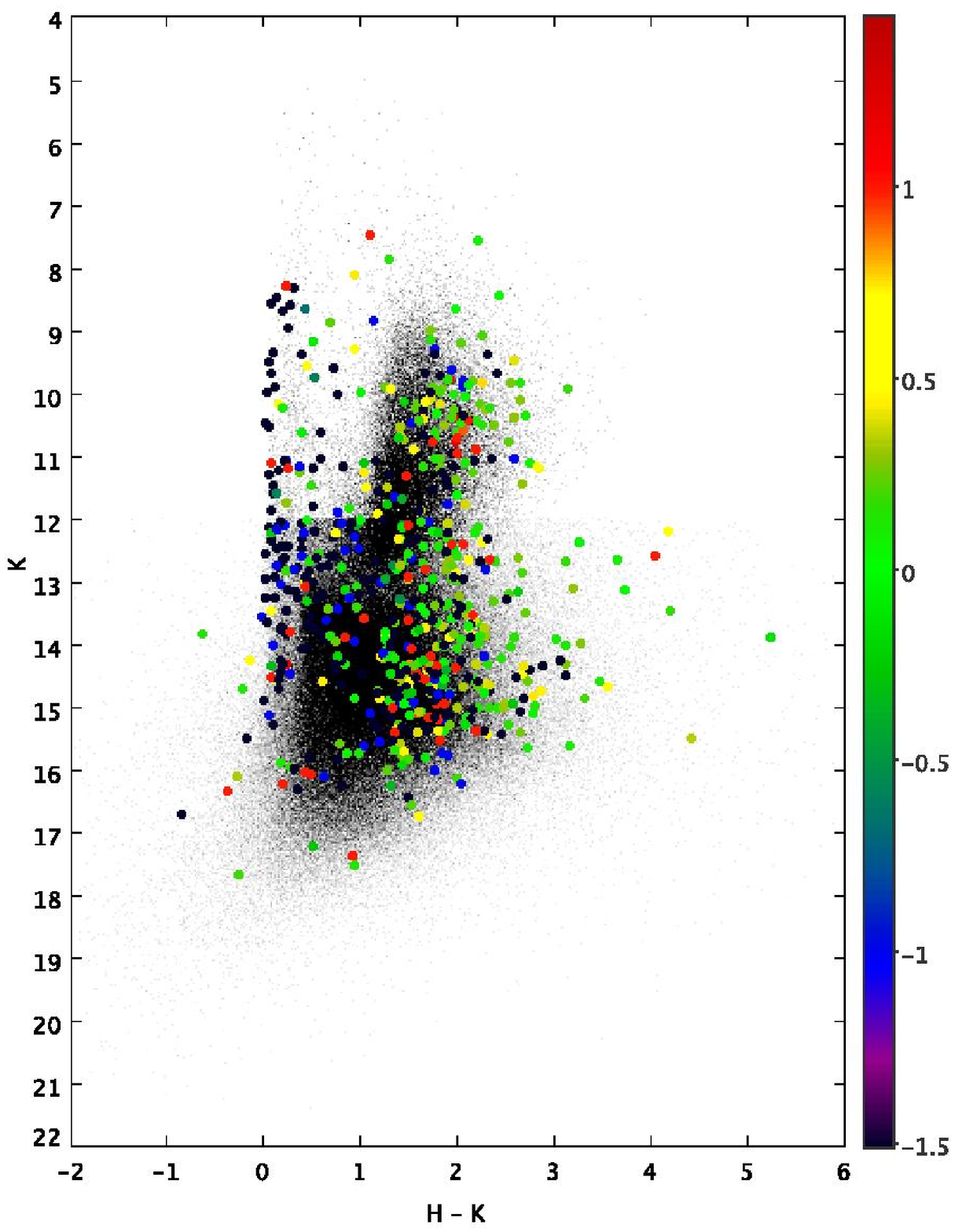}
    \caption[]{Colour magnitude diagrams (CMDs) of the candidate
      counterparts to the \chandra\ \xrss\
      ({\it coloured} points) overlaid on a randomly selected
      sample of the Nuclear Bulge stellar population (small {\it black}
      points) from the \dr\ and 2MASS (used for sources too bright
      for \dr\ where saturation results in inaccurate photometry). A sharp
      decline in source density is seen above the thresholds at m$_H$$<$12.75 
      and m$_K$$<$12.0, owing to the lower source density in 2MASS. All      
      sources are selected from the $2^{\circ} \times 0.8^{\circ}$
      region centred on Sgr A* of the \chandra\ \nb\ X-ray survey. It
      is clear that the candidate counterparts trace the overall
      stellar population. The CMD containing $J$ band data has a higher
      proportion of local dwarf candidate counterparts, indicating this
      could be used to identify more local \xrss. The \xrs\ candidate counterparts are
      coloured representing the \chandra\ hard colour (colour bars
      representing the colour association, see text $\S$4.4 for details of
      \chandra\ hard colour) showing that the local dwarf
      candidate counterparts are more commonly associated with
      \xrss\ with low values of colour, indicating soft spectra,
      low absorption or both. \xrs\ counterparts plotted in 
      in black have no net counts in one of the two energy bands
      (see Muno et al., 2006)}
    \label{f:xrscolmag}
  \end{center}
\end{figure*}

The WFCAM data have uncovered a number of new parts of these flows. Whereas previous surveys 
attributed the driving sources to be within a few arcminutes, the wide field makes it possible to piece 
together outflows which may extend over 30 arcminutes corresponding to parsec scale flows. Two such 
outflows are indicated in figure 19. The data suggest that much of what we see spread
out on the largest scale in $\rho$ Ophiuchi is actually driven from a single
active clump.  One can then determine the true number of outflows and determine their
full extent. With the aid of future epochs of observation it will be possible measure their
velocity and hence estimate their ages and calculate the feedback of mechanical
energy into the cloud. Similar parsec scale flows associated with low mass star formation have been 
identified in several other nearby molecular clouds, (e.g. Bally \& Devine 1994; Devine 1997).
 
In figure 19 the red patches mainly correspond to the locations where the outflows impact with the
environment, creating compressed layers of shocked H$_2$ revealed by
the 2.12~$\mu$m filter. Blue and green correspond to J band and K band respectively. Despite their
large spatial extent, these flows have dynamical lifetimes that are only of order 10$^4$ years,
assuming typical flow velocities of 100-200~km/s. Since the outflows appear to be driven from a single
clump this suggests that only a few protostars are driving them. The fact that most of the pre-main sequence
stars in the cloud do not seem to be generating detectable molecular outflows suggests that the active 
stage of protostellar accretion is extremely short, presumably of the same order as the dynamical timescale
inferred from the extent of the jets detected in these data. This argument assumes that most or all
protostars drive molecular jets when undergoing rapid accretion.

\subsection{X-ray sources in the Nuclear Bulge}

\label{s:nbxrs}

\begin{table*}
  \begin{center}
    \begin{tabular}{c|rrrr|r}
      \hline
      \hline
      X-ray error size & \multicolumn{5}{c}{Number of candidate counterparts} \\
      (arcsec) & 1 & 2 & 3 & $\geq 4$ & Total \\
      \hline
      $0\arcsec - 1\arcsec$ &  523 &  110 &   49 &   27 &  709 \\
      $1\arcsec - 2\arcsec$ &  600 &  153 &   58 &   43 &  854 \\
      $2\arcsec - 3\arcsec$ &  245 &   78 &   44 &   33 &  400 \\
      \hline
      $0\arcsec - 3\arcsec$ & 1368 &  341 &  151 &  103 & 1963 \\
      \hline
      \hline
    \end{tabular}
    \caption[Number of X-ray source counterparts]{The numbers of
    candidate counterparts to the Muno et al.(2006) catalogue of X-ray
    sources from the Wang et al.(2002) \chandra\ observation of the
    Nuclear Bulge. The number of counterparts has been broken down in
    to \xrss\ with positional errors of $0\arcsec - 1\arcsec$,
    $1\arcsec - 2\arcsec$, $2\arcsec - 3\arcsec$ and the total numbers
    on the bottom row. }
    \label{t:ncounterparts}
  \end{center}
\end{table*}

We have performed matching of the \dr {\bf gpsSource} catalogue to the $\sim$4200
\chandra\ \xrss\ of the Muno et al.(2003; 2006) catalogues to search for
the candidate counterparts to the \xrss\ detected on sight lines in the vicinity
of the Galactic centre. The X-ray data covers an approximately rectangular
1.6~deg$^2$ area at -1$<$$l$$<$1, -0.4$<$$b$$<$0.4. There are some gaps in the \dr area
coverage in this region, so the area of the cross match is 1.21~deg$^2$.
The area contains 2.0$\times$10$^6$ infrared sources with an average spatial separation
of 2.8\arcsec. As shown in Bandyopadhyay et al.(2005),
to identify candidate counterparts to \xrss\ in the Nuclear Bulge,
very high resolutions are required to avoid issues of confusion due to
the high stellar density. The UKIDSS survey is the first of its type
to have sufficient resolution, depth and area coverage of the Nuclear Bulge for
such a task to be undertaken. It has sufficient depth to detect giant stars
of all types in the \nb\, as well as a significant fraction of all main sequence 
stars to A or F type. By contrast, the 2MASS survey detects only M giants and
O-B type main sequence stars in the \nb\ (Cox 2000). Even more importantly,
the spatial resolution of UKIDSS is adequate to identify the correct candidate
counterpart to a large fraction of the \xrss, whereas the 2\arcsec\ resolution 
of 2MASS results in severe source confusion and a much greater likelihood of 
mis-matching.

Numerous studies have tried to characterise these newly discovered
\xrss\ in the Bulge based on their X-ray
properties. Pfahl et al.(2002) suggest that a large fraction may be
wind-accreting neutron stars with high mass
companions. Muno et al.(2003; 2006)  and Ruiter et al.(2006) propose that they
could be white dwarfs accreting from main sequence counterparts
(cataclysmic variables, polars and intermediate
polars). Willems \& Kolb (2003) and Liu \& Li (2006) believe them to be neutron
stars with low mass companions and Wu et al.(2007) have speculated that
they could be isolated neutron stars and black holes accreting from
the interstellar medium. Without positive identification of the
stellar counterparts to these \xrss\ we will not be able to
differentiate between all these possibilities.  The extremely high,
variable extinction towards the Nuclear Bulge requires these
observations to be performed in the infrared. As mentioned above, the
UKIDSS GPS is the first infrared survey to have the resolution to allow the
identification of candidate counterparts to the whole population of
\xrss\, and so will be a vital tool in the understanding of this
population.

We used {\sc TOPCAT} to identify candidate counterparts to the
\xrss. The stated positional accuracy of UKIDSS is $\sim$0.1$\arcsec$ as
matched to the 2MASS survey. We used a larger $0.3\arcsec$
positional error for the \dr\ source positions to account for the
possible confusion in the Nuclear Bulge caused by the very high
stellar density. The ``out-of-the-box'' \chandra\ positional error is
$1\arcsec$, though this can be considerably larger (up to $9\arcsec$)
in some cases due to image distortion in the outer regions of the
\chandra\ array and/or to the very low S/N of the faintest sources
detected. For the purpose of this study, we discarded all \xrss\ for
which the X-ray positional errors were greater than $3\arcsec$ radius;
for error circles greater than this size, the number of candidate IR
counterparts is sufficiently high (5-30 stars) to render meaningless
any attempt at astrometric matching between the IR and X-ray
catalogs. We used the $0.3\arcsec$ positional error for the \dr\
source positions, along with the derived positional errors from the
Muno et al.(2003; 2006) \xrs\ catalogues to set a maximum
separation of $3.3\arcsec$ between and \xrs\ and a star that can be
considered a match.

Of the 4256 \xrss\ in the Muno et al.(2003; 2006) catalogues, 3963
\xrss\ have positional errors of less than $3\arcsec$. From these, we
found 3076 candidate counterparts to 1963 of the \xrss. There are 1368
\xrss\ with only one candidate counterpart, 341 with two candidate
counterparts, 151 with three and 103 with 4 or more candidate
counterparts. Table \ref{t:ncounterparts} gives a breakdown of the
numbers of counterparts for \xrss\ with error circles of $0\arcsec -
1\arcsec$, $1\arcsec - 2\arcsec$ and $2\arcsec - 3\arcsec$. As can be
seen, for \xrss\ with positional errors of $0\arcsec - 1\arcsec$, the
majority of these have only one candidate counterpart.
Those with $1\arcsec - 2\arcsec$ errors have a slightly larger
proportion with 2 candidate counterparts. As might be expected, the sources 
with the larger error circles, $2\arcsec - 3\arcsec$,
have a much larger proportion with multiple candidate counterparts.

Figure 20 shows colour magnitude diagrams (CMDs) of the
candidate counterparts compared to a representative sample of Nuclear
Bulge stars from the \dr\ catalogue. The plots include both UKIDSS data and
2MASS data, the latter being used at bright magnitudes where stars
may be saturated in the UKIDSS GPS. The saturation limits in the GPS vary depending 
on the sky background and seeing. Conservatively it is best to assume that sources with
m$_K$$<$12.0, m$_H$$<$12.75 or m$_J$$<$13.25 may be saturated although
the typical saturation limits are $\sim$0.5 magnitudes brighter than this.
The saturation limit for individual fields can be determined with the {\ssq ppErrbits}
parameters in the WSA (see $\S$A2) but here we used the above thresholds for
simplicity. 

From the diagrams it can be seen that the \xrss\ follow the general trend of the 
overall stellar distribution but with some differences to their specific
distribution. In particular, the CMDs containing $J$ band data reveal
a disproportionately high proportion of the candidate counterparts to be in the local,
main sequence arm of the CMDs (see $\S$3.2 for discussion and
analysis of the CMDs of the Bulge). The majority of the remaining
candidate counterparts in all three CMDs are consistent with red
giants at the distance of the Bulge with variable levels of
extinction. Figure 20 also shows the \xrs\ counterparts colour-coded 
by the X-ray hard colour taken from the \chandra\ catalogue (hard colour 
is defined as: $(h-s)/(h+s)$, where $h$ and $s$ are the net counts in the
4.7-8.0\,keV and 3.3-4.7\,keV energy bands; Muno et al., 2006). Sources detected
only in one of these two bands are plotted in black. As
is clearly visible, the colours show that the \xrss\ with local,
dwarf candidate counterparts have a higher proportion of sources
with low values of colour, indicating a soft spectrum and/or low
absorption. The large proportion of single X-ray band detections (black points) 
in the local dwarf branch is mainly due to sources detected only in the 4.7\,keV
band, which is consistent with these being mostly soft X-ray sources.

Based on the identification of candidate counterparts to these \xrss,
we can draw some early conclusions as to the likely nature of the
different populations of \xrss.  We can separate the sources into
those with local dwarf candidate counterparts and those with Bulge
giants when analysing the populations. We noted above that the local
dwarf candidate counterparts have a higher likelihood to be associated
with \xrss\ with low values of hard colour, indicating either an intrinsically
soft spectrum or low absorption. Since local dwarfs make up only a small fraction 
of the total number of sources in this \nb\ field, the fact that
they are disproportionately represented among the candidate \xrs\ counterparts
indicates that these are almost certainly genuine matches, most likely CVs or 
X-ray active stars. 
For the \xrss\ with \nb\ giant counterparts, we see a generally higher
value for the \xrs\ colour indicating a harder or more absorbed X-ray
spectrum in line with the higher level of absorption expected for the
\nb\ \xrs\ population.  The very high stellar density (average
separation of 2.8\arcsec in the GPS and 1.94$\arcsec$ at $K=19$; Gosling et al.2006) 
does mean that a fraction of the Bulge candidate counterparts will be chance
positional coincidences, but the correlation of ``hard'' \xrss\ with
\nb\ giant counterparts suggests that we are detecting true companions
to many of the \xrss.  The colour distribution of the candidate
counterparts suggests the population to be predominantly \nb\ giants,
an older, more evolved population. These systems make up the majority of 
the total population of \xrss counterparts. They are likely to be dominated 
by LMXBs with evolved companions.

\subsection{External galaxies in the Zone of Avoidance}

External galaxies located close to the Galactic plane are hidden from view at optical wavelengths 
by the high extinction. For this reason the Galactic plane is referred to as the ``Zone of Avoidance''
or ``ZoA'' in extragalactic studies. Locating relatively nearby galaxies (within $\sim$100~Mpc) in this region 
is of interest for studies of the structure and kinematics of the Local Group and as well as nearby large-scale 
structures. This includes revealing the nature of the "Great Attractor",
e.g. Dressler et al.(1987); Kolatt et al.(1995); Kraan-Korteweg \& Lahav 2000, Schr\"{o}der et al.(2007); 
Kocevski et al.(2007); and the origin of the dipole in the Cosmic Microwave Background. Near infrared surveys 
such as 2MASS and DENIS have proved their effectiveness in penetrating the ZoA (Jarrett et al. 2000b; 
Rousseau et al. 2000) provided the star density does not exceed a certain value (Kraan-Korteweg \& Jarrett 2005). 
The potential for such studies is greatly improved by the higher sensitivity and spatial resolution of the GPS. 
Furthermore, the star-galaxy separation is improved by the additional detection of the fainter galaxy disks.

We noted in $\S$3.2 that external galaxies form a diffuse but well defined group of spatially
resolved sources ({\ssq pstar$\ll$ 0.9}) at the lower right of the colour-magnitude diagrams for 
fields in the outer galaxy ($l>90^{\circ}$). 
The populations of external galaxies in figures 9 to 13 are located at 1.5$<${\it(J-K)}$<$3.0, 
m$_K>14$. The galaxies appear redder in figures 9 to 11 than in figures 12 to 13, owing to
the greater extinction along the line of sight at $l$=98$^{\circ}$ than near $l$=170$^{\circ}$.
We note that the vast majority of these galaxies are too faint to be detected in 2MASS.

In colour-magnitude diagrams for the outer galaxy that employ the {\it (J-K)} colour 
(see figures 9 to 13) 
there is very little overlap between the stars classified as having ``more reliable'' photometry and 
the spatially resolved extragalactic sources, which are included in our ``less reliable' category 
(marked with blue points). The paucity of point
sources in the extragalactic group demonstrates that WFCAM has spatially resolved the great 
majority of galaxies that are detected in the GPS. Inspection of the parameters in 
{\bf gpsSource} for the objects in the region of the plots occupied by the extragalactic group  
shows that nearly all of them have {\ssq pstar} $\ll$0.9, {\ssq pGalaxy}$>$0.9 and
{\ssq mergedClass=+1}. A significant fraction of the individual sources are not unambiguously resolved 
in a casual visual inspection of the image data, but the classification in the WSA clearly 
benefits from the quantitative measurement of the observed spatial profile in all three passbands that
is provided by the reduction pipeline.
In the WSA, the nature of the profile is described by the parameter {\ssq mergedClassStat},
which describes the extent to which a source appear stellar, resolved, or under-sized (eg.
a cosmic ray artifact affecting only 1 pixel). This parameter has a Normal distribution for sources
in each array, with mean zero and variance unity. Large values correspond to resolved sources
and negative values to under-sized objects.
The {\ssq mergedClassStat} parameter for objects in the region of the plot occupied by the 
extragalactic group typically has a value between 2 and 10.

We note that the classification of source image profiles as resolved or unresolved becomes 
unreliable near the sensitivity limits ($m_K \ga 18$).

These results for the relatively uncrowded fields in the outer Galaxy are highly encouraging.
However, the same cannot be said for much more crowded fields in the direction of the
inner Galaxy. We noted in $\S$3.2 that sources classified as spatially resolved 
({\ssq mergedClass=+1}) are sometimes merely unresolved stellar pairs. Such pairs are very
common in crowded fields.

To test whether it is possible to identify external galaxies in crowded fields using the 
source catalogues, we visually inspected a crowded field located at ($l,b$)=(30.2,1.2)
The area of this field is 
1 WFCAM tile (0.75 deg$^2$). Visual inspection of the 48 images (16 WFCAM arrays in 3 passbands) 
yielded 38 galaxies, 15 possible galaxies and one nebulous region. We refer to this selection
as galaxies for brevity. The nebulous region is not listed in the source catalogues.
Fifty two of the fifty three galaxies have {\ssq mergedClass=+1},
corresponding to {\ssq pGalaxy}$>$0.9. By contrast, the field contains $\sim$1.6 million 
unique sources, of which 18\% have {\ssq mergedClass=+1}.

The {\it(J-K)} vs.$K$ colour-magnitude diagram for this field (not shown) shows no
clear separation between the galaxies and the sequence of reddened red clump giant 
stars in the field, even when the plot is restricted to sources with {\ssq mergedClass=+1}.
The galaxies satisfy the constraints {\it(J-K)}$>1.7$ and m$_K<$16.5, with a further restriction
to {\it(J-K)}$>1.9$ if we include only definite galaxies from the visual inspection.
These cuts agree well with an intrinsic galaxy colour of {\it(J-K)} = 1.0, subjected to
the (total) extinction in this particular field.
 
Two further methods were used to attempt to separate the galaxies from stars: (i) the value of
{\ssq K\_1PetroMag - K\_1AperMag3} (i.e. a measure of resolved (Petrosian) magnitude minus point
source magnitude); and (ii) cuts based on the parameter {\ssq mergedClassStat}. We found that 
both methods can be used to reduce the number of stars with {\ssq mergedClass=+1} and similar colours and 
magnitudes to the galaxies. However, a cut based on either method requires us to throw away
a significant fraction of the galaxies and the remaining contaminants still outnumber the 
the galaxies by a factor of at least 20.
 
We conclude that whilst in the outer Galaxy external galaxies can be easily found using the
parameters in the catalogue, in crowded fields they offer no real advantage over a
laborious visual search of the images. Better results may be achieved in the inner Galaxy
at latitudes further from the mid-plane (up to the survey limit at b=$\pm$5$^{\circ}$), when these data 
become available. We note that the extinction through the Galaxy (as defined by the DIRBE maps of Schlegel 
et al.1998) and the amount of source confusion and reddening varies greatly even within the 
0.75 deg$^2$ field (5.1$<$A$_V<$21.6). Hence it may be possible to improve the discrimination 
between galaxies and red clump giant stars in future by adding a Galactic extinction parameter to 
{\bf gpsSource} for each star (defined in a grid with a scale of a few arcminutes) and subtracting 
this from the measured extinction.

\subsection{Spectrophotometric typing with optical and infrared data from IPHAS and the GPS}

\begin{figure}
\includegraphics[scale=0.5,angle=0]{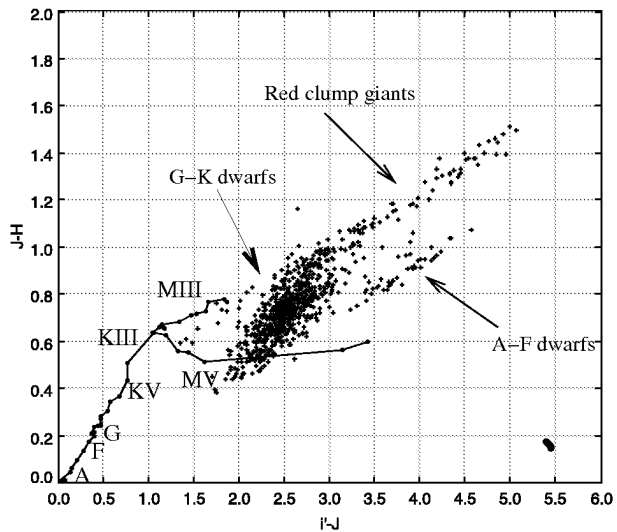}
\small
\caption{A combination of UKIDSS GPS and IPHAS data. This {\it (i'-J)} vs.{\it (J-H)} 
two colour diagram for the field at $(l,b)$=(31,0) allows us to distinguish distant 
early type dwarfs (located on the lower sequence of reddened stars) from nearby late type 
dwarfs (the main clump at {\it(i'-J)}=2.5 and distant red clump giants (the upper sequence
of reddened stars). Data for unreddened stars from Hewett et al.(2006) are plotted at the lower 
left as filled symbols, linked by solid lines. The approximate location of unreddened main
sequence spectral types is marked. Giants are redder than dwarfs in {\it (J-H)} at types K and M, leading
to separate dwarf and giant sequences.}
\end{figure}

A limitation of the GPS data is that while the luminosity class of an individual field star 
can often be determined, it is not possible to estimate the spectral type in fields with 
significant reddening. This is because
the reddening vector in the {\it (H-K) vs.(J-H)} two colour diagram runs almost parallel
to the main sequence of stars with spectral types O-K, introducing a degeneracy between
reddening and the intrinsic colours of the photosphere. The approximate reddening for
individual sources can be estimated using the {\it (H-K)} colour, but only by assuming
the all photospheres have the same intrinsic colour. A more precise reddening value
can be detemined for YSOs by dereddening to the Classical T Tauri locus in the two colour 
diagram, as noted earlier, but the result is only valid for stars that are known to be
YSOs. 

We anticipate that this degeneracy can be broken for many of the brighter sources
in the GPS by using optical data from IPHAS survey (Drew et al.2005; Sale et al., in prep). This 
survey covers the entire northern Galactic Plane down to Dec=-5$^{\circ}$ at $-5<b<5$ in the
Sloan r', i' filters and an H$\alpha$ filter. The survey is now almost complete and the 
IPHAS Initial Data Release (Gonzalez-Solares et al.2008) is now publically available,
containing data from $l$=27 to $l$=117.

Work by members of the IPHAS team (Drew et al.2005) has shown that the IPHAS
{\it (r'-i') vs. (r'-H$\alpha$)} two colour diagram can be used to break the degeneracy
with extinction and estimate the approximate spectral type and distance of most field stars. 
However, this requires detection in the $r'$ and H$\alpha$ filters, which is not usually
achieved in fields with high reddening. Fortunately, $i'$ fluxes will often be available for
GPS sources even in fields with quite high reddening, since the IPHAS 5$\sigma$
sensitivity limit is $i' \approx 20.7$ (Vega system). 

In figure 21 we show an example of an IPHAS/GPS {\it(i'-J)} vs.{\it (J-H)} two colour diagram for the 
field with high reddening at {\it(l,b)}=(31,0) that was illustrated earlier in figure 5. The ``PhotoObjBest''
IPHAS catalogue, which contains sources that were observed in good photometric conditions and detected 
in all 3 IPHAS passbands, was used for the cross match. The scatter
in the diagram has been reduced using cuts very similar to the ``most reliable, low completeness''
cuts used for the near infrared two colour diagrams (see $\S$A3), including a restriction to
sources with uncertainties $<$0.05 mag in $i'$ and $J$. The two sequences of stars
extending to the upper right of the diagram are dwarfs (lower sequence) and giants (upper sequence).
The separation between the two sequences is cleaner than in the two colour diagram in figure 5
because the separation in the {\it (i'-J)} colour is $\sim$1.25 magnitudes, which is
far larger than the typical photometric uncertainties in this colour. That is not the case for 
the {\it(H-K)} colour used in figure 5, even though we restricted the selection to sources with 
uncertainties listed as $<$0.05 mag on each axis. The fundamental reason for the separation remains
the different intrinsic {\it (J-H)} colours of the two populations, which is due to the gravity 
sensitivity of the {\it (J-H)} colour in the giants and the earlier spectral types of the dwarfs.

In addition to the improved dwarf/giant separation, a major benefit of figure 21 is the potential
for spectrophotometric typing. The sources on the reddened dwarf branch are almost entirely distant 
early type stars (spectral types A and F). This can be seen by tracing the dwarf branch back to 
where it would intersect with the unreddened main sequence that is overplotted (using data from 
Hewett et al.2006). The location of the dwarf branch is consistent with either reddened A-F type
stars or M dwarfs. The latter can be ruled out since they are not detected at large enough distances
to incur significant reddening. This interpretation is supported by a Besan\c{c}on model of the field
which confirms that this branch is composed of A and F type stars. (The model uses the Cousins I filter 
rather than Sloan $i'$ so it is not exactly comparable and is therefore not shown).

The reddened dwarf sequence can also be traced back to the lowest part of the main clump of stars near 
{\it (i'-J)}=2.5, as would be expected for these intrinsically blue, early type stars.
The larger population of G and K type dwarfs is contained in the clump of apparently bluer 
stars near {\it (i'-J)}=2.5 since these less luminous stars are not detected at sufficiently large 
distances to suffer such high reddening. The clump appears elongated from lower
left to upper right, and is entirely consistent with a reddened population of G and K type
main sequence stars that mostly have similar extinction. This sequence
is sufficiently inclined to the reddening vector to break the colour-reddening degeneracy and 
distinguish G-type stars from K-type stars. The extinction toward stars in the G-K clump is
A$_V$=4-5 magnitudes and the implied distances are typically $\sim$1-2~kpc. The most distant red clump 
giants in the plot have A$_V$$\approx$9 magnitudes and implied distances of $\sim$3.8~kpc. Stars that are 
1-2 magnitudes fainter in $i'$ than those plotted, and therefore more distant, could be detected by (i) cross 
matching with the IPHAS ``PhotoObj'' catalogue rather than PhotoObjBest; and (ii) including sources
with larger photometric uncertainties. The PhotoObj catalogue does not require detection
in all three IPHAS passbands for a star to be included.


\section*{Acknowledgements}

We wish to thank all the staff of the Joint Astronomy Centre in Hilo and past and present members of
the UK Astronomy Technology Centre who assisted in the construction and commissioning of WFCAM and have made 
the UKIDSS project possible. We also thank the staff of WFAU and CASU who built the data flow system.
Lucas acknowledges the support of a recently completed STFC Advanced Fellowship. Gaspar 
received support from Hungarian OTKA grants TS049872 and T049082.\\
\\

\appendix
\section{Guide to the WFCAM Science Archive for GPS Science}


\subsection{Fundamentals of the GPS data, archive structure and usage}

In this guide to the GPS we do not assume that the reader has previously used the
WSA, nor any knowledge of SQL (the Structured Query Language which is
used to construct and interrogate the WSA). There is extensive general UKIDSS
documentation at the WSA and CASU websites so this guide focuses on information 
that is useful for GPS science.

\subsubsection{Multiframes: the basic unit of data}

WFCAM has four spatially separated $2048 \times 2048$ HAWAII 2 infrared array detectors
laid out in a $2\times 2$ square pattern. Each array covers a 13.65$\times$13.65 arcminute field of view
and the gaps between them are 12.83 arcmin, or 94\% of a detector width. A single
observation therefore covers a partially filled area of sky known as a ``multiframe''. In the 
GPS, a series of four slightly overlapping  multiframes with offsets of 13.24 arcmin in RA and Dec 
is used to produced a 
filled ``tile'' with an area of 0.75~deg$^2$. The multiframe (not the tile) is the fundamental unit
around which the WSA is designed. Every multiframe has a reference number in the WSA. By 
contrast, the tiles are not catalogued. The WFCAM arrays are always oriented
approximately north-south (to within $\sim 1^{\circ}$), so the GPS tiles the Plane
in strips of constant Declination, always slightly overfilling the desired Galactic latitude
boundaries with array corners and edges.

The term multiframe is also applied to a coadded sequence of observations with spatial 
offsets of a few arcsec (see A1.2). Every pipeline processed multiframe is stored in the 
WSA (both individual integrations and a coadded series of observations).
However, only the coadded images have associated source catalogues (see A1.4). 

\subsubsection{GPS observation sequence}

All GPS data are taken with a $2\times2$ ``microstepping'' sequence in order to
fully sample the spatial resolution of the image in the focal plane, which is frequently
$<2$ pixels Full Width Half Maximum, i.e. $<0.8$~arcsec. The microsteps used
in the GPS are spatial shifts of 4.6 arcsec, which is approximately 11.5 pixels
(depending on location within the focal plane). This
``N$+$0.5'' microstepping allows the four separate images to be interleaved 
in order to produce images with a 0.2 arcsec pixel scale. After going through
the microsteps, the telescope is then moved through 3.2 arcsec (8 pixels) in 
RA and/or Dec before repeating the microstep sequence. This jitter procedure
provides independent pixels to aid the removal of any bad pixels that
were not included in the bad pixel mask. From DR1 onward the number of jitters is two,
so the observations comprise eight individual exposures of duration 10~s, 10~s and 5~s
to make up total on source integration times of 80~s, 80~s and 40~s in the J, H and K filters
respectively. In the EDR four jitters were used for the J and H bandpass observations, with 
5~s individual integrations. The number was reduced to two in order to improve observing
efficiency. Observations in the H$_2$ filter used two jitters in all data releases, with
eight individual exposures of 20~s making up the on source integration time of 160~s.

\subsubsection{GPS images in the WSA}

The multiframe images are stored as FITS images with a header unit and four image extensions,
one for each spatially separated array. There are three main types of reduced GPS multiframe 
image. 
(1) Individual integrations, which are listed as {\it normal} multiframes and have a pixel
scale of 0.4 arcsec. (2) Interleaved images, listed as {\it leav} multiframes, which are formed 
from four images taken in the microstepping sequence and have a 0.2 arcsec pixel scale. (3) 
Final coadded images (also with a 0.2 arcsec pixel scale) that are formed by coadding two or 
more interleaved images and are listed as {\it leavstack} multiframes in the WSA. For most 
purposes GPS users will only need to use the {\it leavstack} multiframes. 

The user can make a list of the GPS multiframe images for download, and view high quality jpeg 
preview images, by selecting ``Archive Listing'' at the top level of the WSA and then selecting 
the GPS from the Programme drop down menu. The user will normally use the most recent UKIDSS
release by default but there are also options to use either the earlier
releases (which are a subset of the most recent release but can be searched slightly more quickly)
or ``WSA'' in order to include images which failed quality control but may still be useful
for some scientific purposes. The three types of reduced science image described above will be 
returned if the default settings of ``object'' in the Observation Type drop down 
menu and ``all'' in the Frame Type drop down menu are used. If the latter menu is set to ``stack'' then 
only {\it leavstack} images will be returned. The desired images can then be selected by date, by 
filter and by Equatorial coordinates (FK5 system, defined by a rectangular box). Two 
additional types of GPS image are the H$_2$-K {\it difference} images provided for the 
off-plane TAP region and {\it confidence} images, which exist for all {\it leav} and 
{\it leavstack} images 
and can be used to check for defects in the data. The {\it difference} images are scaled by the filter
width to remove sources with a flat continuum. The {\it confidence} images are only returned
if the Observation Type is changed from ``object'' to ``confidence''. Pixels in the {\it confidence} 
images have a median value of 100. Lower values correspond to either reduced sensitivity
measured in the flat fields, bad pixels, hot pixels detected in the darks or saturated pixels. 
Other image types from the reduction process, such as {\it Sky} images and {\it Dark} images can also be 
obtained with the appropriate setting of the Observation Type. {\it Sky} images are generally
constructed using a sequence of frames taken throughout the night. These may be useful
in order to assess the quality of the background subtraction in Galactic plane regions with
bright nebulosity. 

The coordinates returned by the Archive Listing for a multiframe are those of the centre of the 
WFCAM focal plane, which lies between the four spatially separated arrays and is therefore a 
region with no data! Hence if the full array image nearest a particular location is desired, it is 
necessary to either use the Image tools described below to find the multiframe number or else
search for all images in a box 0.8$^{\circ}$ wide, centred on the target location, to find the
desired one. 

More user friendly image access is provided by the ``GetImage'', ``MultiGetimage'' and ``Region 
Search'' tools at the top level of the WSA, and via links from the {\bf gpsSource} catalogue entries
described below.  The ``GetImage'' tool allows the user to find and 
download cut-out images in all available filters, centred on the desired Galactic or Equatorial
coordinates. The image size is limited by the user-specified dimensions or by the
boundaries of the 13.6 arcminute array field of view. There may be more than one image
in each filter if the coordinates lie in the small overlap region at the edge of an array,
or if the field has been observed more than once. Second epoch observations have not yet occurred 
for the GPS, except when a field has been repeated by the observer to ensure the data quality 
satisfies the survey requirements. The ``MultiGetimage'' tool is similar but it allows a large number of 
images to be downloaded in a single filter, each centred around a list of Equatorial coordinates 
provided by the user. The ``Region Search'' is described below.

The UKIDSS astrometric solutions are determined by fitting to numerous 2MASS stars in the field of each 
separate science array, as noted in $\S$2. The typical precision of the solution is 0.09 arcsec, as determined 
from the r.m.s. residuals to the fit, but there is some variation. The value of the r.m.s. residual for a 
given science detector array is given by the {\sc STDCRMS} parameter. This parameter is given in the image 
headers for each array. It is also accessible via the {\bf CurrentAstrometry} table in the WSA, which can
be viewed using a FreeForm SQL query (see $\S$A1.4).

\subsubsection{GPS catalogues}

Most users will wish to use the SQL source catalogues to search for astronomical objects.
{\bf gpsSource} is the most commonly used catalogue. It is the merged catalogue containing 
most of the 
important data for all sources detected in at least one of the J, H and K passbands. The pairing
scheme starts with the shortest available wavelength (the J band if the data are available, or the
H band if not) and searches for 
counterparts in the H band and then the K band that are located $<$1 arcsec from the J band 
detection. Sources are only merged as a single entry in {\bf gpsSource} if each is the closest
match to the other at all wavelengths. The epoch of observation of a star in {\bf gpsSource}
in each filter can be determined by clicking on the entry in the ``getMFlink'' column in the table, which 
will provide 1$\times$1 arcminute images centred on the star, with associated dates and times. 
The other main table, {\bf gpsDetection}, contains all the information for all sources detected 
in any passband but with no association between detections of the same object in different 
filters. 
While {\bf gpsSource} is more convenient to use, the caveats are: (i) it includes only 
a subset of the photometric aperture sizes that are available in {\bf gpsDetection} or the FITS binary 
catalogues; and (ii) in releases from DR3 onward, most of the extragalactic flux extimates 
such as Petrosian magnitude are no included. The details of how to write an SQL query to extract 
these missing attributes from the {\bf gpsDetection} table are given in the ``SQL Cookbook'' at 
the top level of the WSA.

The available columns or ``attributes'' in each table are seen when conducting a Menu Query (see below), 
which optionally provides a one line description of each attribute. We recommend that the user
reads the more detailed entries for the {\bf gpsSource} and {\bf gpsDetection} parameters
in the ``Schema Browser'' (accessible from the WSA top level) under ``WSA UKIDSS'' and then 
``Tables''. 

The GPS SQL catalogues can be searched using the Region Search, the Menu Query (recommended for 
new users) or the Freeform SQL page, all of which are accessible from the top level of the WSA.
When using these tools, users of the DR2 release and subsequent releases have an option (in the
drop down menu of tables) to view a subset of {\bf gpsSource} which includes only areas which are 
complete in all three filters. This can slightly improve search speeds where the desired 
area is known to be complete (e.g. from an Archive Listing). If previous data releases are being 
searched, the user can similarly search a dataset with full filter coverage by selecting e.g. 
the DR1 database, as opposed to the larger DR1PLUS database.

The simplest way to search the {\bf gpsSource} table is with the Region Search. This 
enables a simple search of a circular area around the desired Equatorial or Galactic coodinates 
and it returns all information in the table for every object therein. In addition, the sources 
listed can be clicked on to provide jpeg and FITS format cut-out images around their location, in 
all available filters (similar to the GetImage results). 

More flexible and sophisticated queries can be made with the Menu Query or a FreeForm SQL
query. Menu Queries are constructed by clicking boxes. They return only the desired attributes and 
restrict their values (e.g. magnitudes, colours, coordinates etc.) in order to refine 
a search. This is especially
important for the GPS because a maximum of only 15 million data entries can be returned
in a single query (e.g. 1.5 million sources with 10 data attributes each), which can quickly be 
exceeded when searching more than a square degree in crowded fields. The {\bf gpsSource} attributes
that are highlighted in the Menu Query list have their own indices in the table, which means that 
searches based on these parameters will proceed much more rapidly than those which do not.
Hence searches based on e.g. specified coordinates (Equatorial or Galactic) and a range of magnitudes
will proceed rapidly if the prferred photometric aperture ({\ssq AperMag3}, see below) is used, but
will take much longer if a different aperture size is used. Again, this is a serious issue
since the size of the GPS catalogues means that searches can often time out (the
maximum duration is presently 80 minutes) when searches are based on non-highlighted attributes. 
Merely including non-highlighted attributes among the data to be returned (as opposed to 
specifying a range of values) does not cause this problem.

FreeForm SQL queries can perform the same tasks as a Menu Query  and also manipulate the data 
returned from the table or operate on two or more tables (e.g. GPS data and Spitzer GLIMPSE data) to 
produce more complicated outputs. Menu Queries are transformed into ASCII format SQL queries before 
processing. A useful feature of the WSA is that this SQL query can be viewed using the ``Show 
and Edit SQL'' button during a Menu Query. This enables the user to learn how to perform basic 
FreeForm SQL queries with no prior knowledge of the language. The SQL Cookbook provides further 
training.

A list of FITS binary source catalogues associated with individual coadded multiframe images 
({\it leavstacks}) is provided along with the image lists described in $\S$A1.3 in an Archive Listing.
They may be viewed with freely available software such as {\sc FV} (FitsViewer).
These catalogues contain only single filter data and source fluxes within the various apertures are given in 
data numbers (normalised to the exposure time of individual frames on sky) rather
than magnitudes. The catalogue headers include the aperture corrections, photometric zero points,
airmass and the exposure times (for the individual frames on sky), so it is straightforward to calculate 
magnitudes or to conduct additional photometry, e.g. on very faint sources which may have been 
missed by the source detection algorithm and are therefore not in the source catalogues.
We note that the FITS binary catalogues have less precise R.A. and Dec coordinates than the SQL
source catalogues, owing to a limitation on the number of data bits which
corresponds to a precision of $\sim$0.36 arcsec. However, the exact coordinates can be recovered
using the precise X, Y array coordinates given for each source and the WCS information in the 
catalogue FITS headers. Details of the columns in the FITS binary
catalogues (which provide additional information relevant to the SQL source catalogues) are 
available from the CASU website (Irwin et al.2005) and in Irwin et al.(2008).

\subsubsection{Simultaneous multi-aperture fitting photometry and {\ssq AperMag3}}

The WFCAM pipeline performs aperture photometry, as opposed to profile fitting photometry. 
However, any overlapping apertures are simultaneously fitted and the fluxes in areas of overlap
are assigned to each target in proportion to the total flux within each aperture. This 
generally produces good photometry even in the crowded GPS fields (see A3), though
there will be cases where the photometry is imperfect, such as faint sources on the wings
of a bright or saturated star. We note that profile fitting photometry may be included
in the final processing of all UKIDSS data when the survey is finished, but tests 
with an algorithm designed for the irregular image profiles often generated by ground
based microstepped data have shown no significant gain in precision.

Photometry is output for a series of circular apertures, for which consecutive aperture numbers 
have a ratio of $\sqrt{2}$ in diameter. The smallest aperture, {\ssq AperMag1}, has a diameter of 1 arcsec.
The default aperture that is used for GPS point source colours in {\bf gpsSource} is {\ssq AperMag3}, 
which has a diameter of 2 arcsec. Only {\ssq AperMag3, AperMag4 and AperMag6} and the much larger
{\ssq Hallmag} aperture (which is a measure of the total flux from each source) are available in 
{\bf gpsSource} in the existing data releases (though
all apertures are available in {\bf gpsDetection}). We recommend that {\ssq AperMag3} is used
for most GPS science unless there is good reason to desire a smaller or larger aperture.
For future UKIDSS releases it is planned to change the available apertures in {\bf gpsSource}
to {\ssq AperMag1, AperMag3 and AperMag4}, with the rationale that the smaller aperture yields
more reliable photometry in the most crowded fields and the slightly larger aperture will sometimes
yield more precise photometry in the off-plane TAP region, where relatively poor seeing is accepted. 

\subsection{Limitations and unusual features of the GPS data}

Here we describe some peculiar features of the GPS data which the user may notice
but whose origin may not be immediately obvious.

\begin{itemize}

\item In many images the profiles of stars in the {\it leav} and {\it leavstack} images may 
appear to be slightly ``speckled'' with departures from the smooth monotonic image profiles usually 
seen in astronomical data. This is due to the changes in the natural seeing during the 
microstepping procedure referred to in A1.2, which can cause adjacent rows and columns to show 
significant differences in flux. Comparison of fields with and without this effect have shown
that it has no significant effect on photometric precision.

\item Electronic cross talk between adjacent array read out channels causes the appearance
of spurious sources located at multiples of 128 pixels (51.2 arcsec) away from all bright stars along the
axis of array readout (which is oriented differently for each quadrant of an array). This
is a general feature of {\sc Hawaii} 2 arrays that is most serious for pixels which are close
to saturation ($>30000$ DN) in the WSA images. Note that the separation  $128$N pixels becomes
$256$N pixels in the coadded and interleaved {\it leavstack} images.
All data in the DR2 release benefit from a cross-talk 
removal procedure which has reduced the problem. Hovewer, the procedure is imperfect so some 
artifacts remain. This is particularly so in regions with spatially variable background, where a 
flaw in the procedure actually generated additional cross talk artifacts that do not always 
lie $128$N pixels from a bright star. Owing to this error some images of regions with
large gradients in the background were removed from DR2, including
several H band images near the Galactic Centre. The error will be fixed from DR3
onward and the lost images will be recovered. 
The archive also addresses the cross talk problem from the DR2 release onward, but in the catalogues rather 
than the images. Possible cross talk artifacts are flagged in the data given in the 
{\ssq ppErrbits} columns in {\bf gpsSource} and {\bf gpsDetection}. This identifies almost all
cross talk artifacts at $128$N pixels from a bright star by using the 2MASS Point Source Catalogue.
Usage of the {\ssq ppErrbits} flags to remove suspect data is described in A3.

The brighter artifacts can be seen to be spurious
since they have positive counts on one side of the profile and negative counts on the other
(relative to the sky level). This is because the cross talk artifacts are related to  the
spatial derivative of the profile of the bright star that generated them, in the readout direction. 
Fainter stars can sometimes produce artifacts
which look more like real stars (and generally have similar colours). It is advisable to guard 
against this by checking the images for bright stars near objects of interest. 
Over-saturated stars produce larger, obviously non-stellar artifacts with a ring-like structure.

\item Saturated stars have $\gtsimeq 35000$ counts in {\it leavstack} images, but the
centre of an oversaturated star has a lower count level (often below the sky background).
This is a general feature of data taken with near infrared arrays in the correlated double sampling
mode (CDS). Saturated stars exist in most GPS fields. They are usually
included in the source catalogues but the fluxes given are not meaningful. Again, the
{\ssq ppErrbits} column in the catalogues can be used to efficiently remove these bad data,
in this case by requiring e.g. {\ssq JppErrbits}$<$65536 to ensure that a source is not saturated
at J band. Conservatively, it is safer to use 2MASS at m$_K$$<$12.0, m$_H$$<$12.75, m$_J$$<$13.25, though
the typical saturation limits are $\sim$0.5 magnitudes brighter than this.
The {\ssq ppErrbits} parameters can be used to find the precise limits for each field in each filter,
using the threshold value quoted above.
Heavily oversaturated stars cause multiple spurious detections in each filter (and
in {\bf gpsSource}), since the saturated region is larger than the size of the stellar image profile 
used by the pipeline source detection algorithm. These are often listed in the archive as
resolved sources rather than stars ({i.e. \ssq mergedclass=+1}).

\item Photometric errors are generally under-estimated.
The photometric errors given in magnitudes and colours are only notional internal errors (excluding
calibration uncertainty) that are
based on source counts in the apertures, the Full-Width-Half-Maximum (FWHM) of the image and the 
standard deviation of the background flux. A plot of errors vs. magnitude yields a single smooth curve for 
fields contained within a single WFCAM array. The WSA is similar to the 2MASS Point Source Catalogue in this 
respect. The given error systematically underestimates the true error by a factor of $\sim$1.2, owing to
the inter-pixel capacitance of the WFCAM arrays. This phenomenon reduces the noise estimate derived from a
measure of the local standard deviation by this factor by causing inter-pixel correlation of the
count values (Irwin et al.2008). Hence sources with photometric uncertainties of 0.20 magnitudes in the 
archive are in fact usually closer to 4$\sigma$ detections than 5$\sigma$ detections.
Additionally, errors specific to individual sources are not included in the error budget: e.g. bad pixels, the 
presence of adjacent stars or cross talk artifacts, or small scale spatial variations in sky background. 
Finally, the CASU pipeline is designed to 
deliver an absolute photometric precision of $\sim$0.02 mag for sources with a high signal to noise ratio.
(The uncertainty in the photometric calibration is discussed below).
Therefore users should not take seriously the millimagnitude uncertainties that are quoted for bright stars.
When studying a particular source users should inspect the image data (and perhaps also the {\it confidence
Images}) to be sure that the given error is meaningful.

\item Nebulosity can also lead to photometric errors, since the sky background is evaluated
with a relatively coarse grid in each field (typically in 25 arcsec grid cells). A more sophisticated method
of measuring the background on smaller spatial scales in nebulous regions has been developed by M.Irwin in order 
to improve the photometry but it will not be implemented in the WSA until DR5.
More seriously, the source detection algorithm fails in regions of very bright nebulosity, since the nebula is 
interpreted as a single extended source. At present this effect has been seen only in parts of the M17 star 
formation region (see $\S$4.2) which is one of the largest regions of bright nebulosity in the sky. In such 
regions, it is necessary at present for users to do their own photometry (see $\S4.2$)

\end{itemize}

Finally, we provide the extinction dependent 2MASS-WFCAM photometric transformations that are
used to generate the photometric zero points for each detector, as noted in $\S$2.

$J = J_{2M} - 0.065(J_{2M}-H_{2M}) + 0.015E(B-V)$\\

\vspace{-2mm} 
$H = H_{2M} + 0.07(J_{2M}-H_{2M}) + 0.005E(B-V) - 0.03$\\ 

\vspace{-2mm} 
$K = K_{2M} + 0.01(J_{2M}-K_{2M}) + 0.005E(B-V)$\\ 

The $E(B-V)$ colour excesses are computed using the prescription of Bonifacio et al.(2000), 
from the data of Schlegel et al. (1998). Since the extinction terms are small ($E(B-V)$=3 corresponds
to 9 magnitudes of visual extinction for a typical interstellar reddening law), differential extinction
across each detector is usually insufficient to contribute significantly to the uncertainty in the zero points,
as measured by the internal scatter amongst the bright 2MASS sources that are used. (The uncertainties in the zero 
points are included in the image headers and the FITS binary catalogue headers.)
However, a small number of fields exist where spatially variable extinction causes significantly increased 
uncertainty in the calculated zero points. This should be borne in mind if the data are used to investigate 
e.g. possible spatial variations in the near infrared extinction law in star formation regions. 


\begin{figure*}
\begin{center}
\begin{picture}(200,250)

\put(0,0){\includegraphics{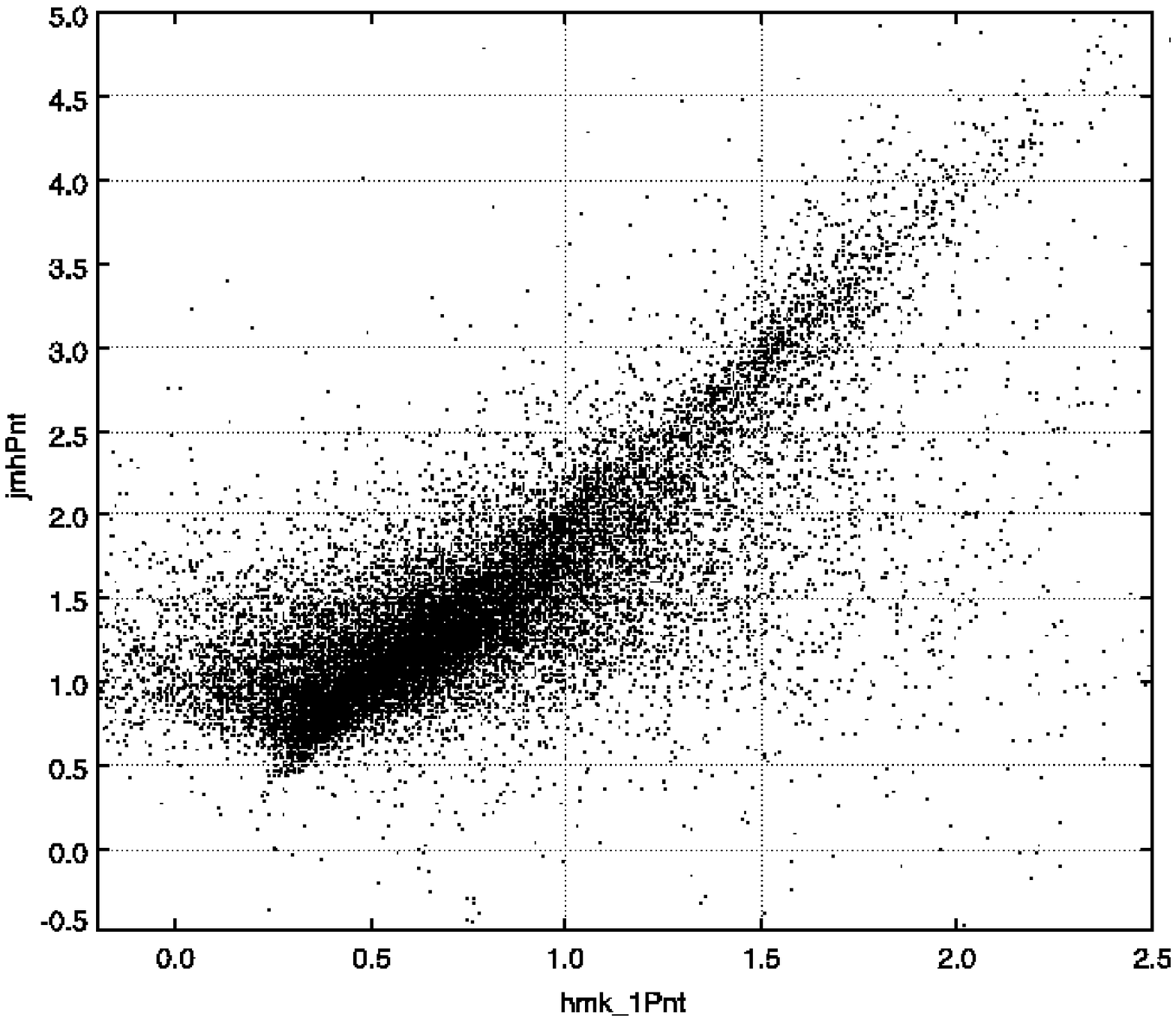}}

\put(0,0){\includegraphics{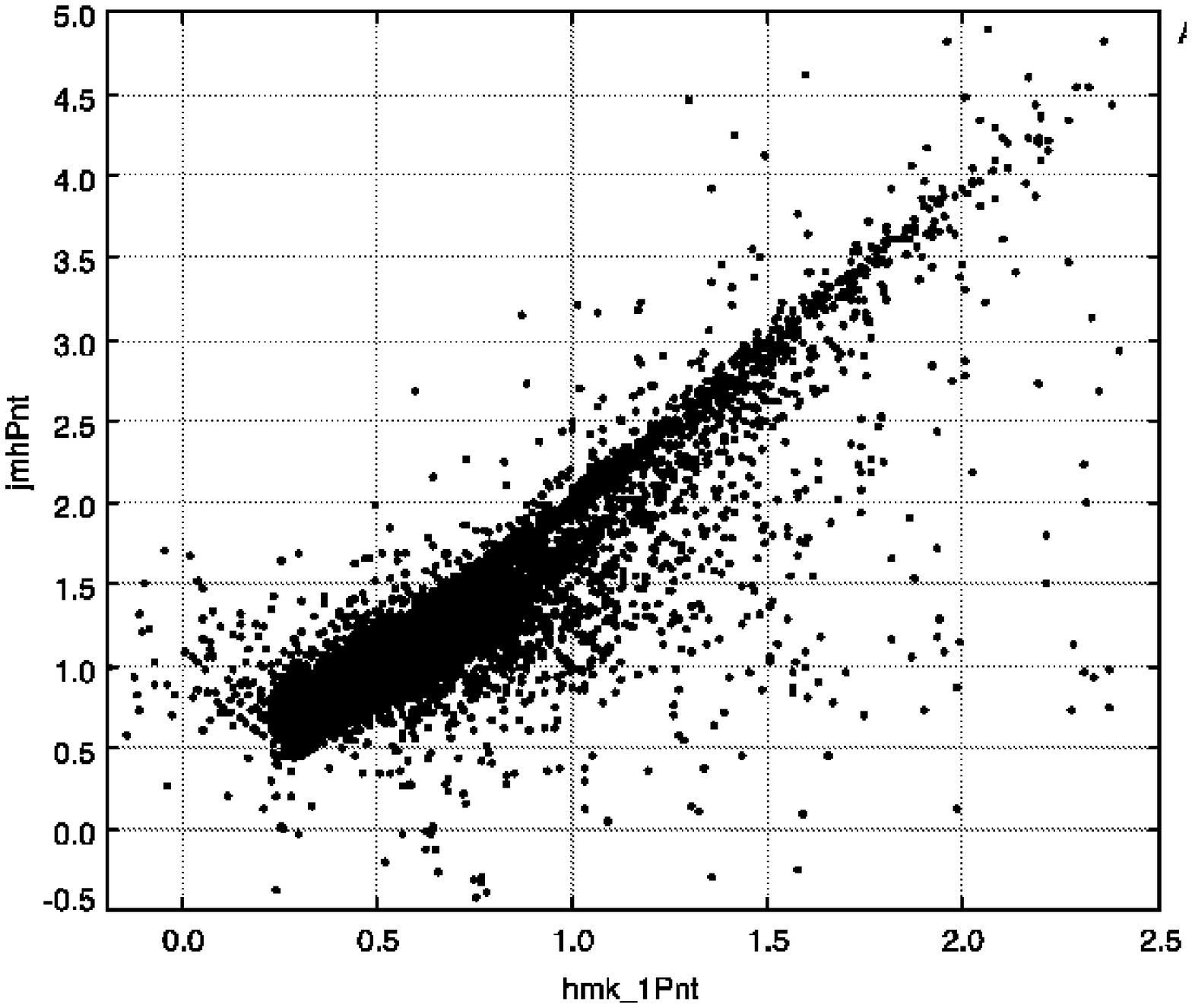}}

\put(0,0){\includegraphics{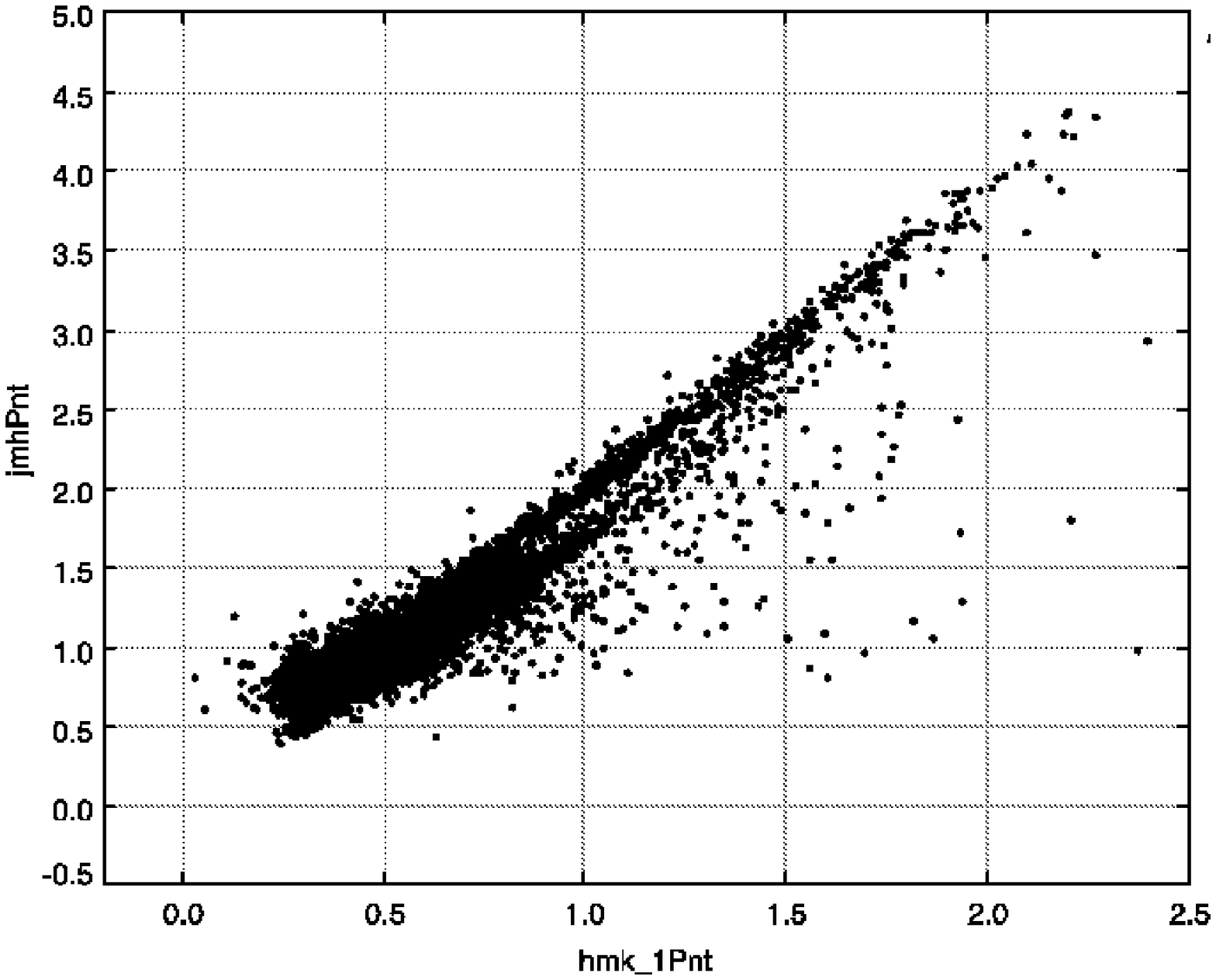}}

\put(0,0){\includegraphics{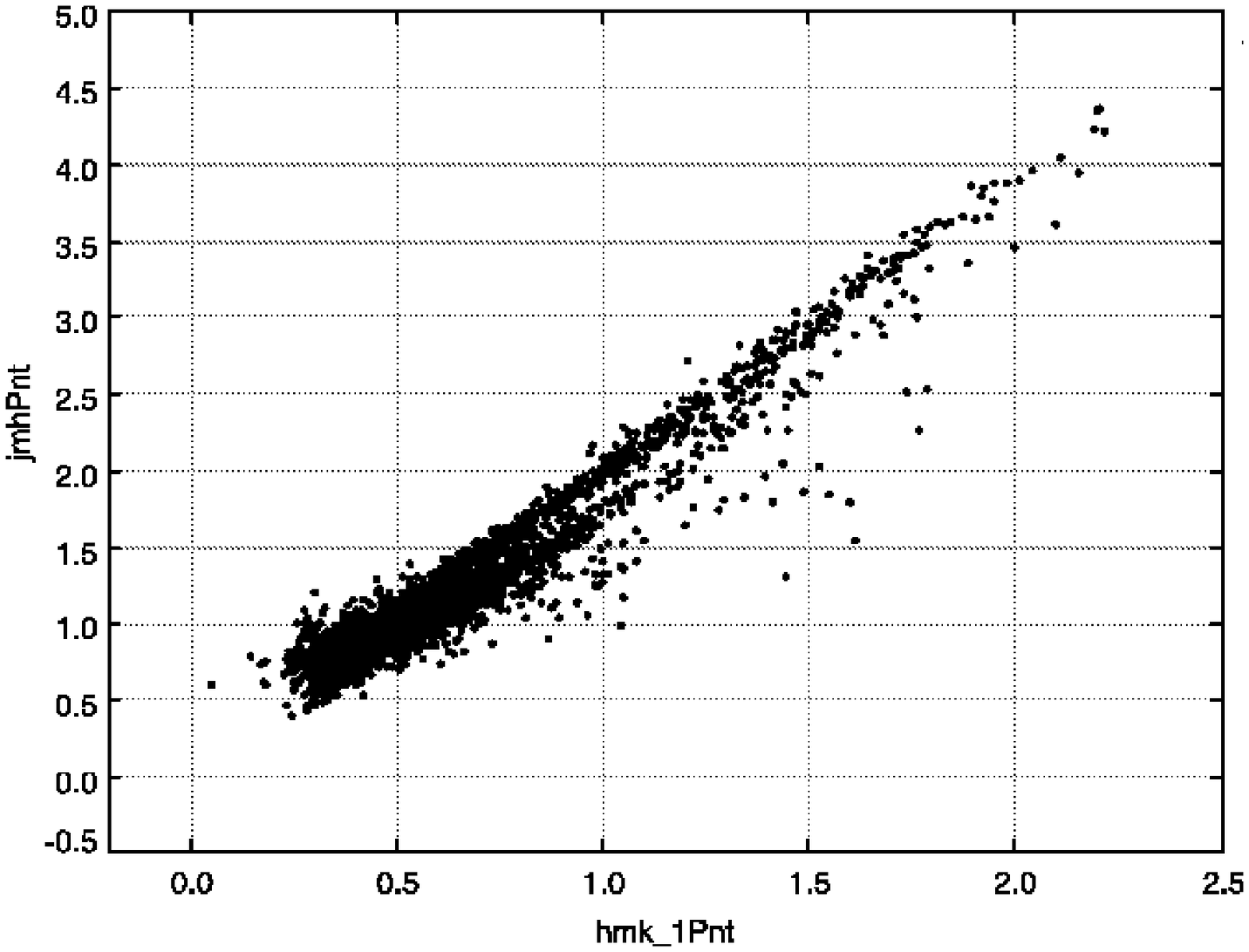}}

\put(-110,220){(a)}
\put(150,220){(b)}
\put(-110,20){(c)}
\put(150,20){(d)}

\end{picture}
\end{center}
\vspace{5.4cm}
\caption{{\it (J-H) vs.(H-K)} two colour diagrams for a 0.04 deg$^2$ area  at ($l,b$)=(31,0) illustrate the trade 
off between completeness and reliability. 
(a) Complete selection of 51780 sources, excluding only ``noise'' sources. (b) Selection of the 8183 sources 
from plot (a) with errors $<$0.05 mag in each colour. (c) Selection of 5914 sources with colour errors 
$<$0.05 mag, stellar image profiles, and more reliable photometry. (d) Selection of 2787 sources with 
colour errors $<$0.05 mag, stellar image profiles, and the most reliable photometry.}
\end{figure*}

\subsection{Optimising the results of GPS catalogue searches}

Here we present three different choices of catalogue parameters that are likely to be useful
to generate different types of source list, ranging from a complete but unreliable selection of the
stellar sources in a region to a highly reliable but seriously incomplete selection. A fourth
option is to use {\bf reliableGpsPointSource} in a Freeform SQL query instead of {\bf gpsSource}.
This ``VIEW'' of the source table is simpler than the methods described below but less effective. Details
of its contents are available under ``Schema Browser'' at the WSA.

Generating a complete list of unique Galactic sources is a little more complicated than simply returning all
sources in a region of the sky. This is for two reasons. (1) {\bf gpsSource} contains sources classified as
``noise'', via the attribute ``{\ssq mergedClass} =0'', in addition to the more numerous real sources 
which are variously classified as stars ({\ssq mergedClass} =$-1$), probable stars ({\ssq mergedClass} =$-2$), 
galaxies ({\ssq mergedClass}=$+1$) and probable galaxies ({\ssq mergedClass} =$-3$), on the basis of the
image profile. The UKIDSS philosophy is to include in the archive all data that might be useful. However,
we recommend that sources classified as noise be excluded from GPS searches, since most of
them are not real sources, and they would inflate the star counts in any given field by $\sim$10\%.  In 
general this will lead to a more accurate star count in any given field than a search on all types of source,
at the price of excluding only a very small number of real sources. 
By contrast, it is necessary to include ``galaxies'' and ``probable galaxies'' in a complete search, since many of 
these (in fact most of them in crowded fields) are marginally resolved pairs of stars which have been 
mis-classified in {\bf gpsSource}.
(2) Sources located in the regions of overlap between different arrays in a WFCAM tile are duplicated
in the WSA. These duplicates can be identified and removed using the {\it PriOrSec} attribute, which describes
Primary or Secondary detections (see Hambly et al.2008). This consideration also affects $\sim$10\% 
of sources.

An illustrative SQL query to return a selection of useful attributes for the complete list of real unique 
sources located near $l=31$, $b=0$ is given below. Comment lines in the query begin with
``/*'' and end with ``*/''.\\

\noindent
\ssq
{\sc select} sourceID, ra, dec, l, b, jmhPnt,\\ 
\indent jmhPntErr, hmk\_1Pnt, hmk\_1PntErr,\\
\indent mergedClass, pStar, jAperMag3, \\ 
\indent jAperMag3Err, hAperMag3, hAperMag3Err,\\
\indent  k\_1AperMag3,k\_1AperMag3Err\\
{\sc from} gpsSource\\
{\sc where} $l$ {\sc between} 30.9 {\sc and} 31.1\\
\indent {\sc and} $b$ {\sc between} -0.1 {\sc and} 0.1\\
/* Exclude detections classified as noise*/ \\
\indent {\sc and} mergedClass != 0\\
/* Exclude multiple detections of the same source /* \\
\indent {\sc and} (PriOrSec=0 {\sc or} PriOrSec=framesetID)\\
\rm

It should be noted that this query must be typed or copied into the Freeform SQL interface, since the last
line cannot be entered with the simple mathematical operators available in the Menu Query interface. 
The two colour diagram for the complete search at these coordinates is shown in figure A1(a).

It is frequently desirable to restrict the sample to sources with small photometric errors, in order 
to identify the main stellar populations more easily in two colour and colour magnitude diagrams.
A restricted search can be done entirely with the WSA but since the user will often wish to vary the search
parameters while investigating the data it is often preferable to download the results of a complete search
and then alter the parameters interactively. This can be done by loading the data into a suitable software
package. We strongly recommend {\sc TOPCAT}, a user friendly software package designed to manipulate and display
data tables written in a variety of Virtual Observatory and text formats.\footnote{{\sc TOPCAT} is available 
from www.starlink.ac.uk/topcat}. 

The following SQL query returns all the parameters needed to construct a more reliable but less complete
photometric sample:\\

\noindent
\ssq
{\sc select} sourceID, ra, dec, l, b, jmhPnt,\\ 
\indent jmhPntErr, hmk\_1Pnt, hmk\_1PntErr, \\
\indent mergedClass, pStar, jAperMag3, \\
\indent jAperMag3Err, jEll, jppErrBits, jXi, jEta, \\
\indent hAperMag3, hAperMag3Err, hEll, hppErrBits, \\
\indent hXi, hEta, k\_1AperMag3, k\_1AperMag3Err, \\
\indent k\_1Ell, k\_1ppErrBits, k\_1Xi, k\_1Eta\\
{\sc from} gpsSource\\
{\sc where} $l$ {\sc between} 30.9 {\sc and} 31.1\\
\indent {\sc and} $b$ {\sc between} -0.1 {\sc and} 0.1\\
/* Exclude noise detections */ \\
\indent {\sc and} mergedClass != 0\\
/* Exclude multiple detections of the same source /* \\
\indent {\sc and} (PriOrSec=0 {\sc or} PriOrSec=framesetID)\\
\rm

Figure A1(b) is the two colour diagram for the subset of this download with errors $<0.05$~mag on each axis.
This was defined in {\sc TOPCAT} by the expression  ``abs({\ssq jmhPntErr})$<$0.05 \&\& abs({\ssq hmk\_1PntErr})$<$0.05'', 
where ``abs'' is the modulus operator and ``\&\&'' is the logical operator AND. (We note that column numbers can
be used in {\sc TOPCAT} instead of parameter names for brevity). The majority of the sources are 
located in a band running
from the lower left to the upper right, caused by the reddening of the main sequence and giant star
populations. However, a large number of photometric outliers are apparent and some sources lie outside the 
boundaries of the plot. A smaller subsample with better photometry and $>50\%$ completeness can be selected 
from the download with the following additional cuts (in {\sc TOPCAT} format):\\

\noindent
\ssq
pstar $>$0.99 \&\& jppErrbits $<$256 \&\& hppErrbits $<$256 \&\& k\_1ppErrbits $<$256 \\
\rm

The ``{\ssq pstar}'' parameter is the WSA estimate of 
the probability that the source is a star, as opposed to a galaxy, or noise. {\ssq pstar}$=$0.9 is the minimum value
for a source to be classified as a star ({\ssq mergedClass}=-1), as opposed to e.g. a probable star or galaxy.
In practice this selection helps to remove close stellar pairs that are resolved in only one or two of the three 
passbands. The ``{\ssq ppErrbits}'' parameters in each waveband can be used to identify sources which are either cross 
talk artifacts (see A2) or stars with less reliable photometry due to saturation, deblending, bad pixels or 
location at the extreme edge of the detector. The {\ssq ppErrbits} parameters are 16 byte integers (values
from 0 to 65535). The least significant byte ({\ssq ppErrbits}$<$256) contains information on minor issues
that are unlikely to greatly affect the photometry: bad pixels within the 2 arcsec {\ssq Apermag3} aperture
and whether the sources is deblended. The more signifcant bytes are populated (i.e.{\ssq ppErrbits}$\ge$256) if 
there are more serious issues such as saturation, cross talk artefacts, or loaction within one jitter position
of an array boundary. A fuller description is given in the WSA.
We note that the alternative ``{\ssq psat}'' parameter for saturation
probability is not functional at present in the WSA. The corresponding plot in figure A1(c) is much cleaner: most of 
the unresolved stellar pairs have been removed by the {\ssq pstar} selection and numerous saturated stars and their 
associated
spurious multiple detections have been removed by the {\ssq ppErrbits} selection. In this case the completeness
of the sample is 72\%, relative to figure A1(b). The vertical separation between the reddened populations of
dwarfs and giant is now readily apparent at {\it(J-H)}$>$1.75. The giants have $\sim$0.25 mag larger {\it(J-H)} 
colours than the dwarfs.

A sample with still more reliable photometry, but with lower completeness, can be defined as follows:\\ 

\noindent
\ssq
pstar $>0.99$ \&\& jppErrbits $<256$ \&\& hppErrbits $<256$ \&\& k\_1ppErrbits $<256$ \&\& \\
sqrt(hXi*hXi + hEta*hEta)<0.3 \&\& sqrt(k\_1Xi*k\_1Xi + k\_1Eta*k\_1Eta)<0.3 \&\& jEll<0.2 \&\& hEll<0.2 \&\& k\_1Ell<0.2
\rm \\

This is our ``most reliable, low completeness'' selection used in the two colour diagrams earlier in $\S$3.
The completeness of this selection is also typically $>50\%$ in the less crowded parts of the GPS but is usually
only $25$-$50\%$ for fields near the mid-plane in the first quadrant.
The ``{\ssq Ell}'' parameters give the ellipticity, which can be used to remove unresolved stellar pairs in a similar 
manner to {\ssq pstar}. The ``Xi'' and ``Eta'' parameters are the distances in arcseconds of each detection from the 
shortest wavelength detection, in RA and Dec respectively. The maximum pair matching radius is 1 arcsec for the 
GPS, but since the typical separation of genuine matches is $\la$0.2 arcsec, these parameters can also be used to 
eliminate unresolved binaries and other false matches. Figure A1(d) shows the corresponding plot, which
is now only 34\% complete relative to figure A1(b). Only a few sources with poor photometry now remain.


{}


\begin{thebibliography}{}

\bibitem[\protect\citename{Allen et al. }2004]{allen04}Allen, L.E., Calvet N., D'Alessio P., Merin B., Hartmann L., 
Megeath S.T., Gutermuth R.A., Muzerolle J., Pipher J.L., Myers P.C., Fazio G.G., 2004, ApJS, 154, 363

\bibitem[\protect\citename{Bally \& Devine }1994]{bally05} Bally J., Devine D., 1994, ApJ, 428, L65

\bibitem[\protect\citename{Bandyopadhyay et al. }2005]{band05} Bandyopadhyay R.M., Miller-Jones J.C.A., 
Blundell K.M., Bauer F.E., Podsiadlowski Ph., Gosling A.J., Wang Q.D., Pfahl E., Rappaport S., 2005, MNRAS, 364, 1195

\bibitem[\protect\citename{Benjamin et al. }2003]{benjamin03}Benjamin R.A., Churchwell E., Babler B.L., Bania T.M., 
Clemens D.P., Cohen M., Dickey J.M., Indebetouw R, Jackson J.M., Kobulnicky H.A., Kobulnicky H.A., Lazarian A., Marston A.P., 
Mathis J.S., Meade M.R., Seager S., Stolovy S.R., Watson C., Whitney B.A., Wolff M.J., Wolfire M.G., 2003, PASP, 115, 953

\bibitem[\protect\citename{Bica et al. }2003]{bica03}Bica E., Dutra C.M., Soares J., Barbuy B., 2003, A\&A, 404, 223

\bibitem[\protect\citename{Bonifacio et al. }2000]{boni00}Bonifacio P., Monai S., Beers T.C., 2000, AJ, 120, 2065

\bibitem[\protect\citename{Cardelli, Clayton \& Mathis } 1989]{cardelli89}Cardelli J.A., Clayton G.C., Mathis J.S., 
1989, ApJ 345, 245

\bibitem[\protect\citename{Casali et al. }2007]{casali07}Casali M., Adamson A., Alves de Oliveira C., and 27 coauthors,
2007, A\&A, 467, 777

\bibitem[\protect\citename{Churchwell et al. }2004]{churchwell04}Churchwell E., Whitney B.A., Babler B.L., 
Indebetouw R., Meade M.R., Watson C., Wolff M.J., Wolfire, M.G. and 15 coauthors, 2004, ApJS, 154, 322

\bibitem[\protect\citename{Devine }1997]{devine97}Devine D.H., 1997, PhD Thesis, University of Colorado at Boulder 

\bibitem[\protect\citename{Dressler et al. }1987]{dressler87}Dressler A., Faber S.M., Burstein D., Davies, R.L., 
Lynden-Bell D., Terlevich R.J., Wegner G., 1987, ApJ, 313, L37

\bibitem[\protect\citename{Drew et al. }2005]{drew05}Drew J.E., Greimel R., Irwin M.J., Aungwerojwit A., Barlow M.J., 
Corradi R.L.M., Drake J.J., G\"{a}nsicke B.T., Groot P., Hales A., Hopewell E.C., Irwin J., Knigge C., Leisy P., 
Lennon D.J., Mampaso A., Masheder M.R.W., Matsuura M., Morales-Rueda L., Morris R.A.H., Parker Q.A., Phillipps S., 
Rodriguez-Gil P., Roelofs G., Skillen I., Sokoloski J.L., Steeghs D., Unruh Y.C., Viironen K., Vink J.S., 
Walton N.A., Witham A., Wright N., Zijlstra A.A., Zurita A., MNRAS, 362, 753

\bibitem[\protect\citename{Dye et al. }2006]{dye06}Dye S., Warren S.J., Hambly N.C., Cross N.J.G., Hodgkin S.T., 
Irwin M.J., Lawrence A.; Adamson A.J. and 37 co-authors, 2006, MNRAS, 372, 1227

\bibitem[\protect\citename{Eisenhauer et al. }2005]{eisenhauer05}Eisenhauer F., Genzel R., Alexander T., Abuter R., 
Paumard T., Ott T., Gilbert A., Gillessen S., Horrobin M., Trippe S. and 11 coauthors, 2005, ApJ, 628, 246

\bibitem[\protect\citename{G\'{o}mez et al. }2003]{gomez03}G\'{o}mez M., Stark D.P., Whitney B.A., Churchwell E.,
2003, AJ, 126, 863

\bibitem[\protect\citename{Gonzalez-Solares et al. }2008]{gonz08}Gonz\'{a}lez-Solares E., Walton N., Greimel R., Drew, J.,
Irwin M., Sale S., \& 35 co-authors, 2008, MNRAS, 388, 89

\bibitem[\protect\citename{Gosling et al. }2006]{gosling06} Gosling, A.J., Blundell, K.M., Bandyopadhyay, R., 2006, 
ApJ, 640, L171

\bibitem[\protect\citename{Gullieuszik et al. }2007]{gullieuszik07}Gullieuszik M., Held E.V., Rizzi L., Saviane I., 
Momany Y., Ortolani S., 2007, A\&A, 467, 1025

\bibitem[\protect\citename{Gutermuth }2005]{gutermuth05}Gutermuth R.A., 2005, PhD thesis, University of Rochester

\bibitem[\protect\citename{Flaherty et al. }2007]{flaherty07}Flaherty K.M., Pipher J.L., Megeath S.T., Winston E.M., 
Gutermuth R.A., Muzerolle J., Allen L.E., Fazio G.G., 2007, ApJ, 663, 1069

\bibitem[\protect\citename{Hambly N. et al.}2007]{hambly08}Hambly N.C., Collins R.S., Cross N.J.G., Mann R.G., Read M.A., 
Sutorius E.T.W., Bond I.A., Bryant J., Emerson J.P., Lawrence A., Stewart J.M., Williams P.M., Adamson A., Dye S., 
Hirst P., Warren S.J., 2008, MNRAS, 384, 637.

\bibitem[\protect\citename{Hewett et al. }2006]{hewett06}Hewett P.C., Warren S.J., Leggett S.K., Hodgkin S.T., 2006,
MNRAS, 367, 454

\bibitem[\protect\citename{Hodgkin et al. }2008]{hod08}Hodgkin S., et al., 2008, MNRAS, submitted

\bibitem[\protect\citename{Indebetouw }2005]{indeb05}Indebetouw R., Mathis J.S., Babler B.L., Meade M.R., Watson C., 
Whitney B.A., Wolff M.J., Wolfire M.G., Cohen M., Bania T.M., Benjamin R.A., Clemens D.P., Dickey J.M., Jackson J.M., 
Kobulnicky H.A., Marston A.P., Mercer E.P., Stauffer J.R., Stolovy S.R., Churchwell E., 2005, ApJ, 619, 931

\bibitem[\protect\citename{Irwin }2005]{irwin05}Irwin M., 2005, on-line, www.ast.cam.ac.uk/vdfs/docs/catalogues.pdf

\bibitem[\protect\citename{Irwin }2005]{irwin05}Irwin M., et al., 2008, MNRAS, submitted

\bibitem[\protect\citename{Jarret }2000]{jarrett07}Jarrett T.-H., Chester T., Cutri R., Schneider S., Rosenberg J.,
Huchra J.P. Mader J., 2000, AJ, 120, 298

\bibitem[\protect\citename{Jiang et al. }2002]{jiang02}Jiang Z., Yao Y., Yang J., Ando M., Kato D., Kawai T., Kurita M., 
Nagata T., Nagayama T., Nakajima Y., Nagashima C., Sato S., Tamura M., Nakaya H., Sugitani K., 2002, ApJ 577, 245

\bibitem[\protect\citename{Kaas }1999]{kaas99}Kaas A.A., 1999, AJ, 118, 558

\bibitem[\protect\citename{Khanzadyan et al. }2004]{khan04}Khanzadyan T., Gredel R., Smith M.D., Stanke T., 2004,
A\&A, 426, 171

\bibitem[\protect\citename{Kocevski et al. }2007]{kocevski07}Kocevski D.D., Ebeling H., Mullis C.R., Tully R.B.,
2007, ApJ, 662, 224

\bibitem[\protect\citename{Kolatt et al. }1995]{kolatt95}Kolatt T., Dekel A., Lahav O., 1995, MNRAS, 275, 797

\bibitem[\protect\citename{Kraan-Korteweg \& Lahav }2000]{kraan00}Kraan-Korteweg R.C., Lahav O., 2000, Astron. Astr. Review, 
10, 211

\bibitem[\protect\citename{Kraan-Korteweg et al. }2005]{kraan05}Kraan-Korteweg R.C., Jarrett, T.H. 2005, in Nearby 
Large-Scale Structures and the Zone of Avoidance, ASP Conf. ~Ser.\ vol. 329 (ASP: San Francisco), 119, eds. K.P. Fairall
\& P.A. Woudt

\bibitem[\protect\citename{Lada }1985]{lada84}Lada C.J., Wilking B.A., 1984, ApJ, 287, 610

\bibitem[\protect\citename{Lawrence et al. }2007]{lawrence07}Lawrence A., Warren S.J., Almaini O., Edge A.C.
Hambly N.C., Jameson R.F., Lucas P., Casali M., Adamson A., Dye S., Emerson J.P., Foucaud S., Hewett P., 
Hirst P., Hodgkin S.T., Irwin M.J., Lodieu N., McMahon R.G., Simpson C., Smail I., Mortlock D., Folger M.,
2007, MNRAS, 379, 1599

\bibitem[\protect\citename{Liu \& Li }2006]{liu06}Liu  X.-W., Li X.-D., 2006, A\&A, 449, 135

\bibitem[\protect\citename{Lockman }1989]{lockman89}Lockman F.J., 1989, ApJS, 71, 469

\bibitem[\protect\citename{Lopez-Corredoira et al. }2002]{lopez02}L\'{o}pez-Corredoira M., Cabrera-Lavers A., Garz\'{o}n F., 
Hammersley P.L., 2002, A\&A, 394, 883

\bibitem[\protect\citename{Luck et al. }2007]{luck07}Luck, R.E., Heiter U., 2007, AJ, 133, 2464

\bibitem[\protect\citename{Mathis} 1990]{mathis90}Mathis J.S., 1990, Ann. Rev. A\&A 28, 37

\bibitem[\protect\citename{Megeath et al. }2004]{megeath04}Megeath S.T., Allen L.E., Gutermuth R.A., Pipher J.L., Myers P.C., 
Calvet N., Hartmann L., Muzerolle J., Fazio G.G., 2004, ApJS, 154, 367.

\bibitem[\protect\citename{Meyer et al. }1997]{meyer97}Meyer M.R., Calvet N., Hillenbrand L.A., 1997, AJ, 114, 288

\bibitem[\protect\citename{Moore et al. }2005]{moore05}Moore T.J.T., Lumsden S.L., Ridge N.A., Puxley P.J., 2005, 
MNRAS, 359, 589

\bibitem[\protect\citename{Muno et al. }2003]{muno03}Muno M.P., Baganoff F.K., Bautz M.W., Brandt W.N., Broos P.S.,
Feigelson E.D., Garmire G.P., Morris M.R., Ricker G.R., Townsley L.K., 2003, ApJ, 589, 225

\bibitem[\protect\citename{Muno et al. }2006]{muno06}Muno M.P., Bauer F.E., Bandyopadhyay R.M., Wang Q.D., 
2006, ApJS, 165, 173

\bibitem[\protect\citename{Naoi et al. }2006]{naoi06}Naoi T., Tamura M., Nakajima Y., Nagata T., Suto H., Murakawa K., 
Kandori R., Sasaki S., Baba D., Kato D., Kurita M., Nagashima C., Nagayama T., Nakaya H., Nishiyama S., Oasa Y., Sato S., 
Sugitani K., 2006, ApJ 640, 373

\bibitem[\protect\citename{Ojha et al. }2000]{ojha00}Ojha D.K., Omont A., Ganesh S., Simon G., Schultheis M., 2000,
J. Astrophys. Astron., 21, 77

\bibitem[\protect\citename{Pfahl et al. }2002]{pfah02}Pfahl E., Rappaport  S., Podsiadlowski, P., 2002, ApJL, 571, L37

\bibitem[\protect\citename{Racca et al. }2002]{racca02}Racca G., G\'{o}mez, M., Kenyon S.J., 2002, AJ, 124, 2178

\bibitem[\protect\citename{Rieke \& Lebofsky }1985]{rieke85}Rieke G.H., Lebofsky M.J., 1985, ApJ, 288, 618

\bibitem[\protect\citename{Robin et al. }1989]{robin89}Robin A.C., 1989, A\&A, 225, 69

\bibitem[\protect\citename{Robin et al. }2003]{robin03}Robin A.C., Reyl\'{e} C., Derri\`{e}re S., Picaud S., 2003,
A\&A, 409, 523

\bibitem[\protect\citename{Rousseau et al.}2000]{rouss00}Rousseau J., Paturel, G., Vauglin I., Schr\"{o}der A., de Batz B.,
Borsenberger J., Epchtein N., Fouqu\'{e} P., Kimeswenger S., Lacombe F., Le Bertre, T., Mamon, G., Rouan, D., Simon, G., 
Tiph\`ene, D. 2000, A\&A 363, 62

\bibitem[\protect\citename{Ruiter et al. }2006]{ruit06}Ruiter A.J., Belczynski K., Harrison T.E., 2006, ApJL, 640, L167

\bibitem[\protect\citename{Salaris \& Girardi }2002]{salaris02}Salaris M., Girardi L., 2002, MNRAS, 337, 332

\bibitem[\protect\citename{Schlegel et al. }1998]{schlegel98}Schlegel D.J., Finkbeiner D.P., Davis M., 1998, ApJ, 500, 525

\bibitem[\protect\citename{Schr\"{o}der et al. }2007]{schroeder07}Schr\"{o}der A.C., Mamon G.A., Kraan-Korteweg R.C., 
Woudt P.A., 2007, ApJ, 466, 481

\bibitem[\protect\citename{Shah et al. }2004]{shah04}Shah R.Y., Rathborne J., Simon R., Jackson J.M., Bania T.M., 
Clemens D.P., Johnson A.M., Flynn E.S., 2004, in ``Milky Way Surveys'', ASP conf. series, vol. 317, p103-105.

\bibitem[\protect\citename{Skrutskie et al.} 2006]{skrutskie06} Skrutskie M.F., Cutri R.M., Stiening R., Weinberg M.D., 
Schneider S., Carpenter J.M., and 25 coauthors, 2006, AJ, 131, 1163

\bibitem[\protect\citename{Ungerechts \& Thaddeus }1987]{ungerechts87}Ungerechts H., Thaddeus P., 1987, ApJS, 63, 645

\bibitem[\protect\citename{Wang et al. } 2002]{wang02} Wang Q.D., Gotthelf E.V., Lang C.C. 2002, Nature, 415, 148

\bibitem[\protect\citename{Warren et al. }2007a]{warren07a}Warren S.J., Hambly N.C., Dye S., Almaini O., Cross N.J.G., 
Edge A.C., Foucaud S., Hewett P.C., Hodgkin S.T., Irwin M.J., Jameson R.F., Lawrence A., Lucas P.W. and 27 co-authors, 2007,
MNRAS, 375, 213

\bibitem[\protect\citename{Warren et al. }2007b]{warren07b}Warren S.J., Cross N.J.G., Dye S., Hambly N.C., Almaini O., 
Edge A.C., Hewett P.C., Hodgkin S.T., Irwin M.J., Jameson R.F., Lawrence A., Lucas P.W., Mortlock D.J.,
and 16 co-authors, 2007, astro-ph/0703037

\bibitem[\protect\citename{Whitney et al. }2004]{whitney04}Whitney B.A., Indebetouw R., Babler, B.L., Meade M.R.,
Watson C., Wolff M.J., Wolfire M.G. and 15 coauthors, 2004, ApJS, 154 315

\bibitem[\protect\citename{Whittet }1990]{whittet90}Whittet D.C.B., 1990, Dust in the Galactic Environment,
IOP Publishing Ltd, London

\bibitem[\protect\citename{Wilking, Greene \& Meyer }1999]{wilking99}Wilking B.A., Greene T.P. \& Meyer M.R., 1999, 
AJ, 117, 469

\bibitem[\protect\citename{Willems \& Kolb }2003]{will03} Willems B., Kolb U., 2003, MNRAS, 343, 949

\bibitem[\protect\citename{Wu et al. }2007]{wu07} Wu  J.F., Zhang, S.N., Lu F.J., Jin Y.K., 2007, 
Chinese Journal of Astronomy and Astrophysics, 7, 81

\end{thebibliography}
\end{document}